\documentclass[a4paper,11pt]{article}
\usepackage{jinstpub} 

\usepackage[mathlines]{lineno}
\usepackage{graphicx}
\usepackage[utf8]{inputenc}
\usepackage{xcolor}
\usepackage{xspace}
\usepackage{ulem} 
\usepackage{hyperref}
\usepackage{amsmath}
\usepackage{accents}
\usepackage{placeins}
\usepackage{siunitx}
\usepackage{tabularx}
\usepackage{booktabs} 
\usepackage{multirow}
\usepackage{enumitem} 
\usepackage[compatibility=false]{caption}
\usepackage{subcaption}
\usepackage{gensymb}
\graphicspath{{graphics/}}

\DeclareSIUnit\clight{\text{\ensuremath{c}}}
\DeclareSIUnit{\atmos}{\text{atm}}
\newcommand{\gfourbeamline}{\textsc{G}4\textsc{beamline}\xspace}
\newcommand{\geantfour}{\textsc{Geant}4\xspace}

                                                                                                                        \newcommand{\m}{$\mu$}

\newcommand{\numu}{\numu}

\newcommand\momentumrange{
\qtyrange{0.4}{1.5}{GeV/c}}

\newcommand\testbeam{Test Beam}
\newcommand\wc{wire chamber}
\newcommand\wcs{wire chambers}
\newcommand\tof{time of flight}
\newcommand\datataking{data-taking}

\definecolor{shutdown}{HTML}{eeeeee}

\let\orgautoref\autoref
\providecommand{\Autoref}[1]
{\def\equationautorefname{Equation}\def\figureautorefname{Figure}\def\subfigureautorefname{Figure}\def\sectionautorefname{Section}\def\subsectionautorefname{Section}\def\tableautorefname{Table}\orgautoref{#1}}
\renewcommand{\autoref}[1]
{\def\equationautorefname{Eq.}\def\figureautorefname{Fig.}\def\subfigureautorefname{Fig.}\def\sectionautorefname{Sec.}\def\subsectionautorefname{Sec.}\orgautoref{#1}}

\title{The NOvA Test Beam Experiment}

\abstract{NOvA is a long-baseline neutrino oscillation experiment designed to study the neutrino mixing parameters, mass ordering, and CP violation in the lepton sector. A key component of the success of the experiment is a robust understanding of the systematic uncertainties associated with detector response and calibration. To address this, NOvA deployed a Test Beam experiment at the Fermilab Test Beam Facility, which collected data from April 2019 through July 2022.
The NOvA Test Beam experiment used a 30-ton segmented liquid scintillator detector functionally identical to the NOvA Near and Far Detectors to analyze tagged particles produced from p--Cu collisions, with instrumentation capable of selecting and identifying electrons, muons, pions, kaons, and protons with momentum ranging from\momentumrange. Analysis of the collected data provides a better understanding of the largest systematic uncertainties impacting NOvA's analyses, which include the detector response, energy calibration, and hadronic and electromagnetic energy resolutions.}

\keywords{Neutrino detectors; Large detector systems for particle and astroparticle physics}

\newcommand{\LErciyes}{a}
\newcommand{\LAtlantico}{b}
\newcommand{\LBandirma}{l}
\newcommand{\LDallasU}{o}
\newcommand{\LImperial}{p}
\newcommand{\LUIowa}{y}
\newcommand{\LFSU}{af}
\newcommand{\LISU}{az}
\newcommand{\LOSU}{ba}
\newcommand{\LFNAL}{d}
\newcommand{\LHomi}{ay}
\newcommand{\LCUSB}{bb}
\newcommand{\LMississippi}{c}
\newcommand{\LJINR}{e}
\newcommand{\LMagdalena}{f}
\newcommand{\LSussex}{g}
\newcommand{\LCincinnati}{h}
\newcommand{\LIndiana}{x}

\newcommand{\LUCL}{ab}
\newcommand{\LVirginia}{ag}
\newcommand{\LIrvine}{i}
\newcommand{\LHyderabad}{j}
\newcommand{\LTufts}{ak}
\newcommand{\LCaltech}{aw}
\newcommand{\LIIT}{ap}
\newcommand{\LPanjab}{m}
\newcommand{\LGuwahati}{n}
\newcommand{\LMinnesota}{k}
\newcommand{\LQMU}{aa}
\newcommand{\LIHyderabad}{q}
\newcommand{\LMSU}{r}
\newcommand{\LCSU}{s}
\newcommand{\LINR}{t}
\newcommand{\LTexas}{u}
\newcommand{\LWisconsin}{v}
\newcommand{\LWandM}{w}
\newcommand{\LDelhi}{z}
\newcommand{\LSMU}{ac}
\newcommand{\LANL}{ad}

\newcommand{\LIOP}{ai}
\newcommand{\LCTU}{as}
\newcommand{\LSAlabama}{aj}
\newcommand{\LPitt}{al}
\newcommand{\LUFG}{am}

\newcommand{\LCarolina}{ar}
\newcommand{\LDuluth}{an}
\newcommand{\LICS}{ao}
\newcommand{\LHouston}{ae}
\newcommand{\LCochin}{at}
\newcommand{\LWSU}{ah}
\newcommand{\LBHU}{av}
\newcommand{\LCharles}{bf}
\newcommand{\LNISER}{ax}
\newcommand{\LSyracuse}{aq}

\author[\LErciyes]{S.~Abubakar}
\author[\LAtlantico]{M.~A.~Acero}
\author[\LMississippi]{B.~Acharya}
\author[\LFNAL]{P.~Adamson}
\author[\LJINR]{N.~Anfimov}
\author[\LJINR]{A.~Antoshkin}
\author[\LMagdalena]{E.~Arrieta-Diaz}
\author[\LSussex]{L.~Asquith}
\author[\LCincinnati]{A.~Aurisano}
\author[\LJINR]{N.~Balashov}
% ...

\collaboration{The NOvA Collaboration}
\date{\today}

\emailAdd{l.asquith@sussex.ac.uk}

\begin{document}
\maketitle
\flushbottom

\section{Introduction}
\label{sec:intro}

The NOvA experiment aims to address such questions as whether the neutrino mass ordering is normal or inverted, whether there may be CP violation in the lepton sector, and whether there is maximal mixing in the $\mu$--$\tau$ neutrino sector, among others. NOvA began data taking in 2014 and is scheduled to run until 2027 when decommissioning of the Neutrinos from the Main Injector (NuMI) beam is required to make way for the construction of the Long-Baseline Neutrino Facility (LBNF). Uncertainties arising from the energy calibration of the detectors are among the larger systematic uncertainties impacting oscillation measurements from NOvA; full understanding of these experimental uncertainties is paramount to ensure these physics measurements are as precise as possible.  

\begin{figure}
\centering
\includegraphics[width=.95\textwidth]{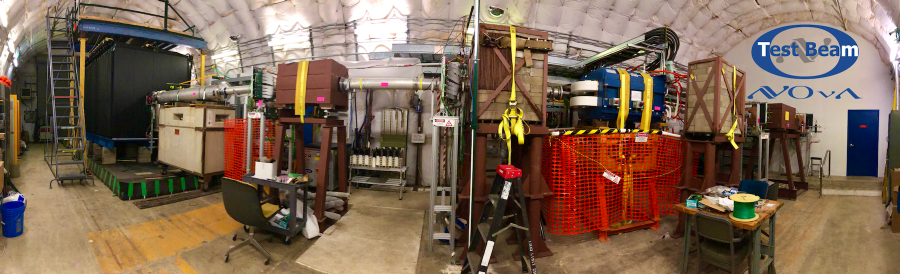}
\caption{Panoramic photograph of the NOvA Test Beam beamline components and detector. Secondary beam particles were delivered to the NOvA target from the right, and the emerging tertiary particles were deflected by the dipole magnet (center-right) and measured by the beamline instrumentation (distributed throughout the center of the image) and by the NOvA Test Beam detector (far left). Detailed discussions of each system and component are provided in the text.}
\label{fig:DetectorAndBeamlinePanorama}
\end{figure}

With this motivation, NOvA developed a \testbeam\ (TB) program at the Fermilab Test Beam Facility (FTBF), using a new tertiary target and instrumentation to tag electrons, muons, pions, kaons, and protons. These particles were measured using a scaled-down NOvA detector functionally identical to NOvA's Near and Far neutrino detectors (ND and FD, sited at Fermilab and in Ash River, MN, respectively)~\cite{NOvA:2007rmc}. This experimental setup, which collected data between 2019 and 2022, enables a detailed understanding of the detectors' electromagnetic and hadronic response and provides real data for the study of particle identification techniques.
\noindent
\autoref{fig:DetectorAndBeamlinePanorama} displays a panoramic image of the NOvA TB experimental setup, showing the components and detectors. 

This paper details the NOvA TB experimental setup, aspects of data collection, and detector calibration using cosmic ray data. The organization is as follows: in \autoref{sec:beamline} 
we discuss beam delivery, starting with Main Injector protons, and the structure of the secondary beamline; in \autoref{sec:tertiary} we describe the secondary target and instrumentation used for particle selection and identification; in \autoref{sec:detector} we describe the TB detector; in \autoref{sec:operations}, we provide an overview of NOvA TB operations and \datataking; in \autoref{sec:calibration} we detail the calibration process and present the results of the dedicated TB calibration; and in \autoref{sec:performance} we summarize the performance of the TB experiment for particle identification and momentum measurement.

\noindent

\section{Beam Delivery from the Main Injector}
\label{sec:beamline}
\noindent
The FTBF \cite{FTBFwebsite} provided particle beams to users for detector R\&D. The facility's beamlines received beam from the Main Injector (MI) rapid cycling synchrotron \cite{MainInjector1994} via the Switchyard transfer line \cite{Switchyard1997, Switchyard2005}. The MI primary beam of \SI{120}{GeV} protons operated at \SI{53.1}{\mega\hertz}, resulting in a separation of \SI{18.83}{\nano\second} between bunches of accelerated particles. Beam was delivered to the facility in batches consisting of 81 such bunches. MI protons collided with the primary \textrm{Cu} target with a typical intensity of $10^9$/spill, where a spill is defined as a directed release of a batch of 81 proton bunches from the Main Injector. The primary target marked the beginning of the secondary beamline, known as the MCenter beamline. 

\begin{figure}[!ht]
    \centering
    \includegraphics[width=1.0\textwidth]{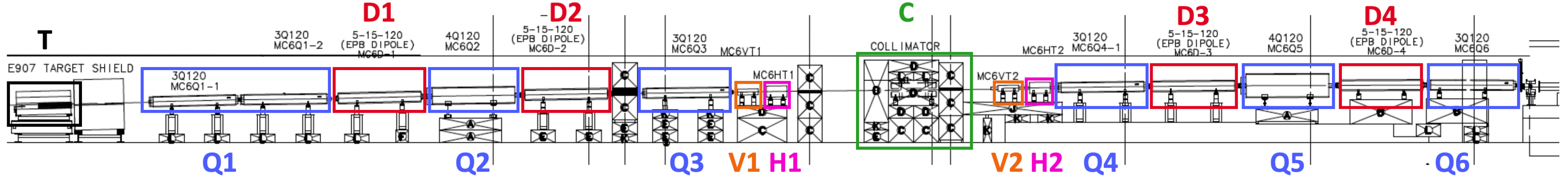}
    \caption{Elevation view of the MCenter secondary beamline. The beam travels from left to right, and the primary target (T) is at the far left of the diagram. The beamline is initially angled slightly upwards, with off-momentum particles absorbed by a variable vertical collimator (C). The four dipoles (D) are used in series to select the desired particle momentum. Beam focusing and trajectory correction are provided by six quadrupoles (Q) and two vertical (V) and two horizontal (H) corrector dipoles.}
    \label{fig:SecondaryBeamline}
\end{figure}

The beam in MCenter consisted predominantly of protons and pions.  The beam momentum could be tuned from \SIrange{8}{64}{GeV/c} in steps of \SI{4}{GeV/c}. The MCenter beam was tuned to the \SI{64}{GeV/c} momentum setting for NOvA TB.
The MCenter beamline comprised four bending dipoles for momentum selection, four corrector dipoles, and two quadrupole triplets; a vertical collimator provided a momentum range of a few percent. A schematic diagram is shown in \autoref{fig:SecondaryBeamline}. The design of the MCenter beamline was driven by the stringent requirements on momentum spread and transverse emittance of its original user, E907, also known as the Main Injector Particle Production experiment (MIPP)~\cite{MIPP}.

\begin{figure}[!ht]
    \centering
    \includegraphics[width=0.9\textwidth]
    {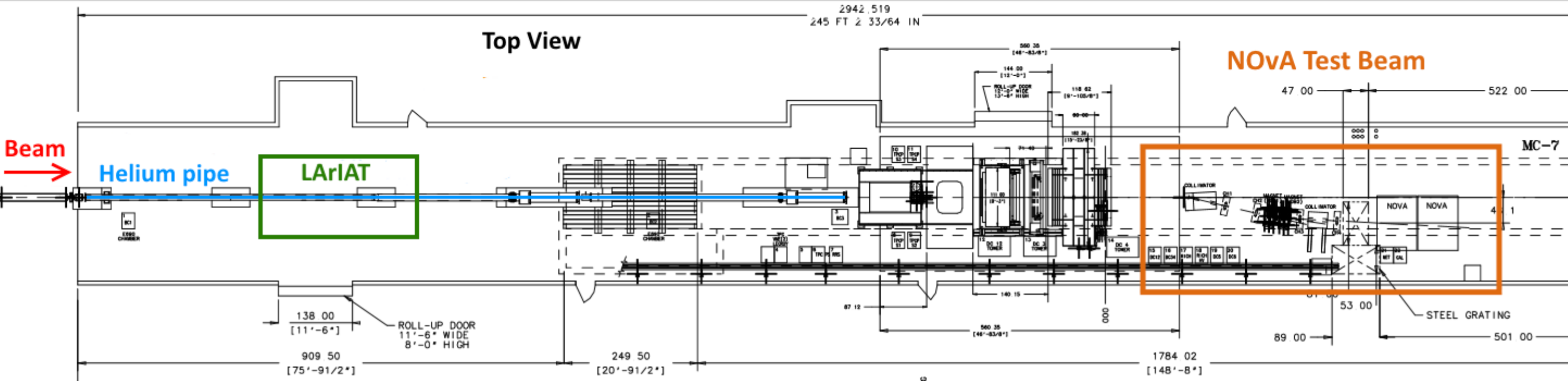}
    \caption{Top view of the MC7 experimental enclosure, showing the helium-filled pipe carrying the secondary beam that travels from left to right, the LArIAT experimental area~\cite{LArIAT:2019kzd}, and the NOvA Test Beam area. The components placed between the LArIAT and NOvA experimental areas were not used. These include the two large magnets, Jolly Green Giant and Rosie, located just upstream of the NOvA Test Beam Hall.} 
    \label{fig:MC7}
\end{figure}

Over the period of \datataking\ for the NOvA TB experiment, beam was delivered to the primary target once per minute over a \SI{4.2}{\second} spill window using MI slow extraction.  Within that spill window, beam particles arrived in \SI{1.6}{\us} batches spaced \SI{11.2}{\us} apart.   The continuation of the MCenter beamline into the MC7 enclosure, where the NOvA TB area was located, is illustrated in \autoref{fig:MC7}.

In the commissioning stage, the NOvA TB detector was found to be subjected to large beam backgrounds, referred to as the ``muon plume," which saturated the readout and data acquisition (DAQ) systems described in \autoref{sec:detdaq}. The NOvA detectors employ continuous-readout electronics that stream data to a buffer to be recorded based on a quasi-online trigger. This was designed for the relatively quiet environment typical of a neutrino experiment. In the Test Beam environment, saturation occurred in the readout system, leading to unavoidable and unrecoverable dead time, compromising the quality of the collected data (\autoref{sec:ops-dataquality}). A full understanding of the muon plume required a detailed secondary beamline simulation described in \autoref{sec:secondary-beamline-sim}. The  mitigation strategy is discussed in \autoref{sec:ops-plume}.

 \subsection{Simulation of the Secondary Beamline}
\label{sec:secondary-beamline-sim}
\noindent
The primary goal of the secondary beamline simulation was to produce qualitative comparisons with data to understand the beam characteristics, particularly the source of the muon plume~\cite{AbhilashThesis}.  To meet this aim, all particles were tracked from the primary target through all the instrumentation and beamline material to the detector. This mode of running required significant computation time and produced large output files, making it unfeasible to produce full Monte Carlo data sets with all beamline components included for physics analyses; alternate techniques were developed for these purposes as discussed in \autoref{sec:tertiary}.

The simulation was built using \gfourbeamline \cite{g4bl} (v3.08) and features a complete model of the beamline elements as documented in engineering specifications and surveys. This is referred to as the ``Full Simulation'' mode in \autoref{sec:tertiary-beamline-sim}. Additional features, such as walls, concrete beamline supports, and leftover parts of the LArIAT experiment, were added to accurately model the material in the MC7 enclosure. \gfourbeamline by default simulates only fields within magnet apertures and not the return fields within the yoke. These yoke returns were a critical addition required for reproducing the effects of the muon plume.  The return fields were added by modeling the magnets in POISSON \cite{superfish}, then including the field maps in the \gfourbeamline model.

 \subsection{Secondary Beamline Background}
\label{sec:ops-plume}
\begin{figure}
    \centering
         \begin{subfigure}[b]{0.48\textwidth}
         \centering
         \includegraphics[width=\textwidth]{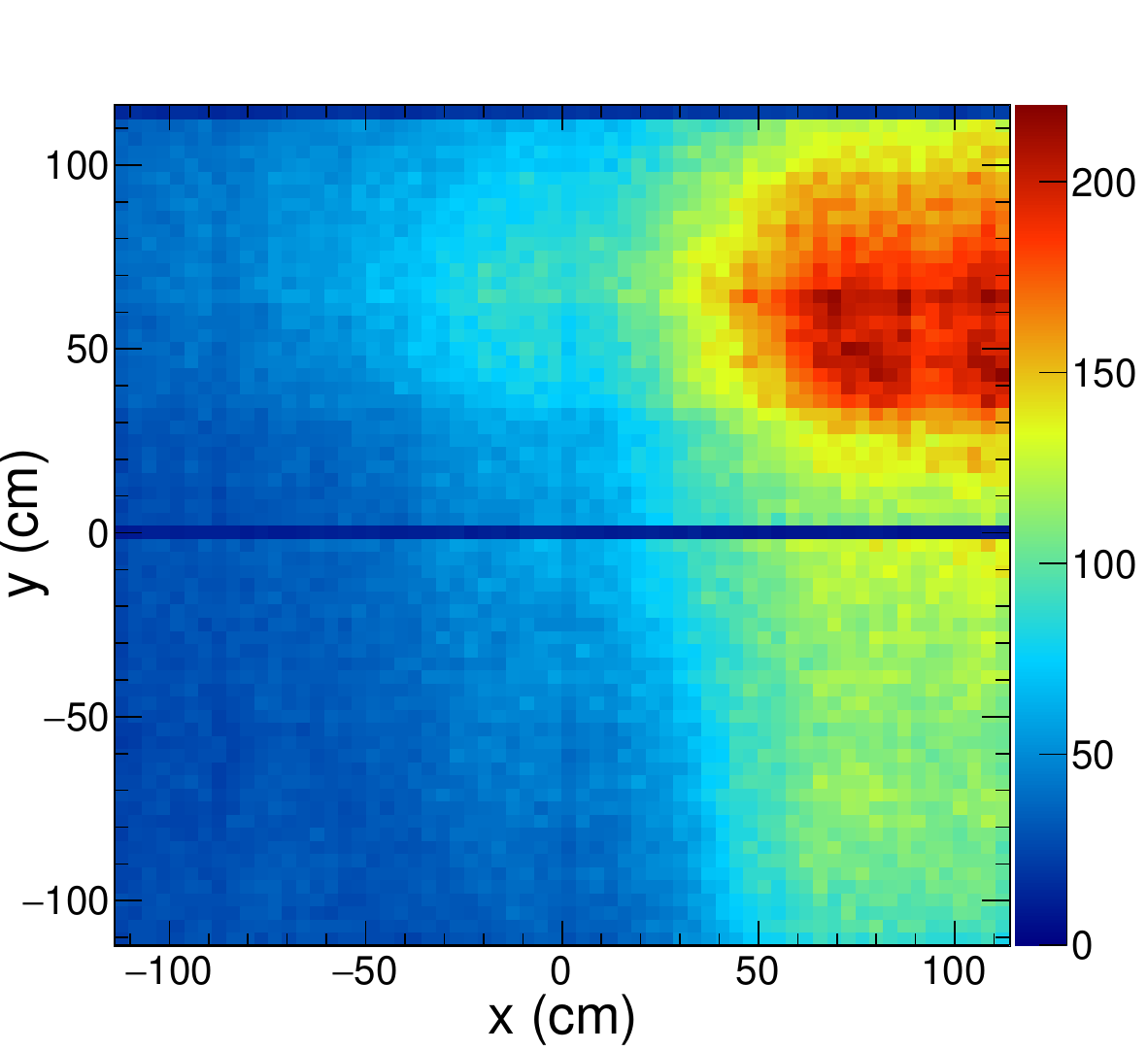}
         \caption{Data without shielding.}
         \label{fig:plume2020}
     \end{subfigure}
     \hfill
     \begin{subfigure}[b]{0.48\textwidth}
         \centering
         \includegraphics[width=\textwidth]{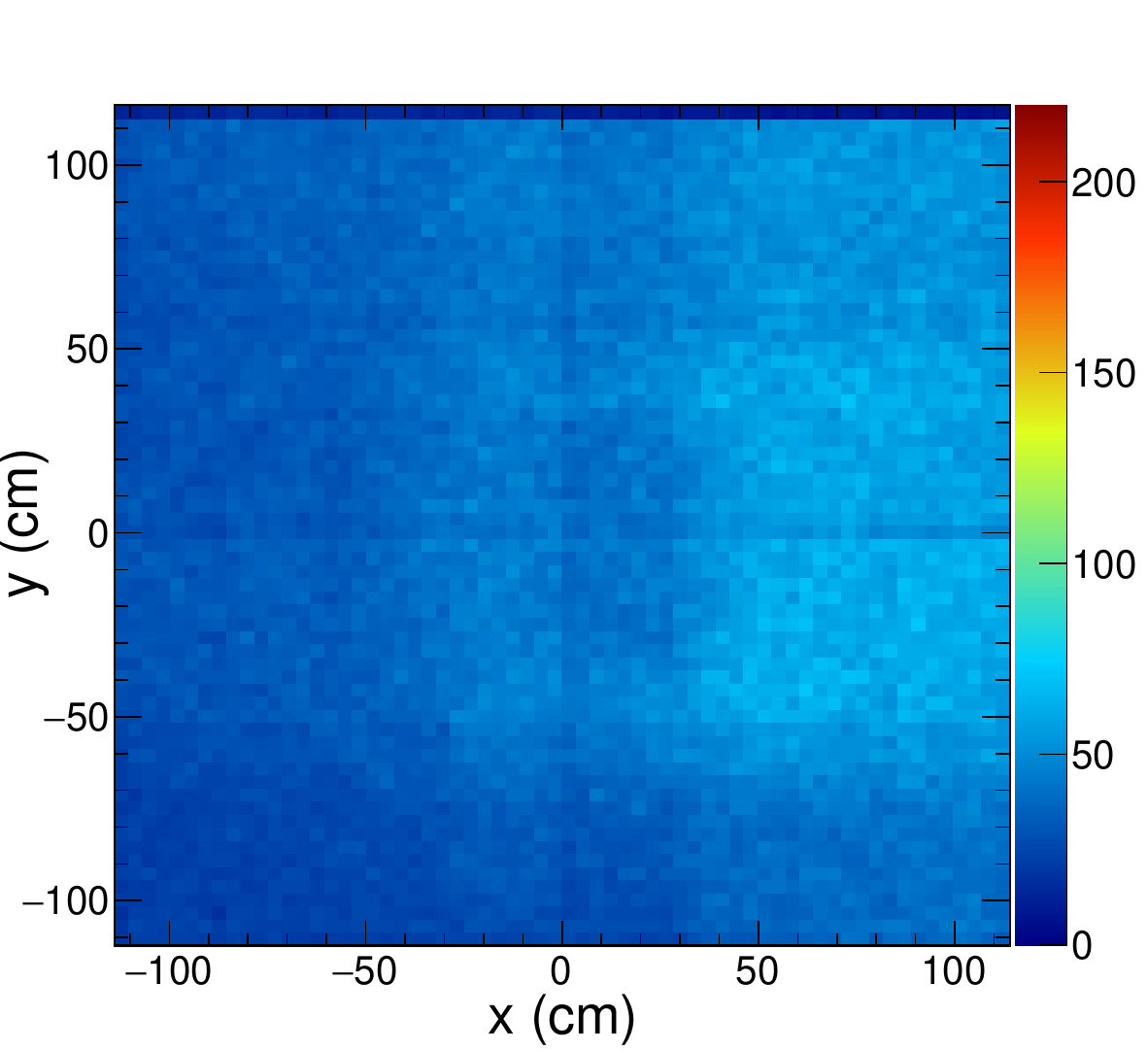}
         \caption{Data with shielding.}
         \label{fig:plume2022}
     \end{subfigure}
    \caption{Profile of the beam on the upstream face of the NOvA TB detector.  The left plot shows the hit rate before the installation of shielding to mitigate the muon plume, while the right plot shows it after. The view is from the detector perspective, with the $z$-axis pointing out of the page (see \autoref{fig:TertiaryModel}). The two low-occupancy horizontal cells evident in the left plot (at the middle and top of the detector) were a result of underfilled detector modules, discussed in \autoref{sec:scint} and remedied prior to the final physics \datataking.  These plots show the profiles for a typical spill during each data collection period: $1\times10^{9}$ protons on the primary target in the left panel and $7\times10^{9}$ in the right panel. They are not normalized by intensity; for comparable beam intensities, the background rates in the shielding case would be even lower.}
\label{fig:PlumeData}
\end{figure}

\noindent
An example profile of the beam measured by the NOvA TB detector is shown in the left panel of \autoref{fig:PlumeData}.  The most striking feature is the intense plume observed on one quadrant of the detector face; however, the backgrounds were high everywhere and impacted data quality. A survey of the enclosure shortly after a beam run found a radiation hotspot at the momentum-selecting collimator in the secondary beamline, pointing to this as the likely source of the plume particles. Scattering of primary particles on material in the collimator resulted in a second beam following a path outside the reference trajectory, interacting further and producing a large level of background downstream. For the profiles in \autoref{fig:PlumeData} the collimator was set at \SI{12}{\milli\meter}, which was a typical setting found to provide a compromise between background shielding and on-axis particle rate.

\begin{figure}
    \centering
     \begin{subfigure}[b]{0.48\textwidth}
         \centering
         \includegraphics[width=\textwidth]{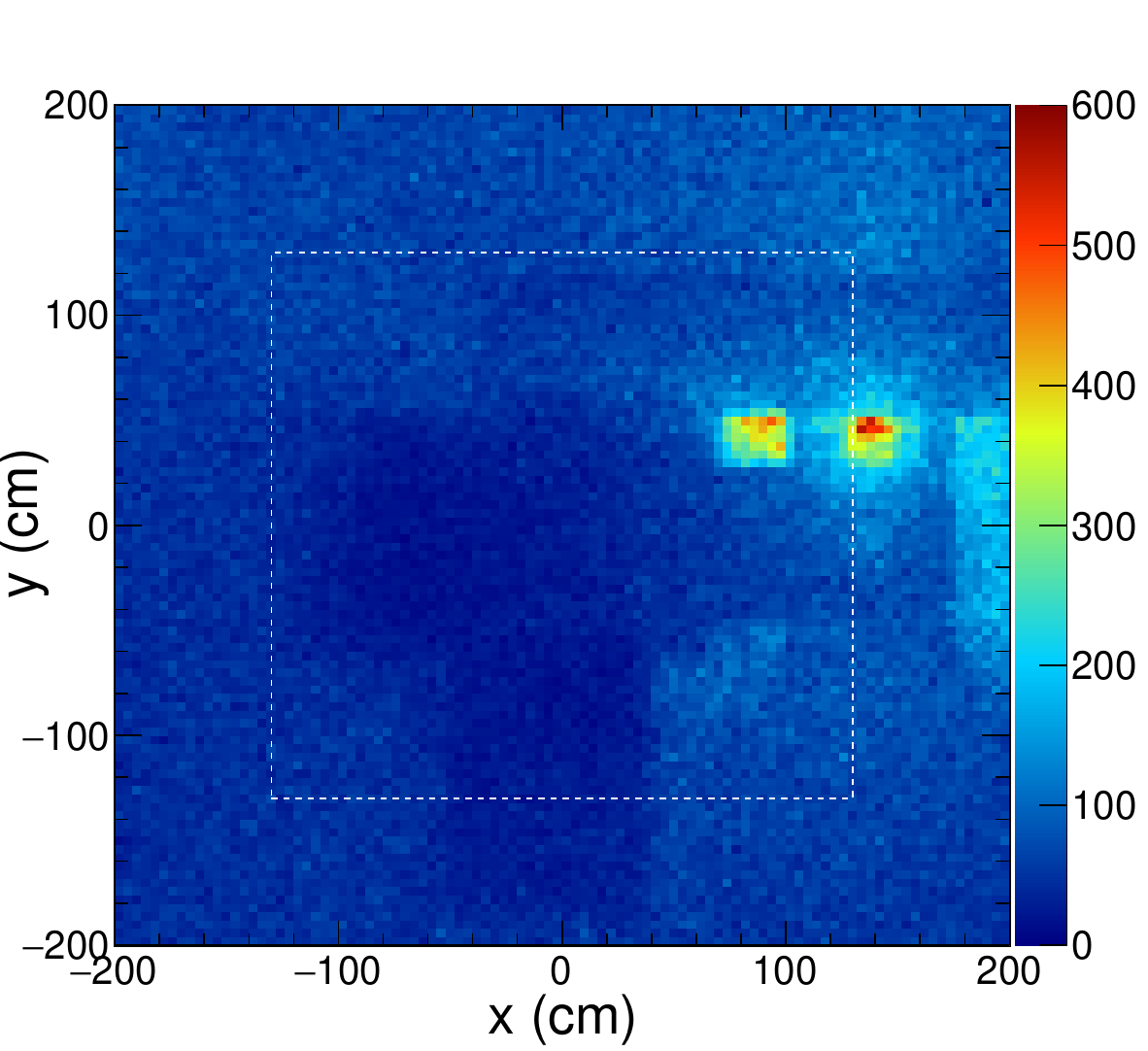}
         \caption{Simulation without shielding.}
         \label{fig:plumesim2020}
     \end{subfigure}
     \hfill
     \begin{subfigure}[b]{0.48\textwidth}
         \centering
         \includegraphics[width=\textwidth]{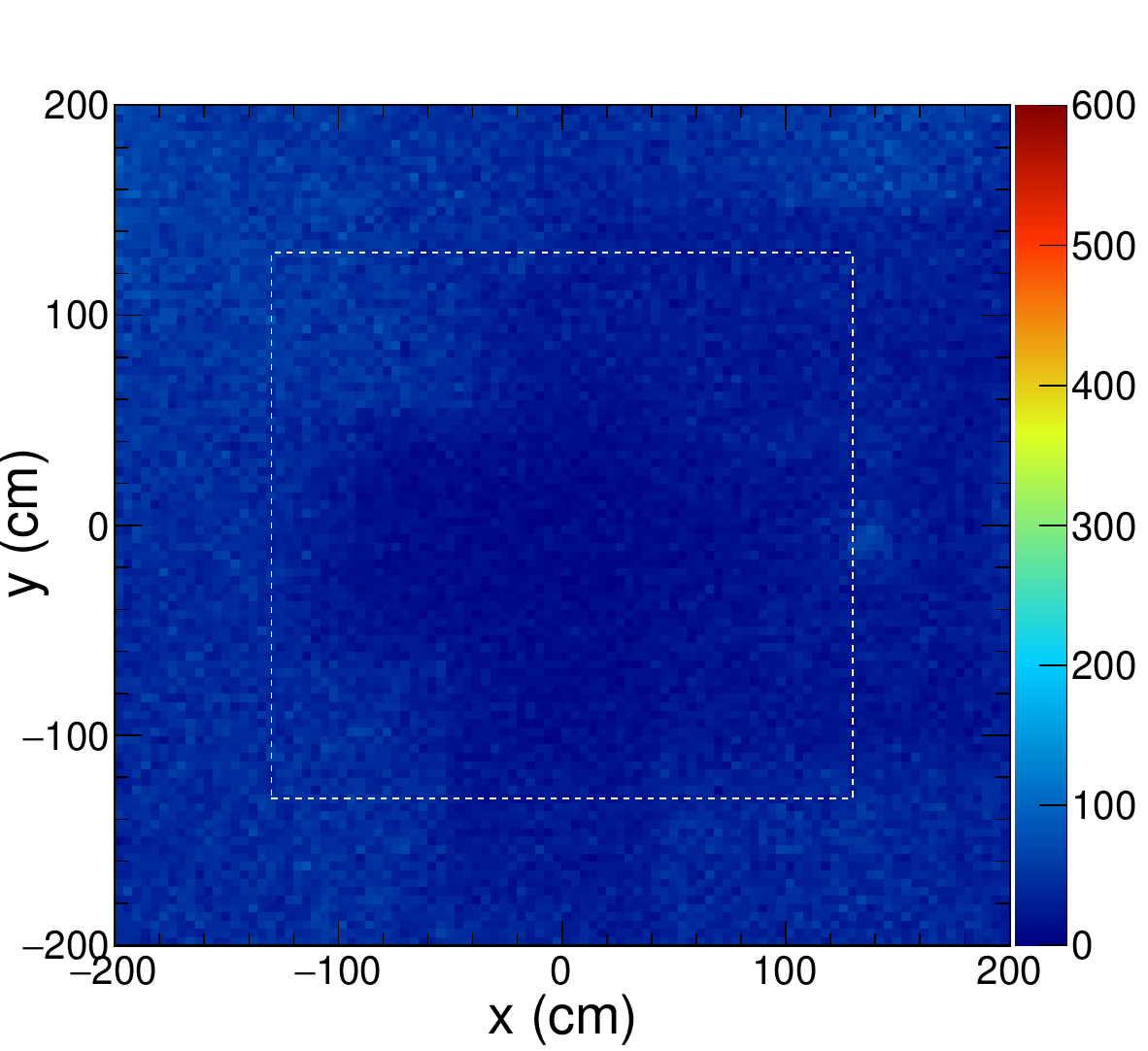}
         \caption{Simulation with shielding.}
         \label{fig:plumesim2022}
     \end{subfigure}
    \caption{Simulations of the beam profile on the upstream face of the NOvA detector using the Full Simulation \gfourbeamline mode (\autoref{sec:tertiary-beamline-sim}) and $10^9$ protons incident on the primary target. The view is from the detector perspective, with the $z$-axis pointing out of the page (see \autoref{fig:TertiaryModel}). The left panel shows the as-found beamline and the right panel shows the impact of the added shielding. The white box shows roughly the cross section of the active detector volume.}
    \label{fig:PlumeSim}
\end{figure}

\begin{figure}
    \centering
    \includegraphics[width=0.6\linewidth]{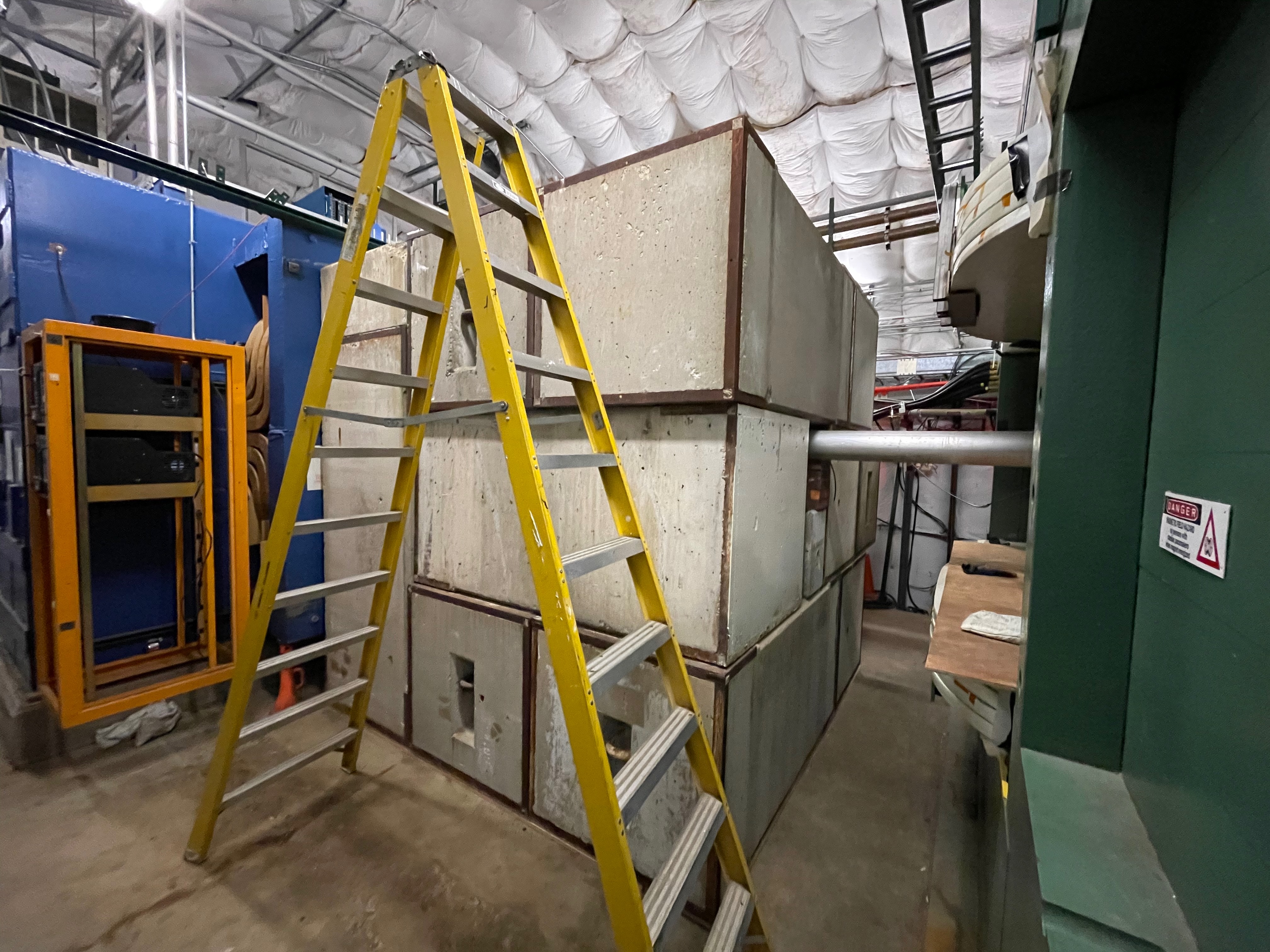}
    \caption{Photograph showing the shielding installed between the Jolly Green Giant magnet (on the right of the picture) and Rosie (on the left).  Beam travels right to left through the pictured beam pipe.}
    \label{fig:Shielding}
\end{figure}

There was no feasible way to prevent the production of the plume; instead the primary focus was finding a shielding solution that could be implemented during the 2020 long accelerator shutdown (spurred by the COVID-19 pandemic). The simulated muon plume, shown in the left panel of \autoref{fig:PlumeSim}, intersected the same region of the detector volume and behaved in a way qualitatively consistent with data, enabling an investigation into possible strategies for reducing the detector background levels. A simulation campaign showed \SI{2.74}{m} of concrete drastically reduced the background rate in the plume region at the NOvA TB detector, and additionally reduced the overall background rate by around a factor of four, as shown in the right panel of \autoref{fig:PlumeSim}. In the simulations the momentum-selecting collimator was set to \SI{127}{\milli\meter} (its physical limit) to show the impact of the shielding wall without the additional impact from the collimator jaws.  
The shielding wall was constructed around the beam pipe in MC7, between two large magnets (called Jolly Green Giant and Rosie), a few meters upstream of the NOvA TB hall. A photograph (\autoref{fig:Shielding}) shows the completed installation. This additional concrete shielding significantly reduced the beam backgrounds, resulting in much improved data quality from 2021 onward, as shown in the right panel of \autoref{fig:PlumeData}.

\noindent

\section{NOvA Tertiary Particles and Beamline Instrumentation}
\label{sec:tertiary}
\begin{figure}
    \centering 
\includegraphics[width=0.8\linewidth]{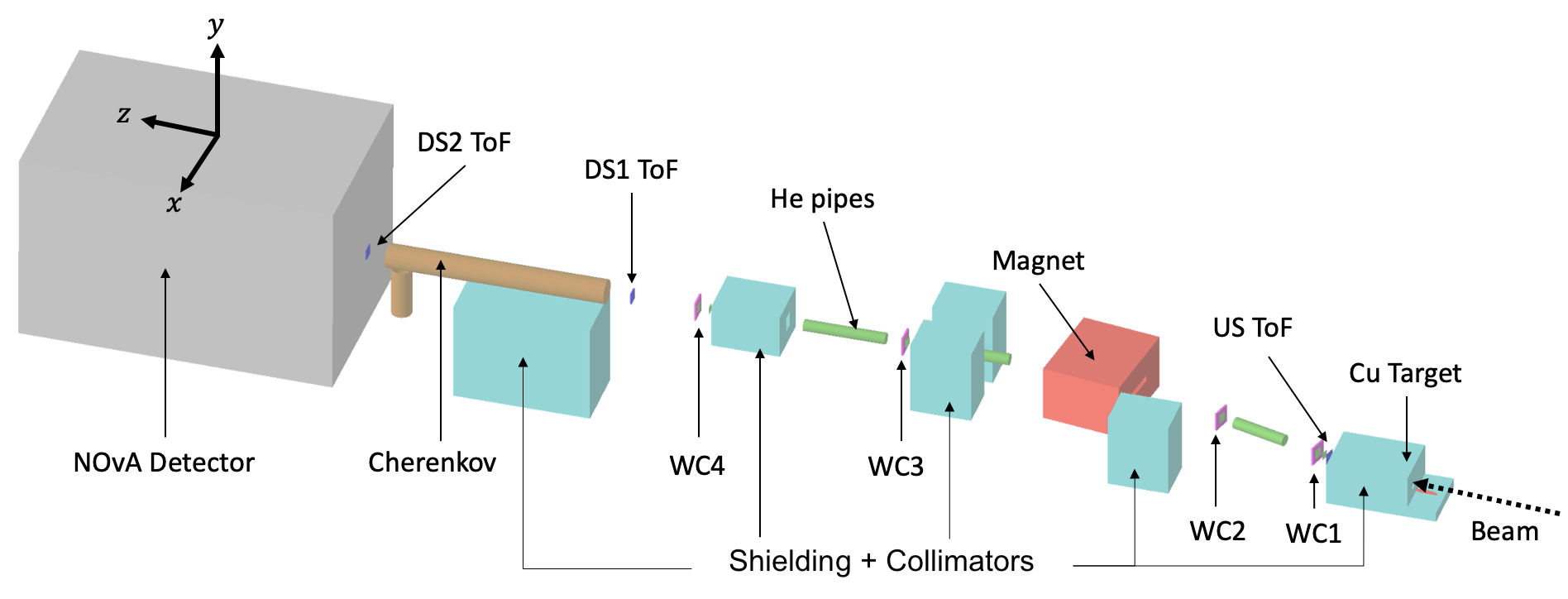}
    \caption{A \gfourbeamline model of the NOvA Test Beam tertiary beamline. Simulation starts at the target (right) and ends on the front face of the NOvA detector. The positions of the various beamline components in the simulation were determined from survey data. The horizontal scale of this illustration is roughly \SI{19}{\meter}.}
    \label{fig:TertiaryModel}
\end{figure}

\noindent
The secondary beam entered the NOvA TB experimental hall and struck a second copper target. The emerging tertiary particles encountered instrumentation to identify electrons, protons, pions/muons, and kaons in the momentum range \momentumrange. Particles were produced and directed using the Target and Collimator Assembly (\autoref{sec:target}). Particle identification, momentum selection, and momentum measurement were enabled by the \tof\ (ToF) System (\autoref{sec:tof}), Wire Chambers (\autoref{sec:wirechambers}), and Dipole Magnet (\autoref{sec:magnet}). Further discrimination between electrons and muons/pions was provided by the Cherenkov Detector (\autoref{sec:cherenkov}). The tertiary instrumentation also included a Beam Trigger (\autoref{sec:trigger}) and DAQ system (\autoref{sec:beamdaq}). The complete set of tertiary instrumentation was modeled in simulation (\autoref{sec:tertiary-beamline-sim}) built in the \gfourbeamline framework \cite{g4bl}, as shown in \autoref{fig:TertiaryModel}. 

Collimated particles exiting the \textrm{Cu} target first encountered the upstream ToF detector (US ToF), then traveled to the upstream multiwire proportional chambers (WC1 and WC2) via a helium-filled pipe to reduce scattering, before entering the dipole magnet aperture. Shielding installed around the magnet absorbed uncollimated backgrounds. Downstream of the magnet, particles in a given momentum range (selected using the magnetic field) passed through the first downstream \wc\ (WC3), the downstream collimator, the second downstream \wc\ (WC4), and the first downstream ToF detector (DS1 ToF). They then entered the Cherenkov detector and passed through the second downstream ToF detector (DS2 ToF) before finally entering the NOvA TB detector (\autoref{sec:detector}). 

 \subsection{Target \& Collimator Assembly}
\label{sec:target}
\begin{figure}
    \centering
\includegraphics[width=0.49\textwidth,clip]{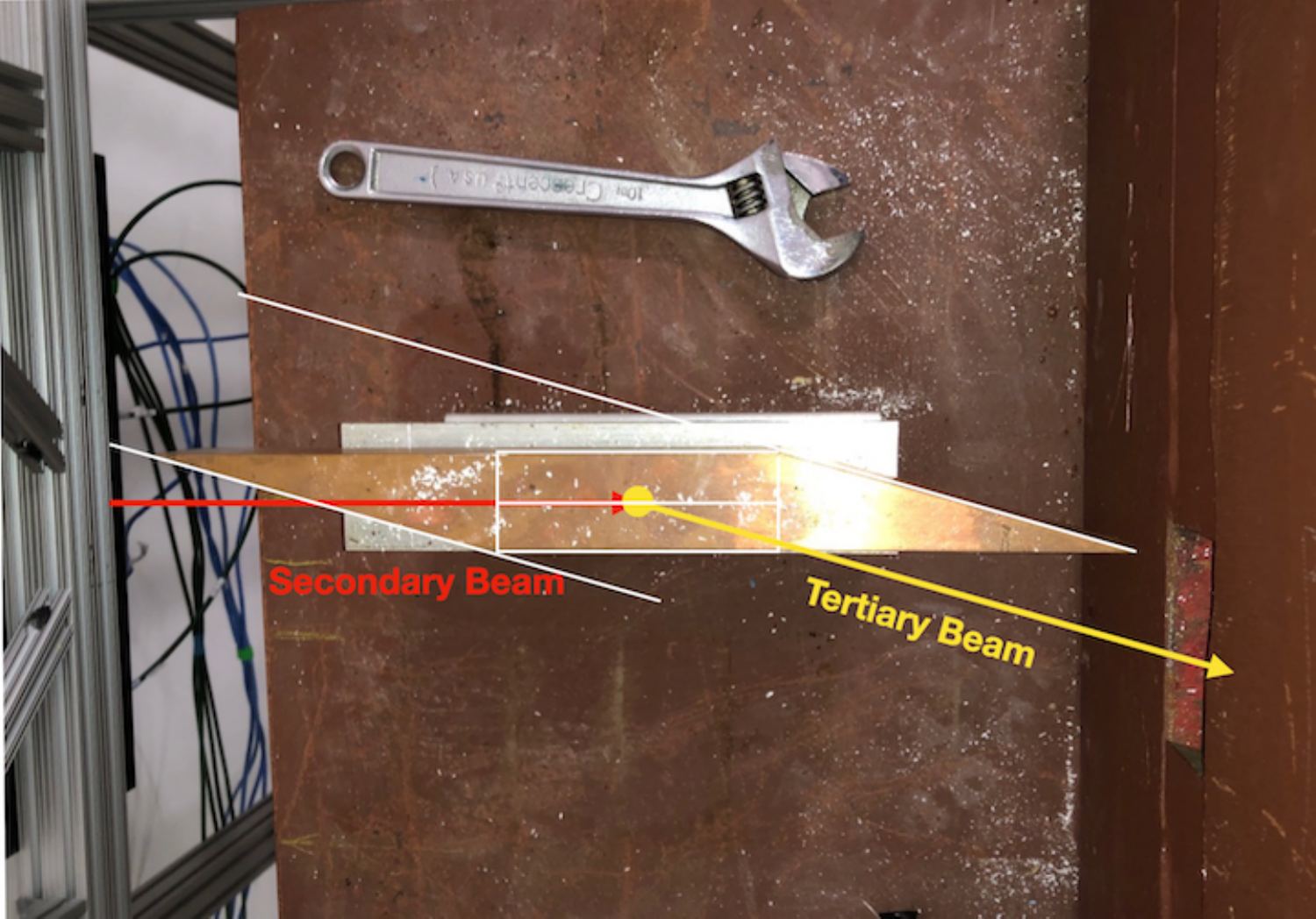}
    \includegraphics[width=0.49\textwidth,clip]{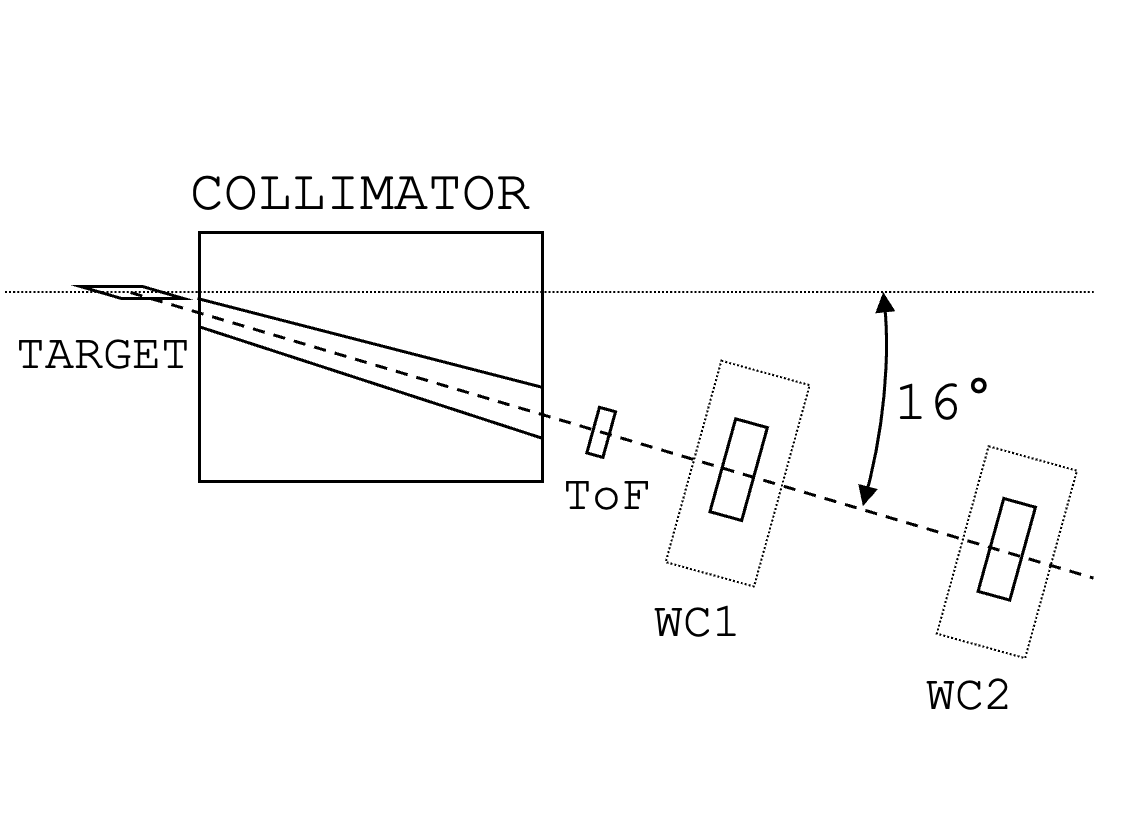}
    \caption{The left-hand side photo shows the overhead view of the copper target used to create the NOvA Test Beam tertiary particles. The secondary beam enters from the left.  Visible at lower right is the collimator opening placed at an angle of 16$\degree$ to the secondary beamline direction (the wrench is included for scale and was removed prior to \datataking). The right-hand side drawing shows the overhead view of the collimator opening and its position with respect to the beamline instrumentation downstream (not to scale).}
    \label{fig:NOvAtarget}
\end{figure}

\noindent
The \textrm{Cu} target was rhombohedral and 28\,cm long $\times$ 2.5\,cm tall, with a maximum thickness of 3.5\,cm. Particles emerging from the target were absorbed or degraded by an iron collimator, with the exception of those traveling through the collimator aperture, at an angle of $16\degree\pm10\degree$ at the aperture entrance and $16\degree\pm3\degree$ at the aperture exit, in the ($x$,$z$) plane with respect to the secondary beamline, as shown in \autoref{fig:NOvAtarget}. 

\begin{figure}
    \centering
\includegraphics[width=0.55\textwidth,clip]{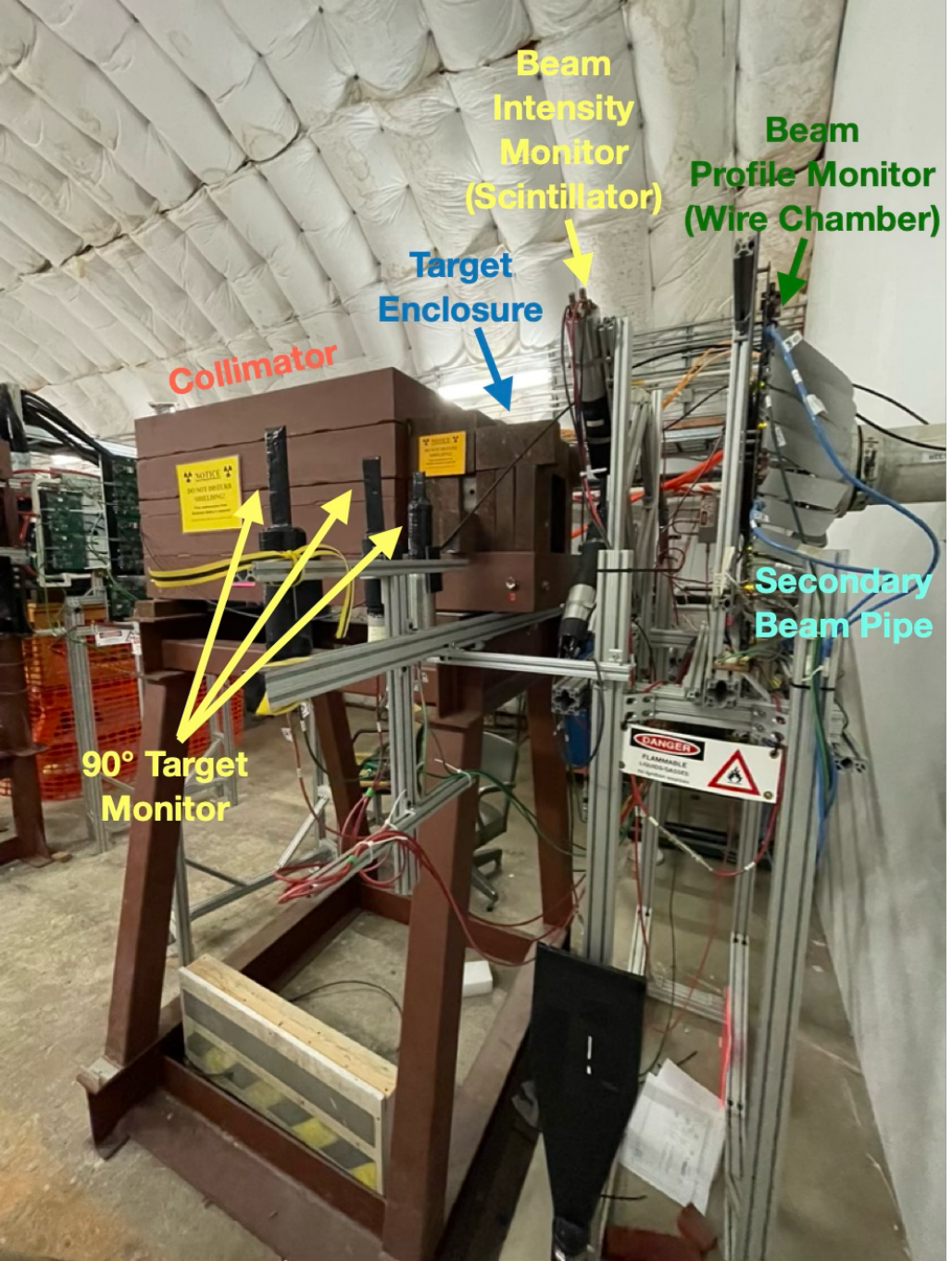}
    \caption{Collimator and target enclosure assembly following the end of the secondary beam pipe. Also identified in the figure are the wire chamber that monitors the secondary beam position profile, the scintillation counters that monitor the secondary beam intensity before target interaction, and the 90$\degree$ target activity monitor. Beam travels from right to left.}
    \label{fig:collimator}
\end{figure}

The target and collimator enclosure also housed the Secondary Beam Monitor, consisting of large scintillation counters placed between the helium-filled beam pipe and the copper target, and the 90$\degree$ Target Monitor, consisting of three small scintillator counters positioned perpendicular to the beam. These enabled measurements of particle rates through a small aperture in the target shielding enclosure. The assembly is shown in \autoref{fig:collimator}. 

 \subsection{Time-of-Flight System}
\label{sec:tof}
\noindent
The ToF system consisted of three detectors placed along the path of the tertiary particles. Their positions relative to the other tertiary beamline instruments are illustrated in \autoref{fig:TertiaryModel} with the labels US ToF, DS1 ToF, and DS2 ToF. The first two were read out by photomultiplier tubes (PMTs), and the third, DS2 ToF, by silicon photomultipliers (SiPMs). All three ToF detectors were constructed from square plastic scintillator plates measuring \qtyproduct{15 x 15}{cm^2} in the transverse plane. Eljen Technology EJ-232Q (0.5\%) fast timing plastic scintillator was chosen for its combination of light output (19\% anthracene) and fast rise time ($\SI{110}{ps}$) \cite{eljen:ej232}.

\begin{figure}
    \centering
\includegraphics[width=0.65\textwidth,clip]{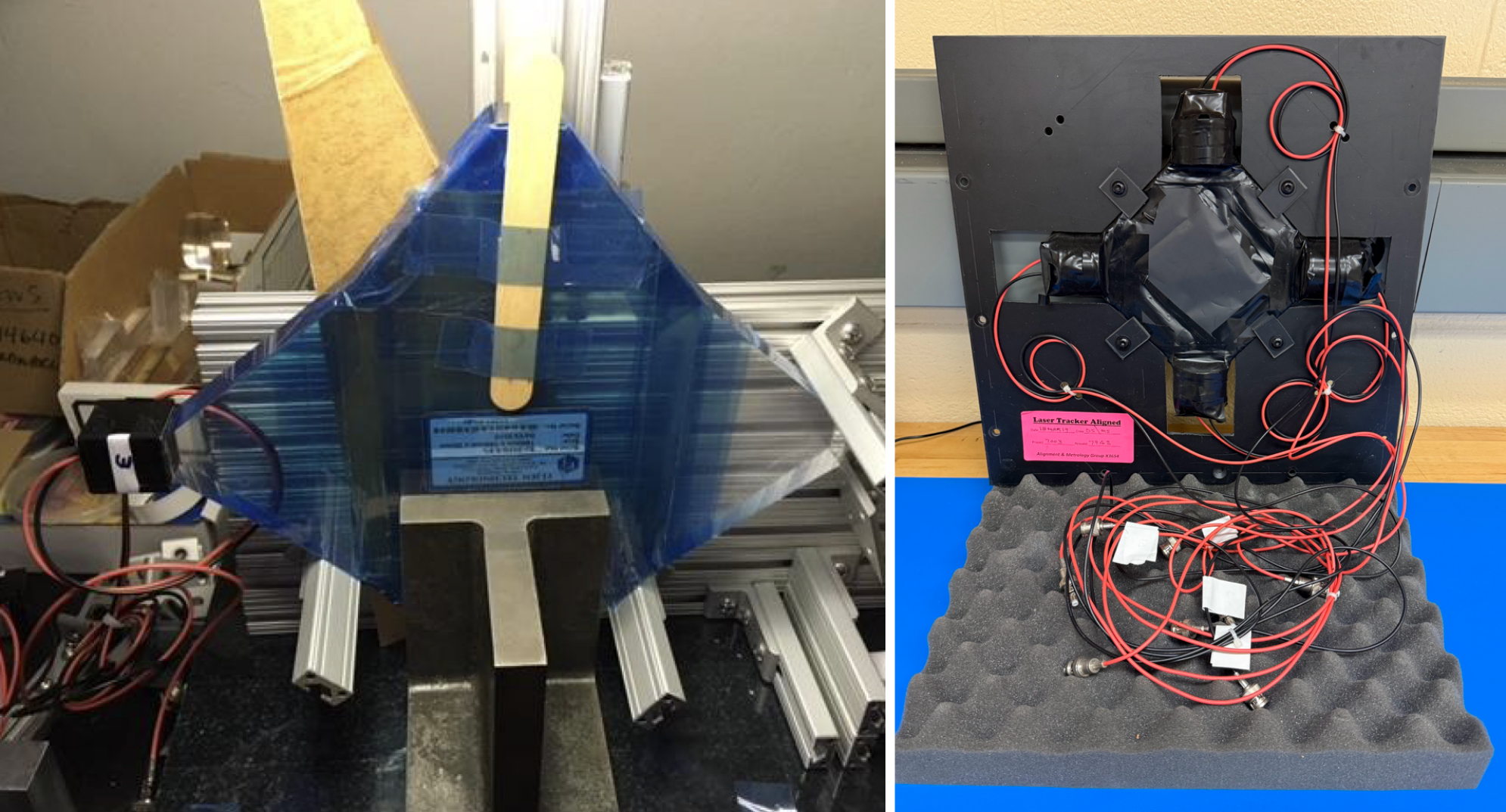}
    \caption{One of the ToF detectors shown during assembly (left), displaying the first PMT connected to the scintillator plate, and after assembly completion (right), displayed in the mounting frame used to deploy it in the beamline.}
    \label{fig:tof-photos}
\end{figure}

The first two ToFs had PMTs glued to each of the four corners of the plates via light guides that tapered to \qtyproduct{2 x 2}{cm^2}. \autoref{fig:tof-photos} shows one of the ToF detectors, both during its assembly and within the support mount used for beamline deployment. The upstream (US) ToF plate was \SI{0.6}{\cm} thick and was positioned at the collimator exit. The first downstream (DS1) ToF was thicker at \SI{2}{\cm} and was placed just before the Cherenkov detector, \SI{3.3}{\meter} upstream of the NOvA detector.  Hamamatsu H11934-200 PMTs were chosen for their fast rise time ($\SI{1.3}{ns}$), high quantum efficiency (43\%), and peak wavelength sensitivity ($\SI{400}{nm}$) close to the wavelength of maximum emission of the scintillator ($\SI{370}{nm}$) \cite{hamamatsu:h11934}.
 
DS2 ToF was of the same design and dimensions as the thinner US ToF, but was read out by SiPMs rather than PMTs, and was placed directly in front of the NOvA detector. The SiPM ToF was originally used in addition to the PMTs because it was not known which device would work more reliably. The PMTs were found to be more reliable, but the SiPM ToF was kept for redundancy and proved useful for providing an alternative ToF measurement. The Onsemi J-Series SiPMs~\cite{onsemi} had similar gain and quantum efficiency to the PMTs (between 37\% and 50\% based on bias voltage) at peak wavelength sensitivity ($\SI{420}{nm}$).

The four upstream (downstream) PMTs operated at a nominal voltage of \SI{-900}{\volt}~(\SI{-820}{\volt}) and were gain-equalized for their respective scintillator plate. The lower operating voltage for the downstream PMTs was selected to avoid saturating the digitizer with the large number of photons produced in the thicker scintillator plate.
The ToF signals were available to the trigger logic~(\autoref{sec:beamdaq}) and were digitized and recorded for offline analysis~(\autoref{sec:tracking-tof}).

 \subsection{Wire Chambers}
\label{sec:wirechambers}
\noindent
The NOvA TB experiment used four multiwire proportional chambers (WCs),
provided by FTBF and previously used by LArIAT~\cite{LArIAT:2019kzd}. Each \wc\ consisted of one $x$-plane and one $y$-plane, each with 128 sense wires. The gold-tungsten sense wires were \SI{10}{\um} in diameter and spaced at \SI{1}{\mm}. Flanking each sense plane were \SI{12.7}{\um} aluminum foil cathode planes. These planes were separated from their neighbors by G10 spacers, giving a cathode-to-sense-plane spacing of \SI{3.175}{\mm}. Aluminum foil windows with a thickness of \SI{12.7}{\um} and an area of \qtyproduct{15.24x15.24}{cm^2} completed and sealed the chambers. The cathode planes were held at $\approx$\SI{-2.4}{\kilo\volt}, with each WC optimized individually. The chambers were filled with a gas mixture of 85\% argon plus 15\% isobutane that flowed constantly via a pressurized gas distribution system. The two upstream chambers were plumbed in series, as were the two downstream chambers. The upstream chambers were also in series with another monitoring wire chamber placed just before the \textrm{Cu} target. The two sets of serially plumbed wire chambers were then connected in parallel. 

When a charged particle passed through a \wc, it produced a charge cloud that was accelerated toward the sense wires. Signals from each sense wire were then digitized and saved when a trigger was recorded. Due to the cathode voltages used, the charge cloud was amplified, and hits were often observed on multiple adjacent wires induced by the passage of a single particle as its cloud spread out. Based on LArIAT's work~\cite{LArIAT:2019kzd}, NOvA TB employed a DBSCAN-based~\cite{dbscan} clustering algorithm to link hits likely produced by a single particle. DBSCAN works by determining the number of neighbors within a given distance of a hit; if a hit's number of neighbors meets or exceeds a predefined threshold, then it and its neighbors are added to a cluster. 

The NOvA TB particle momentum measurement used \wc\ tracks, described in \autoref{sec:tracking-mom}, reconstructed from hits. The earliest hit from the DBSCAN cluster was identified as the ``good hit.'' This good hit was used to determine the position of the particle as it passed through the \wc. A good hit was required on at least one vertical and at least one horizontal wire within a chamber to provide the spatial information needed for reconstruction. In cases where multiple good hits were found in one or both planes, the total number of combined hits was $n_{hit} = n_{x} \times n_{y}$ for any given wire chamber.

 \subsection{Dipole Magnet}
\label{sec:magnet}

\noindent
The spectrometer dipole magnet, known as M1, was used to control the momentum range and charge sign of particles entering the downstream instrumentation. The downstream instrumentation was parallel to the secondary beamline, while the upstream instrumentation was at $\theta$ =\SI{-16}{\degree} (see \autoref{fig:collimator}). The $z$-axis of the magnet was at $\theta=$\SI{-8}{\degree}, such that particles with momenta matching the magnet settings exited parallel with the secondary beamline; an illustration of this is provided in \autoref{sec:tracking-mom}, \autoref{fig:bfield-angles}.

\begin{figure}
\centering
\begin{minipage}[l]{0.49\textwidth}
\includegraphics[width=\textwidth,trim={10in 3in 10in 3in},clip=true]{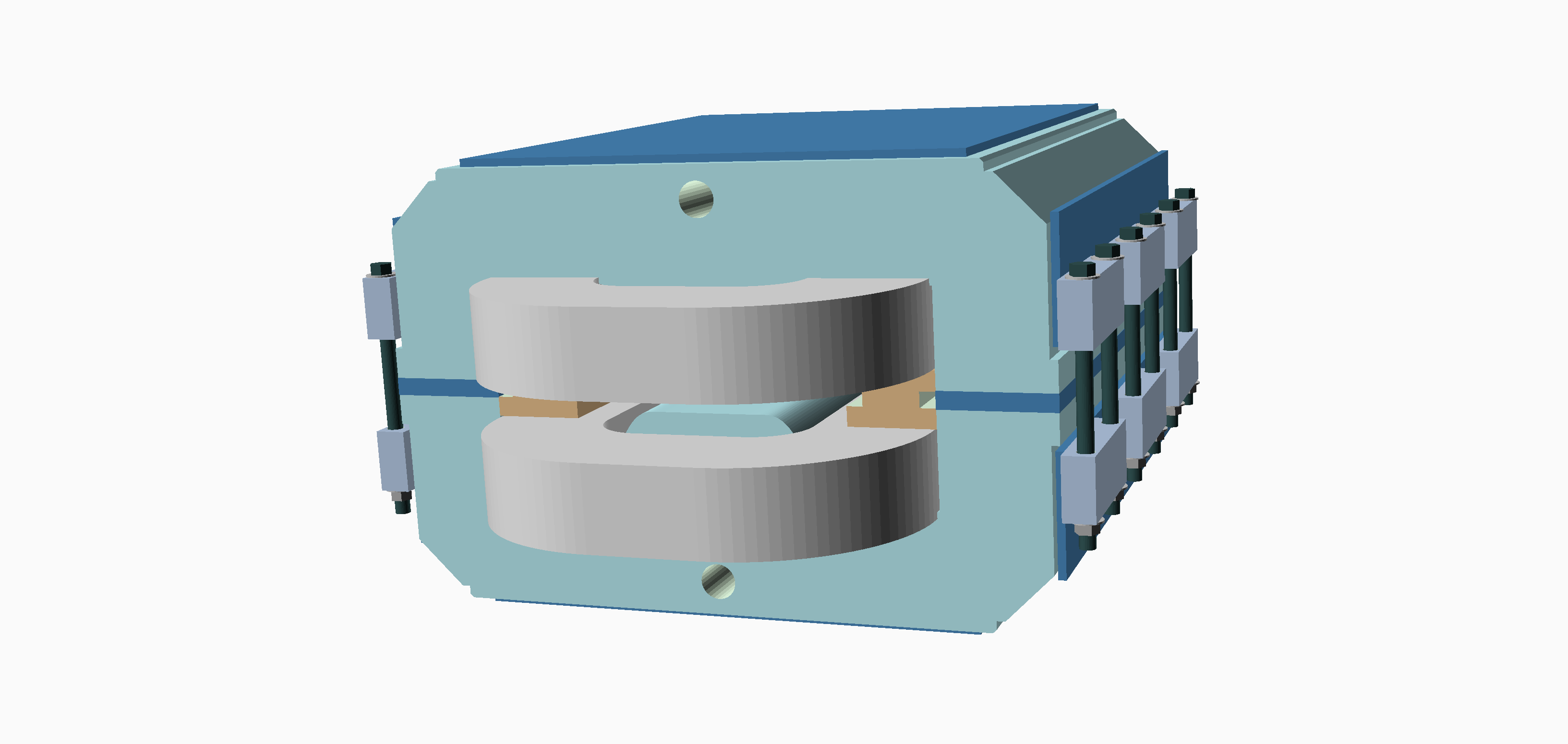}
\end{minipage}
\begin{minipage}[r]{0.49\textwidth}
\includegraphics[width=\textwidth]{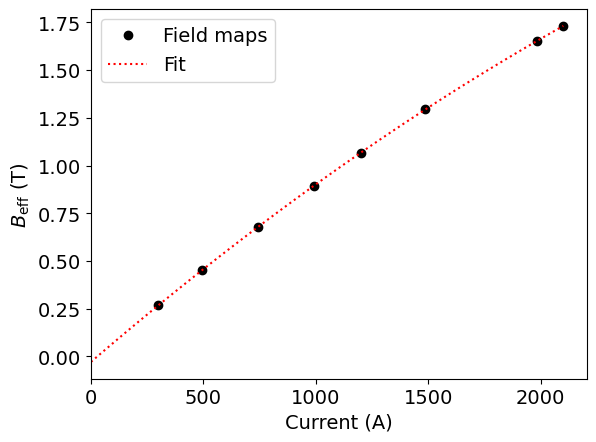}
\end{minipage}
\caption{Left: Model of the M1 spectrometer magnet. The aperture was \SI{45.1}{\cm} wide and \SI{8.9}{\cm} high. The field was in the vertical direction and was assumed active over the entire length of the iron yoke, $L_{\rm eff}=$~\SI{106.7}{\cm}. Right: The effective strength of the field within the active area of the dipole magnet, $B_{\rm eff}$, as a function of the magnet current. The effective field strength $B_{\rm eff}$ is defined as the magnitude of the average vertical component of the field. }\label{fig:M1scadgeneralview} \label{fig:beff}
\end{figure}

A model of the magnet is shown in \autoref{fig:M1scadgeneralview}. The aperture was \SI{45.1}{\cm} wide and \SI{8.9}{\cm} high, and the field was assumed active over the entire length of the iron yoke, $L_{\rm eff}=$~\SI{106.7}{\cm}. The magnetic field strength in this region was controlled and stabilized using a DC power supply with a current in the range \qtyrange{400}{1500}{A}. The magnetic field was directed vertically downward (upward) to select positively (negatively) charged particles, with the magnet current inversion switch operated remotely so that physical access was not necessary to change the magnet polarity.

M1 was previously used as part of the spectrometer for the FOCUS experiment~\cite{LINK2002174}. Prior to being used by NOvA TB, the magnet was refurbished and surveyed, and the strengths of the 3D field components were measured at \SI{0.3175}{cm} ($1/8$ inch) intervals throughout the active volume, for three magnet current settings $I =$ \SI{300}{A}, \SI{1200}{A}, and \SI{2100}{A}. The uniformity of the field was within 1\% in all cases. The field maps constructed from these measurements were compared with the predicted field through full 3D calculations, with excellent agreement. The predicted field calculations were then used to generate interpolation maps for five additional magnet current settings.

The effective field strength $B_{\rm eff}$ is defined as the magnitude of the average vertical component of the field. The average is calculated over the effective length $L_{\rm eff}$ of the active field region. The effective field strengths from the measured and interpolated field maps were used in the fit shown in~\autoref{fig:beff}, providing a parameterization 

\begin{equation}\label{eqn:baverage}
B_{\rm eff}  = - 0.0294 + \dfrac{|I|}{985.3}\; - \left(\dfrac{|I|}{3451.2}\right)^2.    
\end{equation}

\noindent This parameterization was compared with calculations using the full field map, and the agreement was generally better than 30 parts per million.

 \subsection{Cherenkov Detector}
\label{sec:cherenkov}
\DeclareSIUnit \atm {\ensuremath{\text{atm.}}}

\begin{figure}
  \centering
 \begin{minipage}[l]{0.49\textwidth}
  \includegraphics[width=\linewidth]{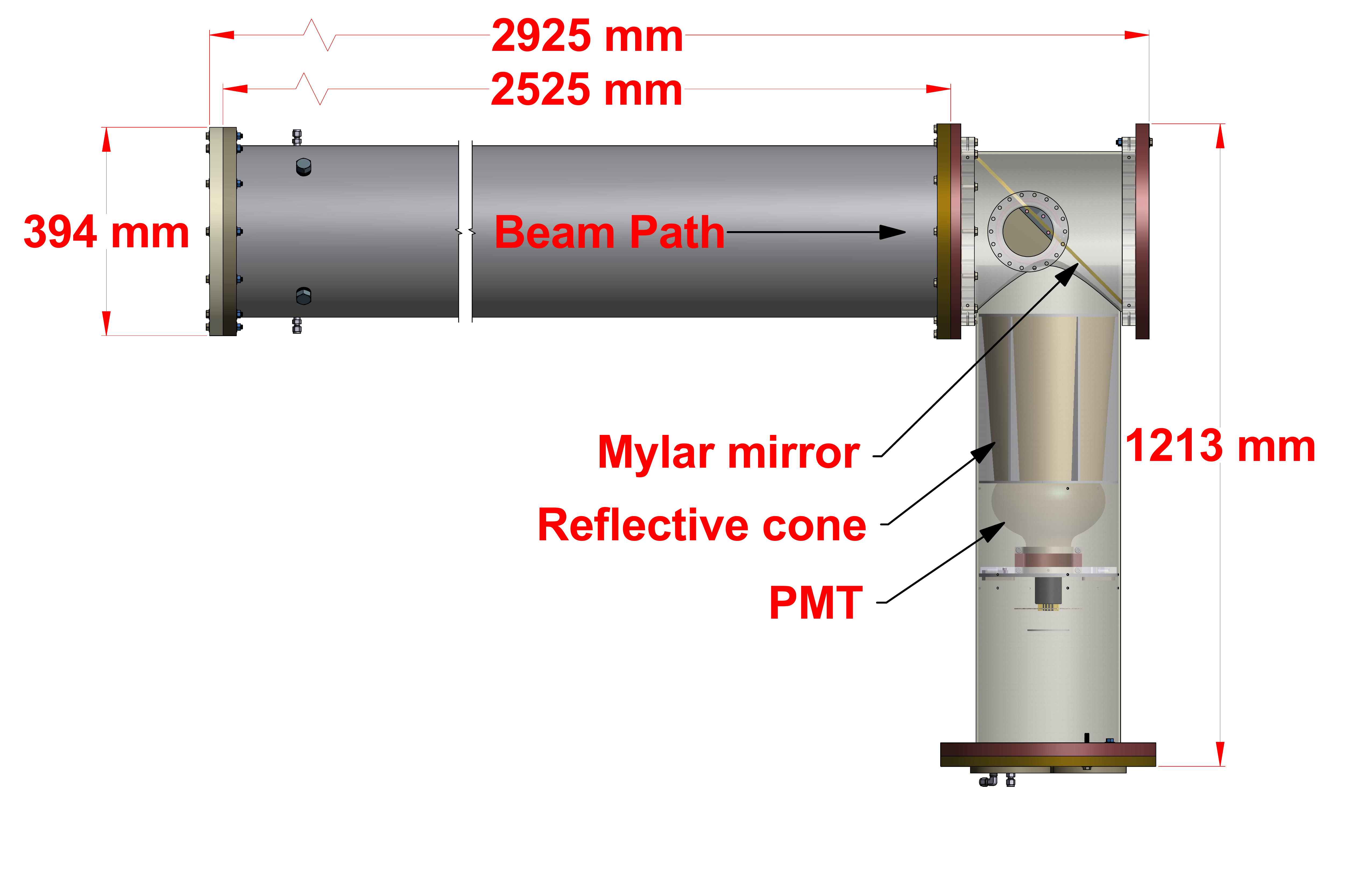}
  \end{minipage}
 \begin{minipage}[l]{0.49\textwidth}
 \includegraphics[width=\linewidth]{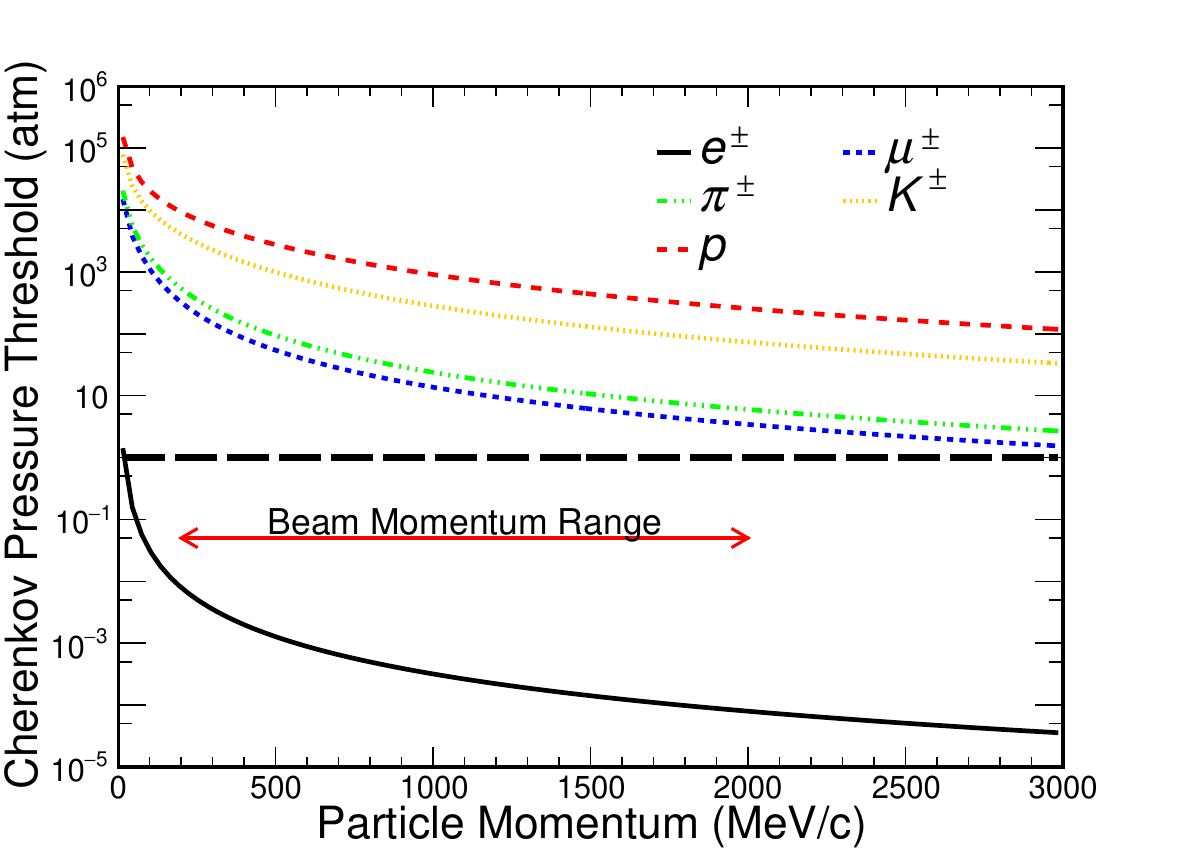}
 \end{minipage}
  \caption{Left: A diagram of the NOvA Test Beam Cherenkov counter. The long arm of the counter (shown with an artificial break to enhance details in the short arm) was coaxial with the beam direction (left to right as drawn) while the short arm was perpendicular to and below the beam direction. Right: The relationship between momentum and Cherenkov pressure threshold for the particle species produced in the NOvA TB experiment. The dashed black line is the \SI{1}{\text{atm}} pressure at which the Cherenkov counter operated. This value is above the threshold pressure only for $e^\pm$ in the expected beamline momentum range (shown with the red arrow).}
  \label{fig:CherenkovCounterAssembly} \label{fig:CherenkovPressureThreshold}
\end{figure}

\noindent
The Cherenkov counter, schematically depicted in  \autoref{fig:CherenkovCounterAssembly}, was made of two stainless-steel pipe sections joined to form an ``L'' shape with the long arm coaxial with the direction of the tertiary particles and the short arm pointing downwards. It was filled with $\text{CO}_2$ at a pressure of \SI{1}{\atmos}. The long arm of the counter enclosed the gas medium with \SI{152.4}{\mathrm{\mu} m}-thick vinyl windows on either end, allowing tertiary particles to pass through with minimal scattering. At the bottom of the short arm, there was a flange to allow access to the Hamamatsu R5912-03MOD2 PMT and pass-through electrical connections for HV and readout (\autoref{sec:beamdaq}). The PMT was operated at +\SI{1600}{\volt}. Inside the joining section of the two arms was a \SI{76.2}{\mathrm{\mu} m}-thick mylar foil mirror angled at 45$\degree$ with respect to the direction of the downstream tertiary particles. Light emitted roughly along the axis of the long arm was reflected downward into the short arm, where a conical mirror guided the light towards the readout PMT.

The Cherenkov detector was designed to tag electrons~\cite{DaltonThesis}. Cherenkov light is emitted by particles whose velocity exceeds $c/n$, where $c$ is the speed of light in vacuum and $n$ is the index of refraction of the gas medium. The index of refraction of the CO$_2$ changes linearly with changes in the pressure of the gas and is modeled by

\begin{equation}\label{eq:refrac}
n = 1 + kP,
\end{equation}

\noindent where $k = \qty[per-mode = symbol]{4.1e-4}{\per\atmos}$ is a constant and $P$ is the pressure of the gas medium. \Autoref{fig:CherenkovPressureThreshold} shows the pressure needed for a particle to make Cherenkov light as a function of momentum for each NOvA TB particle species, showing that only electrons emit Cherenkov light for the momentum range of interest.

 \subsection{Beam Trigger} 
\label{sec:trigger}
\noindent
A total of four trigger counters were placed in the beamline, adjacent to the four \wcs. Each of the trigger counters consisted of a PMT connected to a plastic scintillator plate of thickness \SI{3.1}{mm} and area \qtyproduct{10.1x10.1}{cm^2}. A beam trigger was issued if there was a signal in the trigger counters during the beam window and the beamline readout electronics were available. For most of the \datataking\ periods, the trigger required hits in all four trigger counters. For the final run, a three-fold coincidence was used to improve trigger efficiency.

A pulse covering the \SI{4.2}{s} beam window was formed using a NIM gate generator with input signals from the Fermilab accelerator clock denoting the beginning and end of the beam spill. 
The readout electronics were deemed to be available when the digitizer used to process data from the TOF and Cherenkov detectors was not in its “busy” state, indicating it had finished digitizing the previous trigger. In addition, a hold-off period of \SI{10}{\mathrm{\mu} s} was enforced after a trigger, to ensure the stability of front-end components, particularly the \wc\ controller.
The trigger rate varied from 1--10 triggers per beam spill over the \datataking\ periods.

\noindent

\noindent
 \subsection{DAQ \& Timing Systems} 
\label{sec:beamdaq}
\noindent
The beamline DAQ collected and saved data from all tertiary instrumentation upstream of the NOvA detector. It was fully decoupled from the NOvA detector DAQ, which is described in \autoref{sec:detdaq}. The data from the two systems were synchronized through the use of a common clock on NOvA timing units.

\begin{figure}
    \centering
    \includegraphics[width=0.7\linewidth]{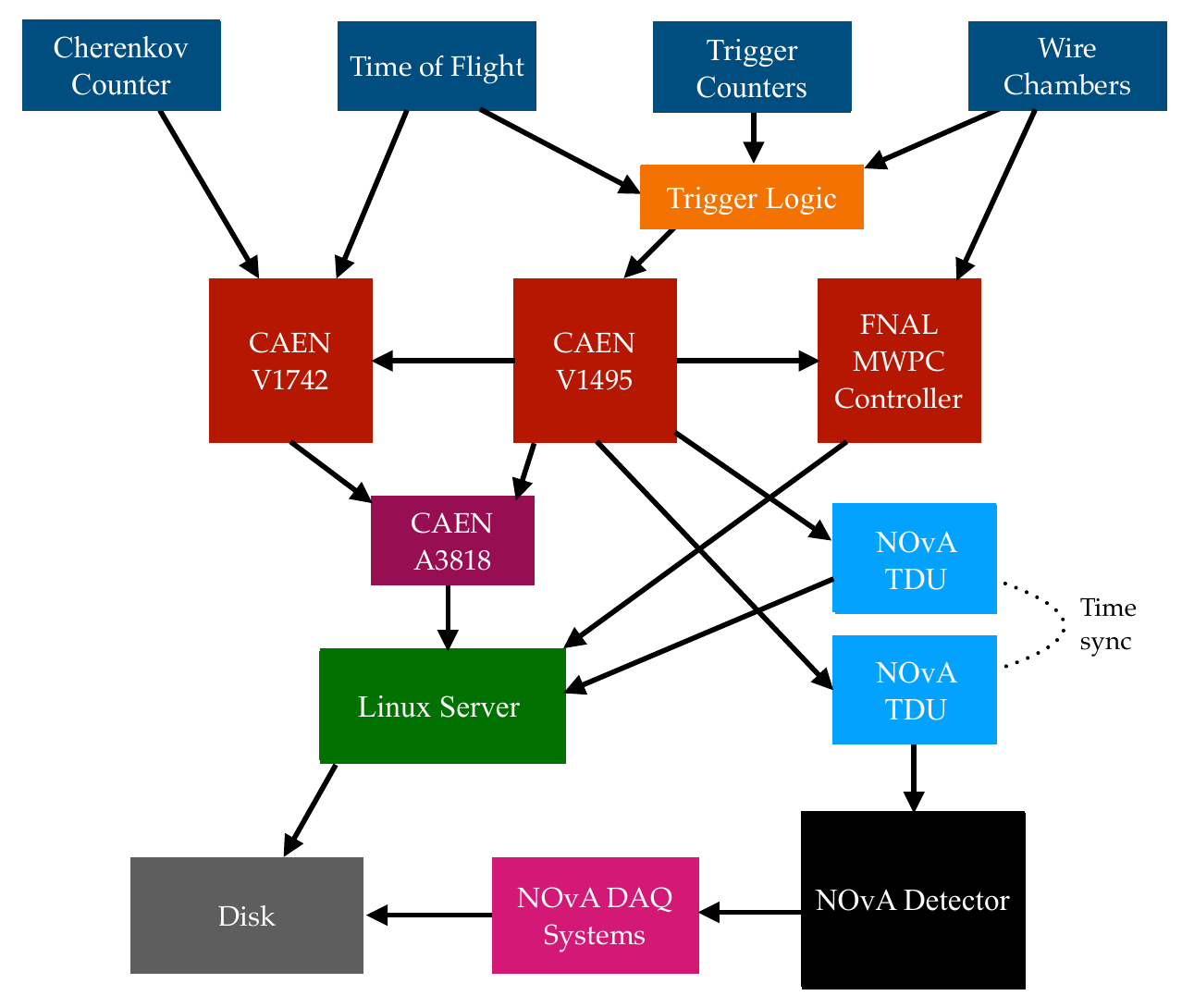}
    \caption{Schematic of the beamline DAQ. CAEN V1742 is a digitizer, CAEN V1495 is a trigger board, and CAEN A3818 is a PCIe card.}
\label{fig:BeamlineDAQ}
\end{figure}

A diagram of the beamline DAQ is shown in \autoref{fig:BeamlineDAQ}.
A VME crate housed a CAEN~\cite{caen} V1495 trigger board, a CAEN V1742 digitizer, and a CAEN V2718 crate controller that interfaced with the DAQ Linux server via a CAEN A3818 PCIe card.
The digitizer was used to process data from the ToF detectors and Cherenkov counter; data from the \wcs\ were collected using a Fermilab-custom wire chamber controller~\cite{fnal-mwpc}. 
Signals from all detectors were collected using RG-174 LEMO cables and pre-processed using NIM electronics. Upon a trigger decision, an electronic signal was sent from the trigger board to the digitizer and the wire chamber controller to initiate readout. The same signal was sent to two custom NOvA Timing Distribution Units (TDUs)~\cite{Norman:2012zzc}. One TDU was used to trigger readout of the NOvA TB detector, while the second TDU provided a timestamp for the data stream from the instrumentation components in the path of the tertiary particles. The timestamps saved in the NOvA trigger and the tertiary beam instrumentation data were identical, ensuring the two data streams could be matched during offline processing. 

The beamline DAQ software used artdaq \cite{artdaq}, a Fermilab package providing standard DAQ tools for handling each of the required processes.  Each trigger was saved as a separate artdaq event in a custom data structure shared with the offline processing tools.

\subsection{Simulation}
\label{sec:tertiary-beamline-sim}
\noindent
Various modes of simulation using the \gfourbeamline code were employed for different purposes during and after data-taking. The two main modes, ``Full Simulation'' and ``Tertiary Simulation'', are briefly described here along with the ``Data-Seeded'' simulation, which does not use  \gfourbeamline.

\begin{itemize}

\item ``Full Simulation'' is a \gfourbeamline simulation of the entire secondary and tertiary beamline components, interfaced to a \geantfour~\cite{GEANT4,GEANT4dev,GEANT4devapp} simulation of the NOvA detector (\autoref{sec:nova-simulation}). The Full Simulation starts just upstream of the primary target shown to the left of \autoref{fig:SecondaryBeamline}. This simulation was used to understand backgrounds and optimize shielding deployment, notably the ``muon plume'' described in \autoref{sec:secondary-beamline-sim} and shown in \autoref{fig:PlumeSim}.

\item ``Tertiary Simulation'' is a \gfourbeamline simulation of the tertiary beamline components interfaced to a \geantfour simulation of the NOvA detector. The Tertiary Simulation assumes \SI{64}{GeV/c} protons interacting in the secondary \textrm{Cu} target, and starts just downstream of the target \& collimator assembly described in \autoref{sec:target} and shown on the right of \autoref{fig:TertiaryModel}. This simulation mode was used for a preliminary estimate of energy loss in the tertiary beamline components, described in \autoref{sec:tracking-mom} and shown in \autoref{fig:pion-momcor}.

\item ``Data-Seeded'' simulation is a \geantfour simulation of the NOvA detector, seeded with data using the momentum estimated using the \wc\ track directions on either side of the magnet (\autoref{sec:tracking-mom}), and using preliminary particle identification which additionally relies on the measured ToF (\autoref{sec:tracking-pid}). This mode of simulation, which does not use \gfourbeamline, is described in \autoref{sec:nova-simulation}.
\end{itemize}

In the \gfourbeamline modes (Full Simulation and Tertiary Simulation), the positions of the modeled components were obtained from survey measurements. To improve simulation speed, a cutoff of \SI{20}{\mega\electronvolt} on the particle kinetic energy was applied, and particles of interest were required to have a momentum of at least \SI{200}{MeV/c}. The simulation included not only the particles of interest, but also intermediate particles such as gammas and neutral pions.  Simulations were conducted under several instrumentation configurations, including various analyzing-magnet field strengths. 

\section{The NOvA Detector}
\label{sec:detector}
\begin{figure}
\centering
\includegraphics[width=0.85\textwidth,clip]{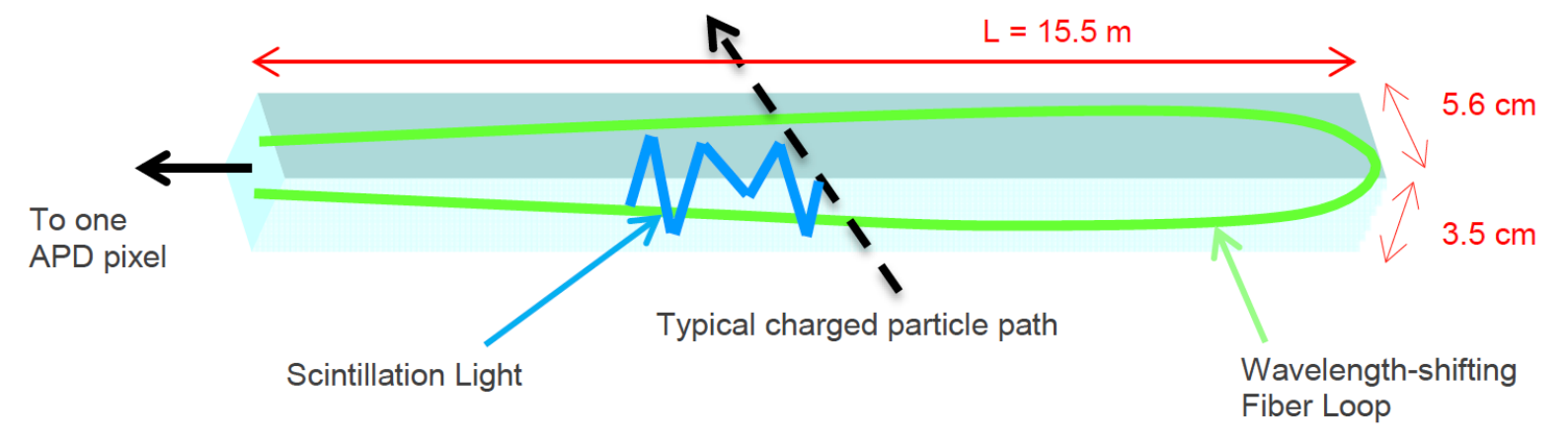}
\caption{Depiction of a NOvA cell, reproduced from~\cite{Talaga:2016rlq}. Ionizing particles traverse liquid scintillator inside a PVC extrusion cell and produce light, which reflects off the PVC walls multiple times until captured by a wavelength-shifting fiber-optic loop. Light within the fiber is detected by an APD. The \SI{15.5}{\m} cell length in the figure refers to the FD; the TB detector cells were \SI{2.6}{\m} long. 
}
\label{fig:tech}
\end{figure}

\noindent 
The NOvA detectors (FB, ND, and TB) are segmented tracking sampling calorimeters employing PVC extrusions for the dual purposes of containing liquid scintillator and providing the segmentation. Approximately 38\% of the detector mass is in the PVC structure and 62\% is in the liquid scintillator, both low-$Z$ materials, with an average radiation length of \SI{40}{cm}. The nuclear interaction length is $\approx\SI{80}{cm}$.
\autoref{fig:tech} illustrates the particle detection technology for one cell of a NOvA FD 16-cell PVC extrusion.

\subsection{Detector Geometry}
\label{sec:det-geo}
\noindent
The configuration and dimensions of the NOvA TB detector were originally determined through studies conducted in 2012, during early construction of the NOvA FD and ND. The primary criterion was energy containment of simulated pions and electrons. The adopted configuration contained 100\% of particle energy for electrons with momentum\momentumrange, and over 99\% for pions in the same momentum range. The width of the detector in the direction transverse to the tertiary beam was particularly important for pion energy containment, as pion reinteractions can lead to large transverse momentum. 

The NOvA TB detector was formed of two blocks assembled from PVC modules fabricated for the NOvA ND and cut to two-thirds of their original length, thus sharing precisely the same materials, segmentation, and overall design with the NOvA FD and ND. Each PVC module contained 32 cells (see \autoref{fig:tech}). Each plane was formed by two PVC modules. The planes alternated between vertical and horizontal cell orientations. Each block consisted of 31 planes glued together. 
The last plane in the upstream block and the first plane of the downstream block both had vertically oriented cells, so an additional plane with horizontal cells (plane 31) was sandwiched between the two blocks to preserve alternating cell orientation, resulting in a total of 63 planes, each with 64 cells. In the adopted numbering scheme, even-numbered planes had vertical cells, and odd-numbered planes had horizontal cells. The cells were~\SI{261.64}{cm} long, with an internal transverse width of~\SI{3.46}{cm} and internal depth (in the $z$-direction) of~\SI{5.56}{cm}.
The origin of the detector coordinates was the point where the tertiary beam was expected to enter the front face of the detector, such that:
\begin{enumerate}[label=(\roman*)]
\item The $x$-axis pointed west, from the center of cell zero at $x=-124.71$ cm to the center of cell 63 at $x=+124.70$ cm.
\item The $y$-axis pointed upwards, from the center of cell zero at $y=-123.33$ cm to the center of cell 63 at $y=+126.08$ cm.
\item The $z$-axis pointed north and ran from the center of plane zero at $z=+3.38$ cm to the center of plane 62 at $z=+418.03$ cm. 
\end{enumerate}

The physical parameters of the NOvA Test Beam detector are summarized in \autoref{tab:detector}.

\begin{table}[ht]
\centering
\caption{Physical parameters of the NOvA Test Beam detector.}
\begin{tabularx}{0.65\textwidth}{l  l }
\toprule
\textbf{Dimension} & \textbf{Value}\\
\midrule
Width (m)  & 2.6 \\
Height (m) & 2.6  \\
Length (m) & 4.2 \\
Scintillator (gallons/plane) & 86\\
Total scintillator volume (gallons) & 5418\\
PVC mass (kg/plane) & 171.2\\
Adhesive mass (kg/plane) & 3.1\\
Total empty mass (kg) & 10,981\\
Total scintillator mass (kg) & 17,601\\
Total detector mass (kg) & 28,582\\
\midrule
\bottomrule
\end{tabularx}
\label{tab:detector}
\end{table}

\begin{figure}[ht!]
    \centering
\includegraphics[width=0.49\textwidth,clip]{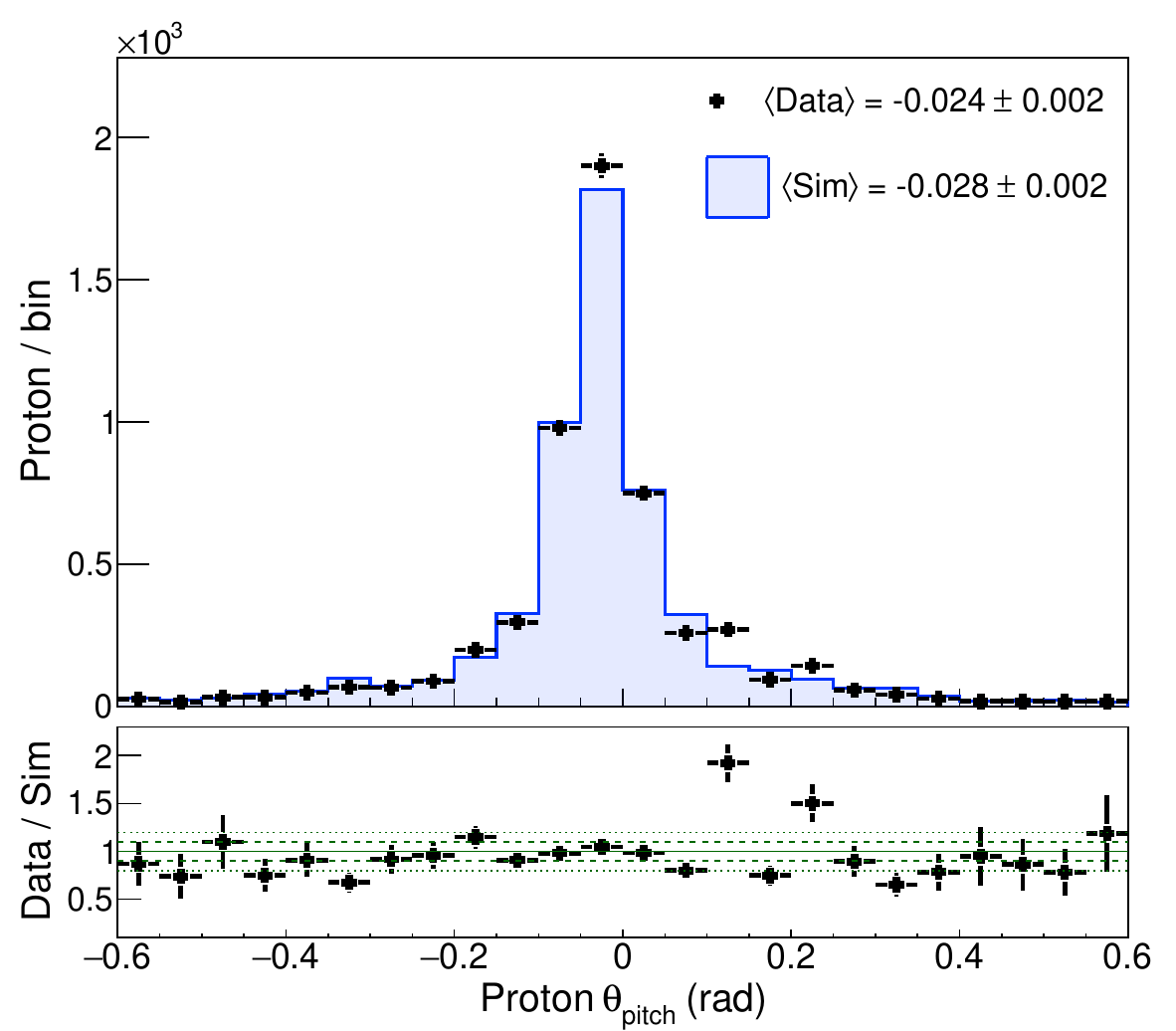}
    \includegraphics[width=0.49\textwidth,clip]{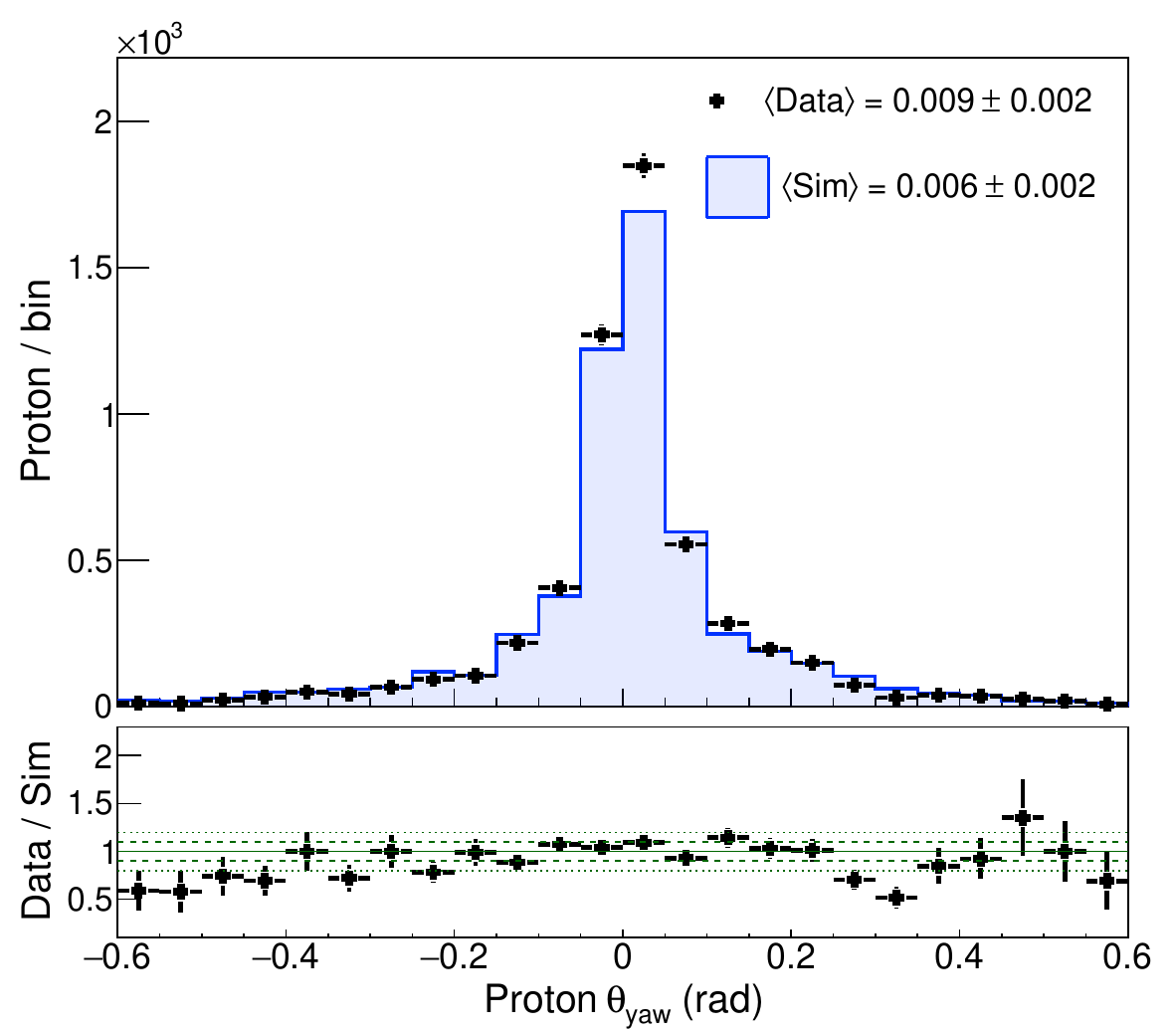}
    \caption{Comparisons of data and Data-Seeded simulation (\autoref{sec:nova-simulation}) showing the pitch and yaw angles of the detector with respect to the beamline, where $\theta_{pitch} = \arctan{\frac{dy}{dz}}$ and $\theta_{yaw} = \arctan{\frac{dx}{dz}}$. The mean values are calculated from the histograms over the range \SIrange{-0.6}{0.6}{\radian}. These angles were calculated using the detector hit positions for beamline proton candidates identified using loose selection criteria similar to those given in \autoref{tab:detsel}.     }
    \label{fig:placeholder-geobl}
\end{figure}

A full survey of the detector and beamline instrumentation took place in 2019 and informed the simulation of the detector geometry and beamline component positions and rotations during data collection. A final survey was conducted before decommissioning. This survey, along with detailed comparisons of data and simulation, led to final updates to the simulated detector geometry. The geometry description was validated by comparing the data and Data-Seeded simulation (\autoref{sec:nova-simulation}) in terms of the orientation of the detector planes with respect to each other and of the detector with respect to the beamline. The hit positions of proton candidates in the detector were used to calculate the rotation of the detector with respect to the beamline in terms of the pitch, yaw, and roll angles, which describe the relative rotation of the detector about the $x$, $y$, and $z$ axes respectively. An example of the excellent agreement between the data and simulation is given in \autoref{fig:placeholder-geobl} for the pitch and yaw angles. The relative pitch (yaw) angle is calculated as $\theta_{pitch} = \arctan{\frac{dy}{dz}}$ ($\theta_{yaw} = \arctan{\frac{dx}{dz}}$), where e.g. $dy = y_f - y_i$ is the $y$-displacement between the final and initial detector hits produced by a proton candidate. The mean pitch angle in simulation differs by \SI{0.004}{\radian} (\SI{0.2}{\degree}) from that measured in data. This corresponds to a difference of \SI{1.4}{cm} in the height of the center of the back face of the detector relative to the front face, which is roughly one third of the transverse size of a cell.

 \subsection{Extruded PVC Modules}  
\label{sec:PVC}
\noindent
The PVC extrusions provided mechanical structure and served as containers for the liquid scintillator, just as for the NOvA FD and ND. To capture the scintillation light for readout, the extrusion cell walls had a high reflectance,  obtained through a custom formulation using 15\%~TiO$_2$ by weight to significantly boost reflectance over that found in commercial PVC products, while preserving the necessary mechanical strength~\cite{Talaga:2016rlq}.

Each cell in a module was read out using one channel of a 32-channel avalanche photodiode (APD, \autoref{sec:apd}).  A 0.7 mm-diameter Kuraray Y-11 (200 MJ) wavelength-shifting fiber looped within the cell transported light created in the scintillator to the APD~\cite{NOvA:2007rmc}.  An injection-molded cover sealed one end of the module and supported the optical connector that coupled the fibers to the APD, while the other end of the module was sealed by a load-bearing PVC plate.

\begin{figure}[ht!]
    \centering
\includegraphics[width=0.85\textwidth,clip]{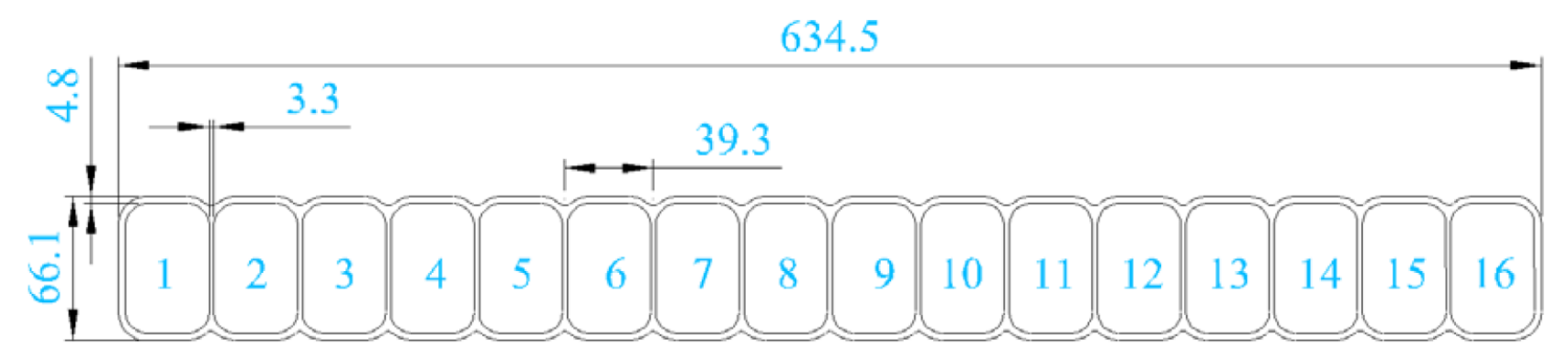}
    \caption{NOvA PVC extrusion cross section (dimensions in mm). Reproduced from Ref.~\cite{Talaga:2016rlq}.}
    \label{fig:extrusion}
\end{figure}

The horizontal planes of cells were deliberately oriented with a slight inclination of $\approx\SI{0.6}{\degree}$ to ensure the oil fully filled the cells.  Unfortunately, during manufacturing of the blocks, the fill ports were incorrectly installed on the lower end of the modules, resulting in air being trapped in the cells at the top of each of the two modules forming each horizontal plane. The top of the upper module is not critical for analyses, but the top of the lower module is located in the center of the detector, close to where the tertiary beam enters, and is therefore crucial for data quality. The underfilled cells resulted in a reduced hit rate, visible in \autoref{fig:PlumeData}, which was rectified midway through operations using an ``overfilling'' procedure in which extensions were attached to the fill ports on the east side of the detector and small holes were drilled in the west side, to allow the air to vent, before being resealed. The data collected after this intervention had a consistent hit rate across all cells in each plane.

\subsection{Liquid Scintillator}
\label{sec:scint}
\begin{figure}
    \centering
\includegraphics[scale=0.7]{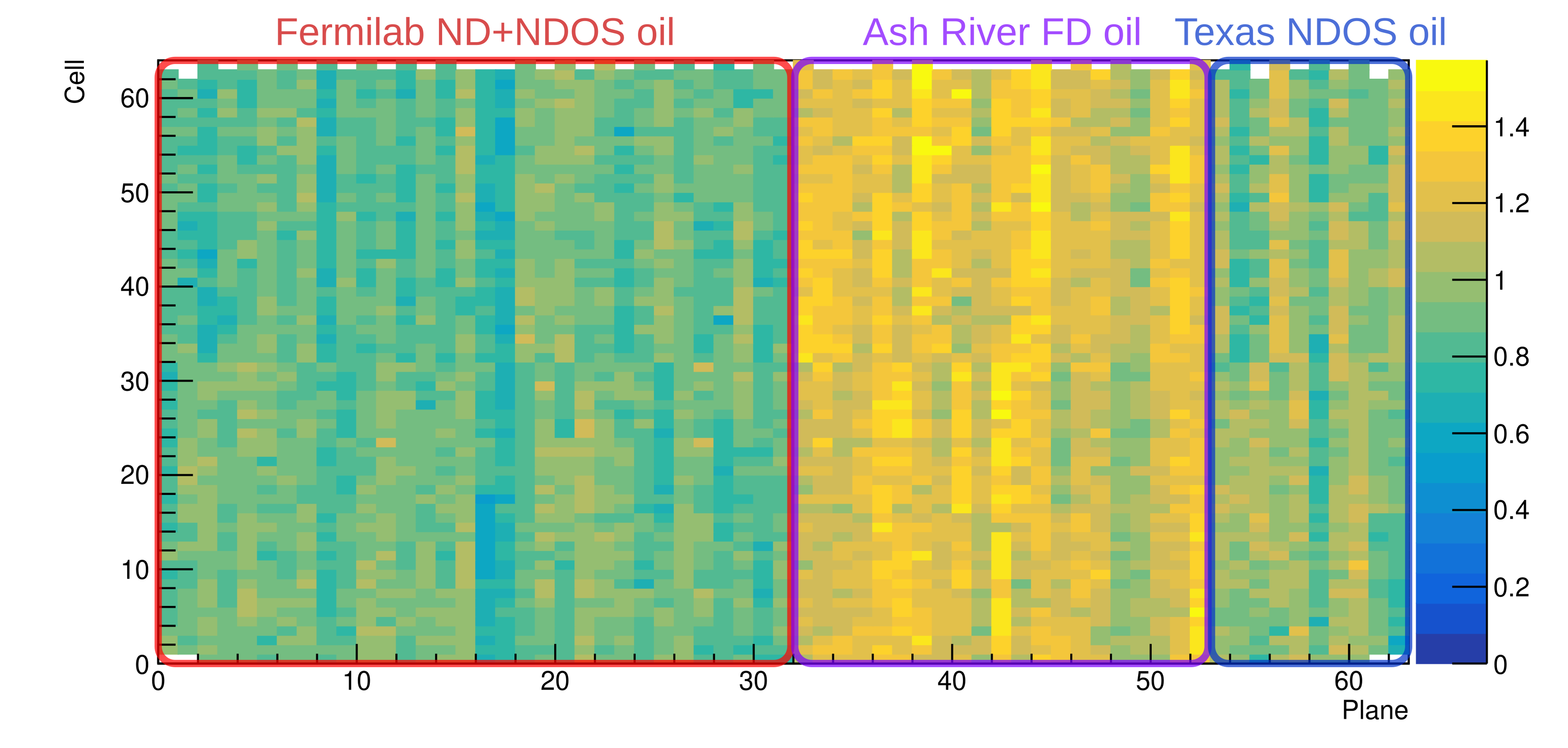}
    \caption{Uncorrected brightness levels using scintillator from three different sources: Fermilab ND + NDOS scintillator~\cite{Mufson_2015}, Ash River scintillator, Texas scintillator.}
    \label{fig:scint-placeholder}
\end{figure}

\noindent
The liquid scintillator used in the NOvA detectors was composed primarily of food-grade mineral oil, with pseudocumene serving as the scintillating agent and PPO and Bis-MSB as wavelength shifters. The mixture was produced in two slightly different blends, shown in \autoref{tab:scint_blend} and described in detail in~\cite{Mufson_2015}. The NOvA TB detector was filled with a combination of the two scintillator blends. The first block reused scintillator stored at Fermilab after decommissioning of NOvA's NDOS (Near Detector On the Surface) prototype. The second block reused some of that same blend that had been stored at University of Texas at Austin, and also used a second blend of leftover FD scintillator stored in Ash River. The differences in brightness between the two blends are illustrated in \autoref{fig:scint-placeholder}; these differences were corrected with the calibration procedures described in \autoref{sec:calib-intro}. Before the filling of the detector blocks, the different blends of scintillator were sampled at the filling nozzle and tested for optical transmission; all samples passed the NOvA-required $>95$\% purity threshold.

\begin{table}[ht]
\centering
\caption{The composition of the NOvA scintillator blends, showing mass fractions.}
\begin{tabular}{c | c | c c}
\toprule

\multirow{2}{*}{\textbf{Component}} & \multirow{2}{*}{\textbf{Purpose}} & \multicolumn{2}{c}{\textbf{Mass fraction (\%)}} \\
& & \textbf{Blends 1--2}  &\textbf{Blends 3--25} \\

\midrule
Mineral oil & Solvent & 94.91&  94.63  \\
Pseudocumene & Scintillant & 4.98&  5.23\\
PPO & Waveshifter & 0.11&   0.14 \\
Bis-MSB & Waveshifter & 0.0016&   0.0016\\
Stadis-425 & Antistatic & 0.001&  0.001 \\
Vitamin E & Antioxidant & 0.001&  0.001\\
\midrule
\bottomrule
\end{tabular}

\label{tab:scint_blend}
\end{table}

\subsection{APD Readout}\label{sec:apd}
\noindent
Each 32-cell detector module was read out by a custom 32-channel Hamamatsu APD with 85\% mean quantum efficiency over \SIrange[]{500}{550}{\nano\meter}. Both ends of each looped wavelength-shifting fiber were optically coupled to a single pixel on the APD. Each APD pixel had a unique hit threshold, determined from a measurement of the noise envelope for the relevant channel. The voltage-to-gain conversion of each APD was determined by a bench test prior to installation. A voltage of \SI{425}{\volt} was supplied, and an on-board voltage regulator was used to set a nominal gain of 100.

Noise was reduced by cooling the APDs to \SI{-15}{\celsius} with a \SI{3.5}{\volt} Peltier thermoelectric cooler (TEC). Heat was removed from the TECs via a water cooling system. Chilled water was routed along four parallel paths: two for the vertical modules and two for the horizontal modules. Each path was further split into four more parallel routes, each containing 7--8 APDs in series. To prevent condensation, nitrogen gas flowed through the volume containing the APD interface with the ends of the optical fibers, with a single line supplying all 126 APDs in series. 

\subsection{Detector DAQ \& Timing Systems}\label{sec:detdaq}
\noindent
The DAQ system for the TB detector was a slightly modified version of the DAQ used for the FD and ND. This system was designed for continuous readout buffered on several computing nodes, allowing trigger decisions to be made after data collection. While a continuous readout is crucial for collecting all of the interesting data from a neutrino beam, it posed challenges for a detector operated in a high background environment.

Connected to each APD was a front-end board (FEB) that performed the signal amplification, integration, shaping, and digitization of the APD signals. A custom 32-channel application-specific integrated circuit (ASIC) performed the signal processing and then multiplexed the channels, which were sent to an ADC~\cite{norman_2012}. NOvA uses differing FEB designs for the FD and ND. The FD, which has a low pileup rate, uses a multiplexing ratio of 8:1 with the four output groups being sent to a single quad (4 input) ADC. Due to its proximity to the NuMI neutrino source, the rate of pileup at the ND is much greater, necessitating a 2:1 multiplex ratio and four quad ADCs. The multiplexing occurs at \SI{16}{\mega\hertz}, resulting in a sample every \SI{500}{\nano\second} for FD-type FEBs (v4.1) and a sample every \SI{125}{\nano\second} for ND-type (v5.2). To allow studies of potential differences between the FD and ND electronics, the TB detector was outfitted with both FEB versions. FD-types made up 118 of the 126 FEBs with ND-types being used for the remaining eight. 

The digitized signals were passed to a field-programmable gate array (FPGA) on the FEB, which determined whether the signal was high enough to be a potential hit. Rather than apply a hit threshold directly on the recorded ADC value, dual-correlated sampling (DCS) was used to remove low frequency noise and the impact of a possible wandering baseline. The DCS was performed with three time steps, where the value for sample $n$ was the ADC difference with respect to the $n-3$ sample. If the DCS value was above the threshold, then the sample and the prior three samples were all recorded. The timing resolution achieved was 
\SIrange{10}{20}{\nano\second} (\SIrange{5}{10}{\nano\second}) for the FD (ND) FEBs. Using bench measurements of the gain of the APDs, the peak ADC value as determined from the DCS
was converted into an uncorrected estimate of the number of observed photoelectrons (PE) for each energy deposit. These estimates were then converted into corrected MeV units via the calibration process described in \autoref{sec:calibration}.

Signals from up to 64 FEBs were collected by Data Concentrator Modules (DCMs)~\cite{norman_2012}. Each DCM had an FPGA that received the signal packets and grouped them into \SI{50}{\micro\second}-long slices, which were further packaged into \SI{5}{\milli\second} slices, known as millislices. The DCMs sent their outputs to a set of computing nodes arranged as a ring buffer where the final decision to keep or drop the data was made by data-driven triggers (DDTs) in addition to the beam-based triggers (\autoref{sec:trigger}). 

To accurately group the FEB hits in time, the DCMs were connected to a precise timing system~\cite{Norman:2012zzc} consisting of a GPS-based external clock and a timing distribution unit (TDU). The DCMs were connected in series to the TDU with the final DCM having a loop-back feature, which was used to measure the cable delays for each link in the system. The TDU sent a pulse down the timing chain and measured the time difference between the initial signal and the echo passed back up the chain by the loopback feature. Time syncing across the DCMs was achieved via a TDU signal declaring the value of a future timestamp to be set on the receipt of a sync signal. When the signal was sent down the timing chain, each DCM counted down its delay time. Once that countdown ended, the DCM clocks took over, starting at the declared timestamp.

The NOvA DAQ allowed multiple triggers to run independently, with each streaming data from the front-end buffers to disk in separate data files.  It was thus possible for the same event to exist in multiple data streams if it matched more than one trigger condition.  The events were timestamped using a \SI{64}{\mega\hertz} clock.  The primary data stream for TB data was the beamline trigger, described in \autoref{sec:trigger}.  Activity-based DDTs were used to collect out-of-spill data, such as from interactions of cosmic particles used in the calibration procedures discussed in \autoref{sec:calibration}.  Finally, the ``beam spill" data stream recorded all data from the beam window and was used to determine the dead time in the front-end system as described in \autoref{sec:ops-dataquality}. 

\subsection{Simulation of the NOvA TB Detector}
\label{sec:nova-simulation}
\noindent
The NOvA TB detector was simulated using \geantfour~\cite{GEANT4,GEANT4dev,GEANT4devapp} version 4.11.0 with the  \textsc{qgsp\textunderscore{}bert\textunderscore{}hp\textunderscore{}emz} physics list. The particles used as input to the detector simulation were simulated in various modes, described in \autoref{sec:tertiary-beamline-sim}. In all modes, the tertiary particles were propagated through the NOvA detector by \geantfour to produce energy deposits in the detector's active material. The list of energy deposits was then passed to a parameterized front-end simulation that converted energy deposits into scintillation light,
transported scintillation light to the APD, and simulated the readout electronics response~\cite{Aurisano:2015oxj}. The final output was formatted like raw data.

In the ``Data-Seeded'' simulation mode, 
the simulated tertiary particles were generated with the four-momenta of real data particles. This relied on the measured momentum estimated using the \wc\ track directions on either side of the magnet (\autoref{sec:tracking-mom}), and preliminary particle identification which additionally relied on the measured ToF (\autoref{sec:tracking-pid}). 

The uncorrected Data-Seeded simulation did not account for energy losses in the downstream instrumentation (\autoref{sec:tracking-mom}). Prior to the final estimates of downstream energy loss, which are particle-specific and are not included in this paper, the uncorrected Data-Seeded simulation was used to check the alignment and rotation of the simulated tertiary beamline components with the NOvA detector (\autoref{fig:placeholder-geobl}), to explore the processes included in the \geantfour simulation, and to examine the detector reconstruction in the absence of energy loss corrections. Future analyses will derive full energy loss corrections in the tertiary beamline components by tuning the Tertiary Simulation described in \autoref{sec:tertiary-beamline-sim}. These corrections will be combined with the data momentum measurements to produce a Corrected Data-Seeded simulation, which will be compared with NOvA detector data, allowing for 
direct measurements of the detector response to particles of known momenta.

\begin{figure}[ht]
    \centering

    \includegraphics[width=0.32\textwidth]{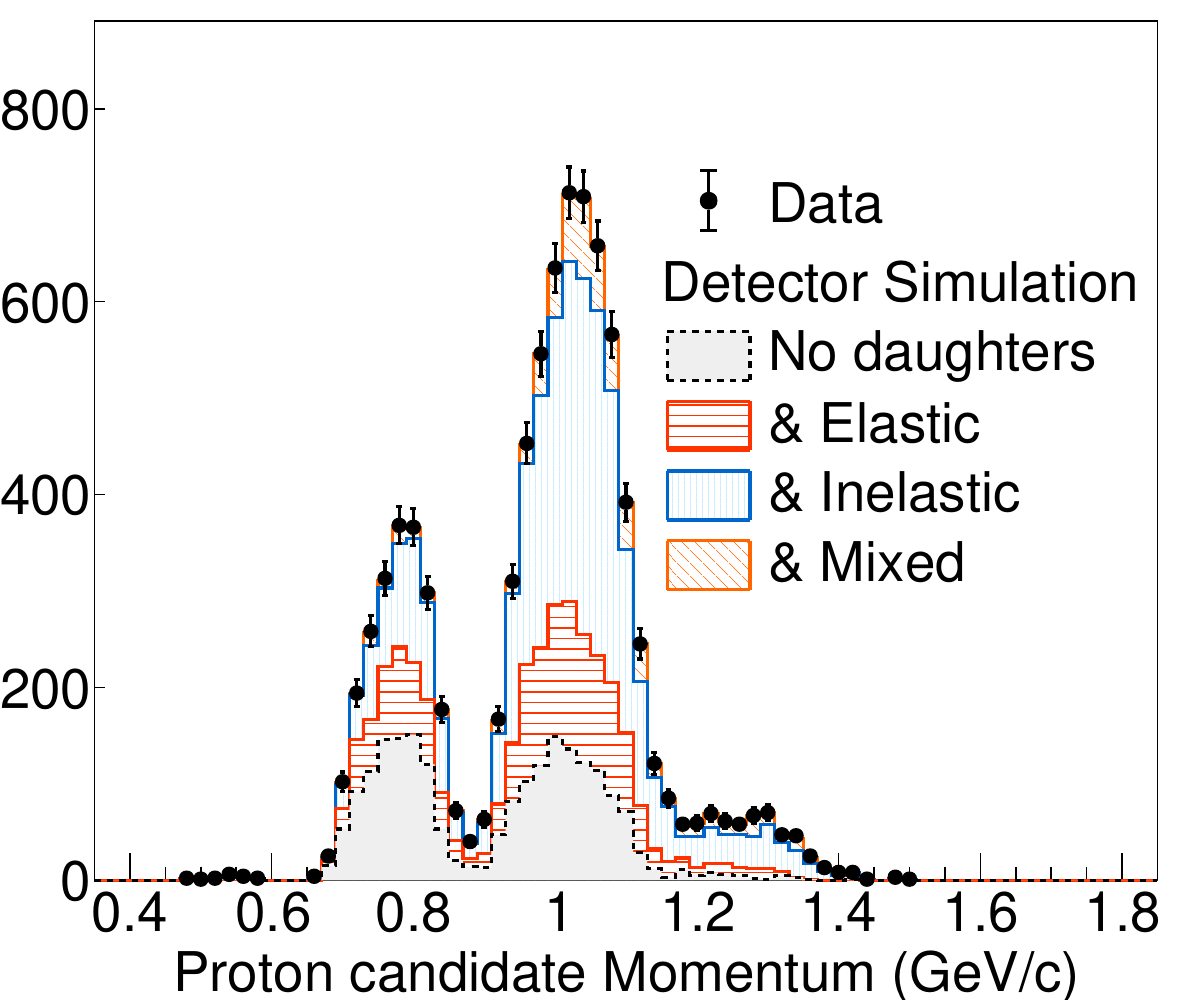}
\includegraphics[width=0.32\textwidth]{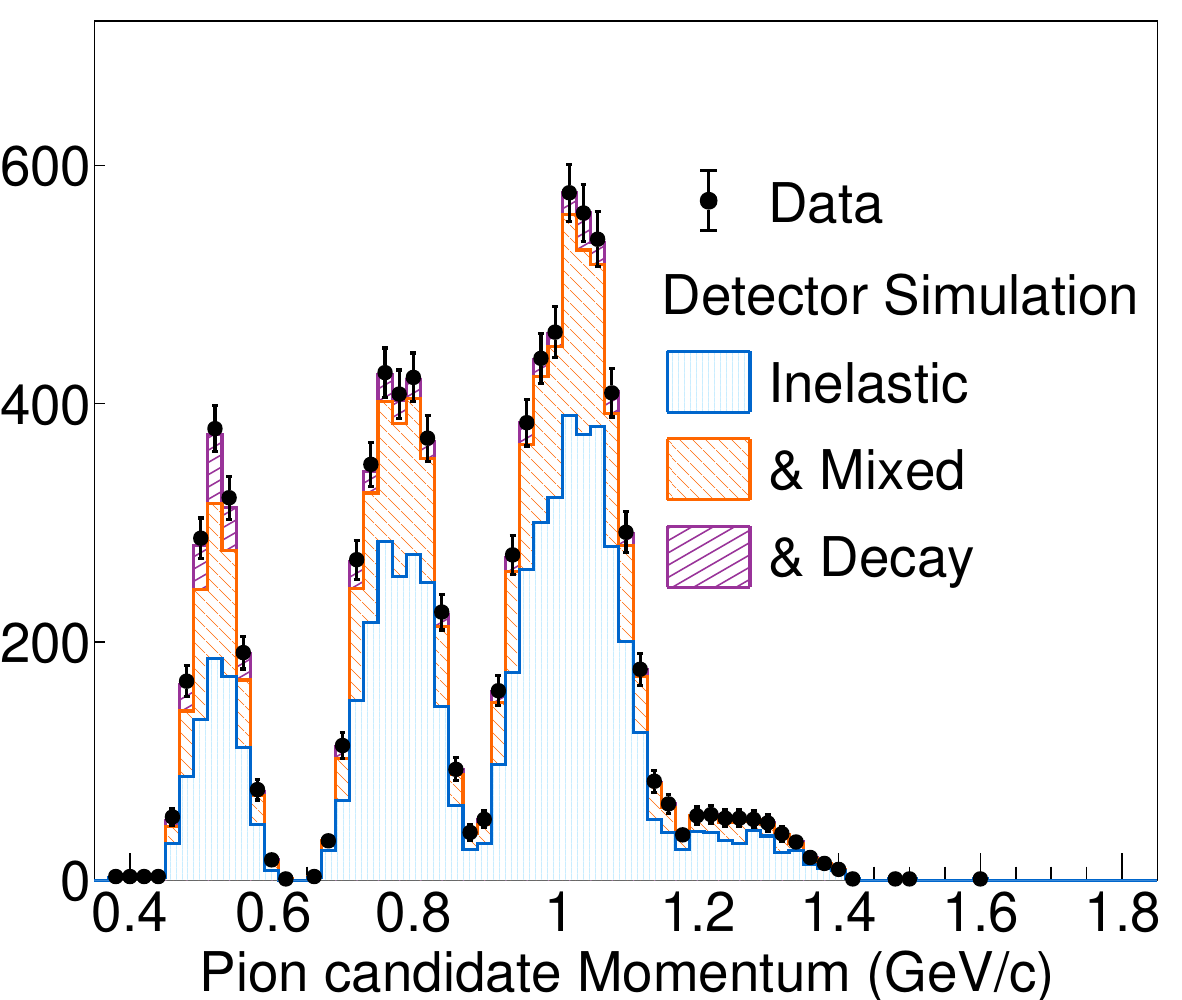}
\includegraphics[width=0.32\textwidth]{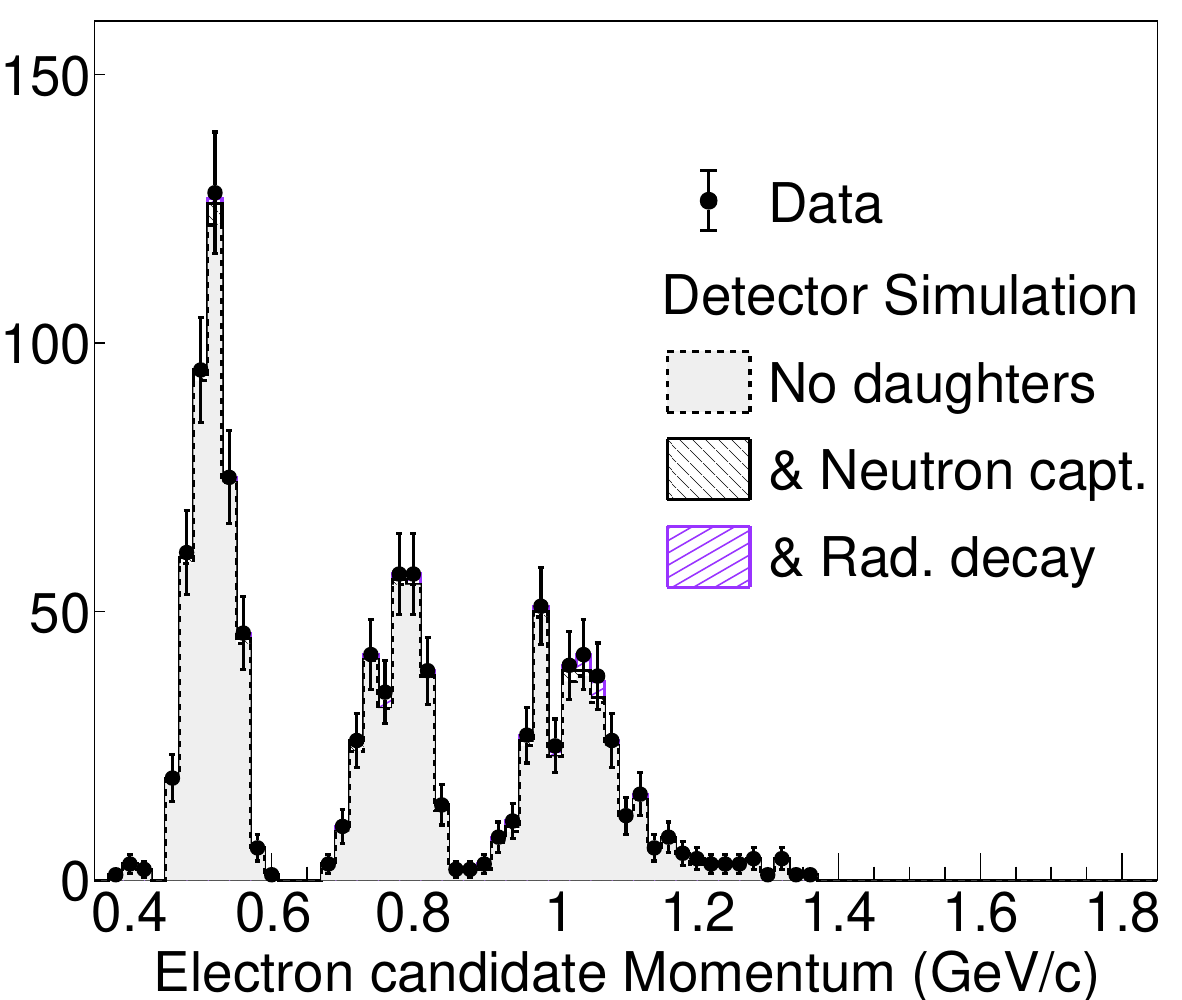}

    \caption{The particle momentum in data compared with uncorrected Data-Seeded \geantfour simulation for (left) proton, (center) pion, and (right) electron candidates. The numbers of entries in each \SI{20}{\mega\electronvolt}/c momentum bin are shown on the $y$-axes. The peaks in momentum correspond to the magnet current settings used, described in \autoref{sec:ops_data}. The simulation processes given in the legends are those provided by \geantfour, and are detailed in \cite{GEANT4}. Only processes that contribute $\geq1\%$ of the distribution have been given legend entries. Proton, pion, and electron candidates are identified using loose selection criteria similar to those given in \autoref{tab:detsel}.
    }
    \label{fig:detsim}
\end{figure}

\begin{figure}[ht]
    \centering
 \includegraphics[width=0.32\textwidth]{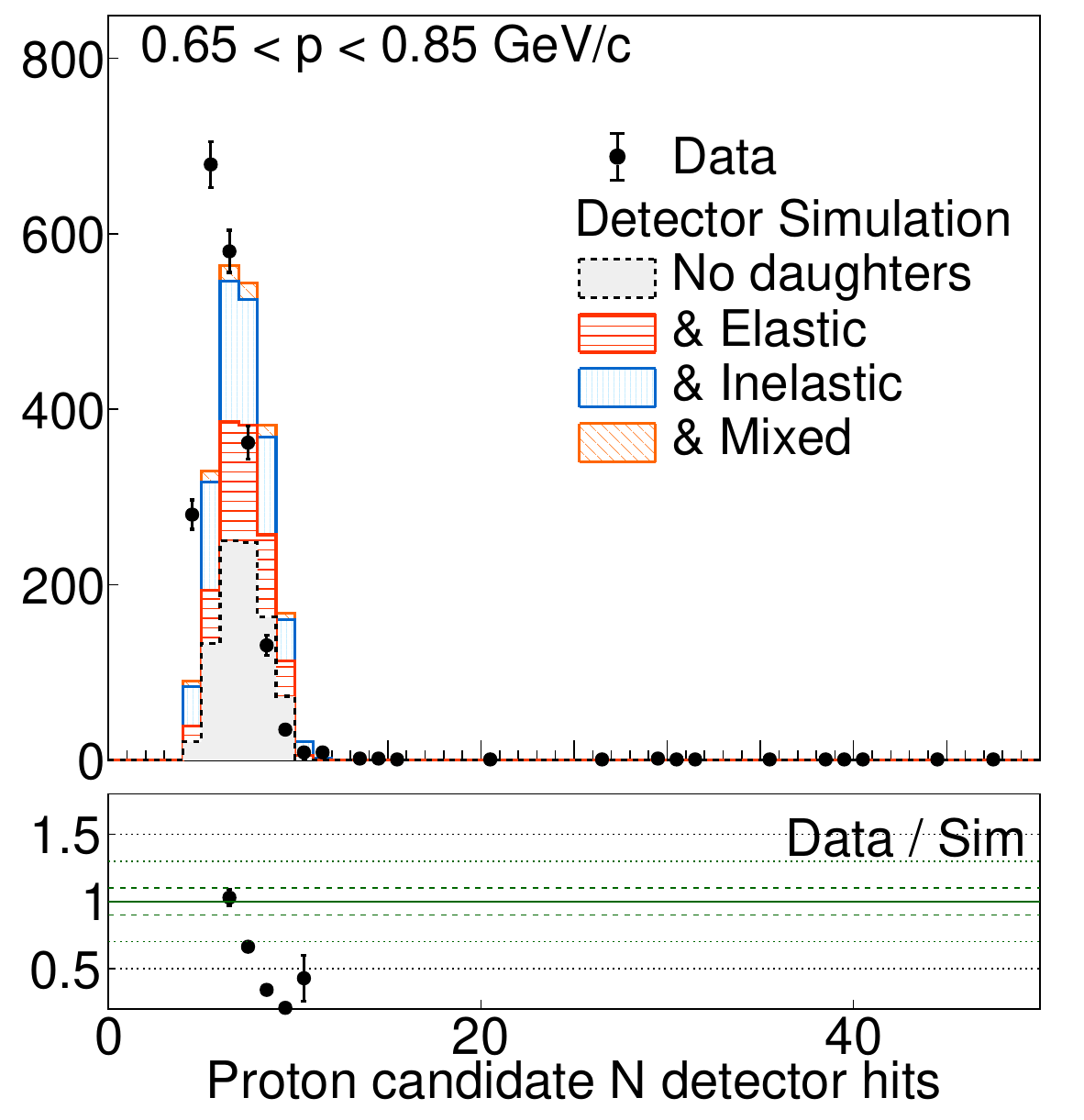}    
\includegraphics[width=0.32\textwidth]{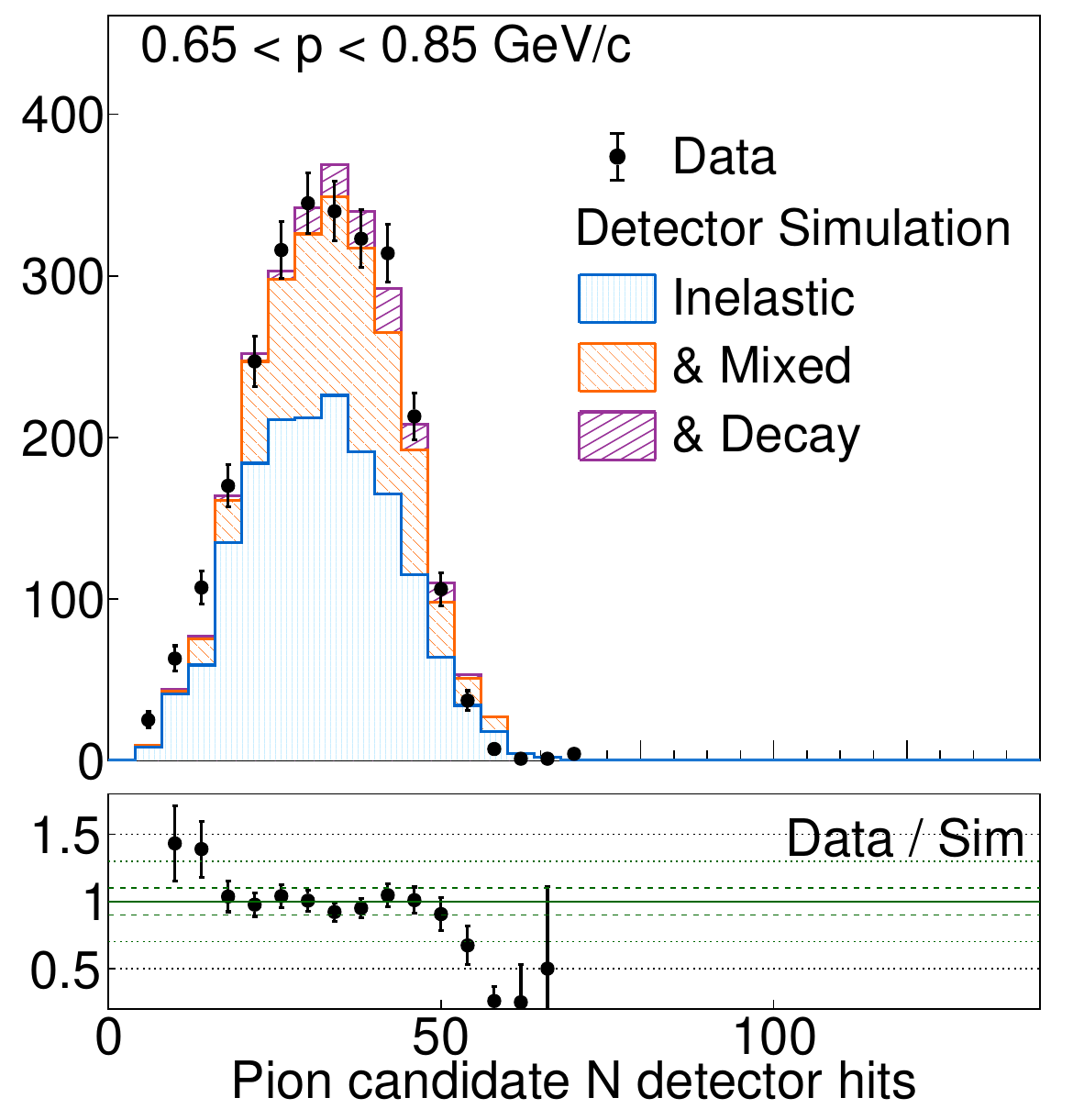}
\includegraphics[width=0.32\textwidth]{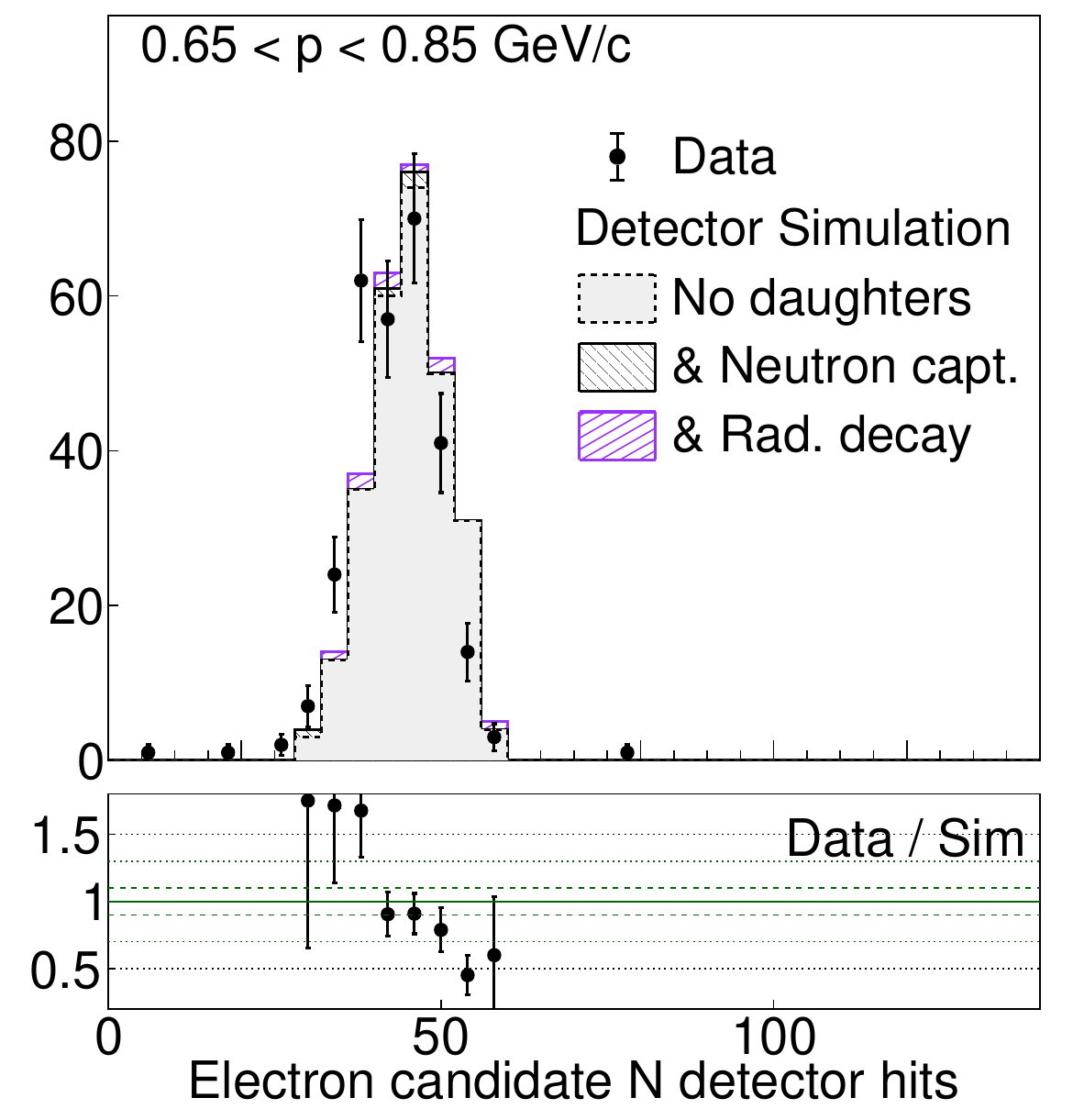}

     \includegraphics[width=0.32\textwidth]{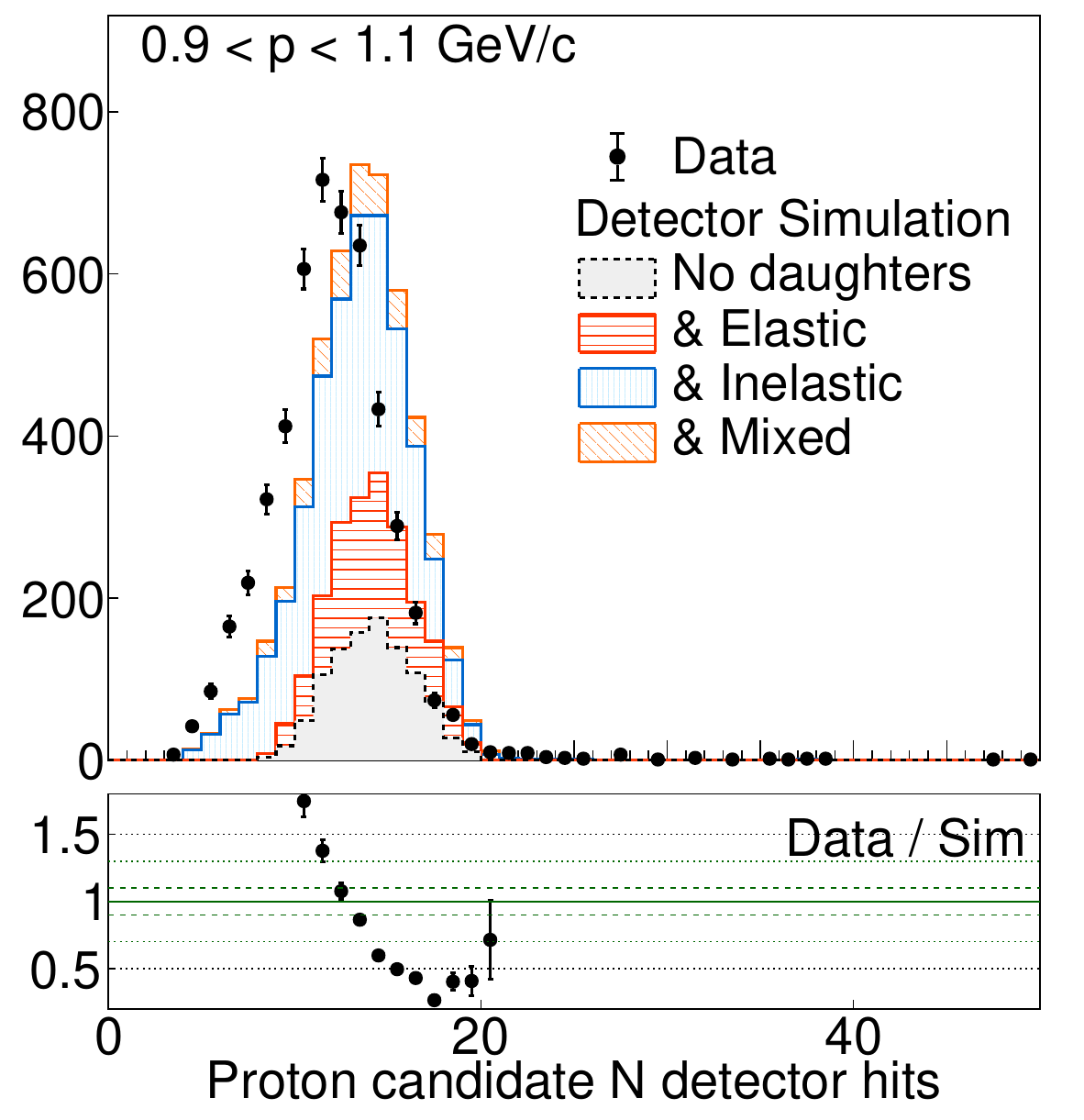}    
\includegraphics[width=0.32\textwidth]{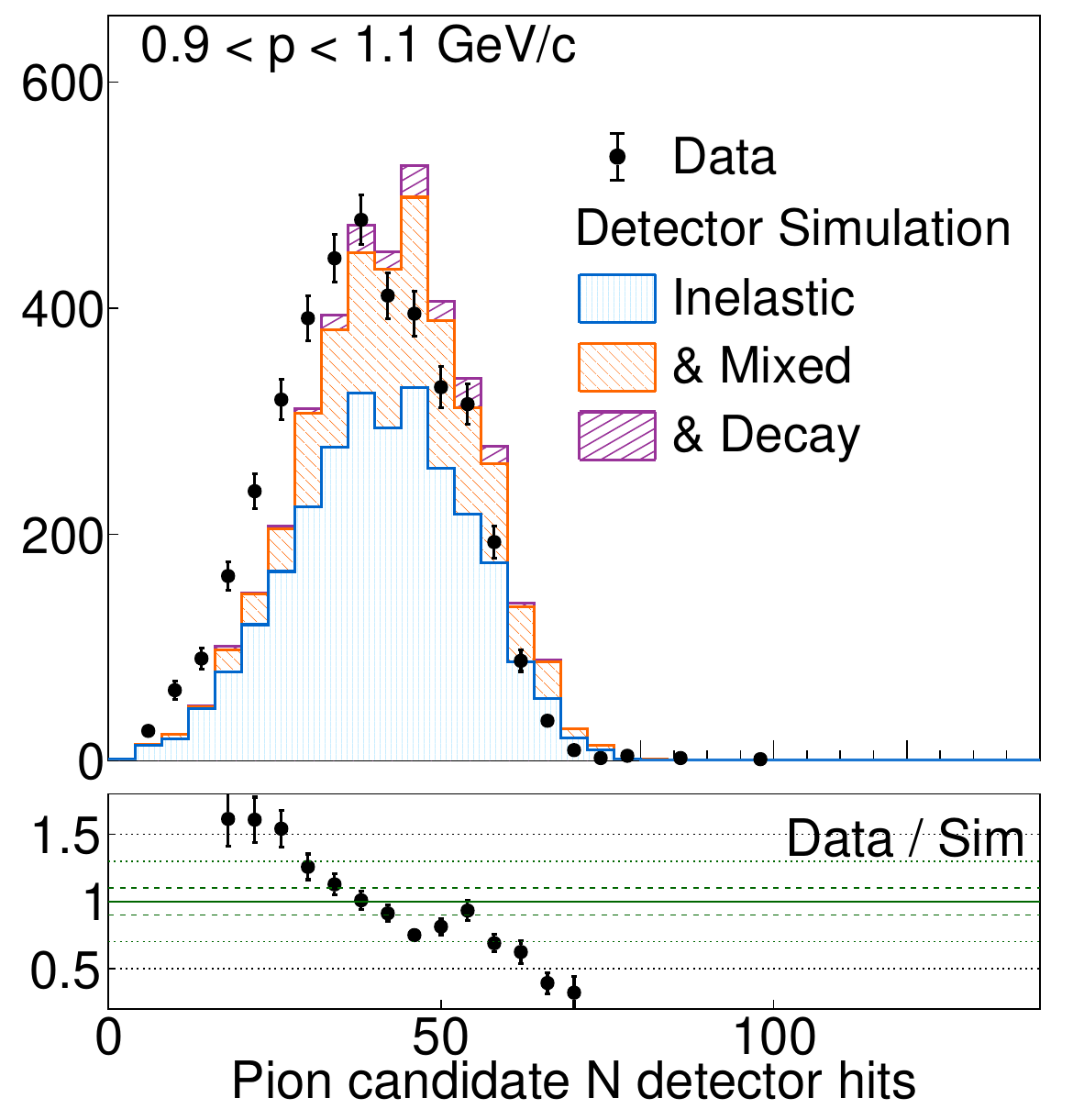}
\includegraphics[width=0.32\textwidth]{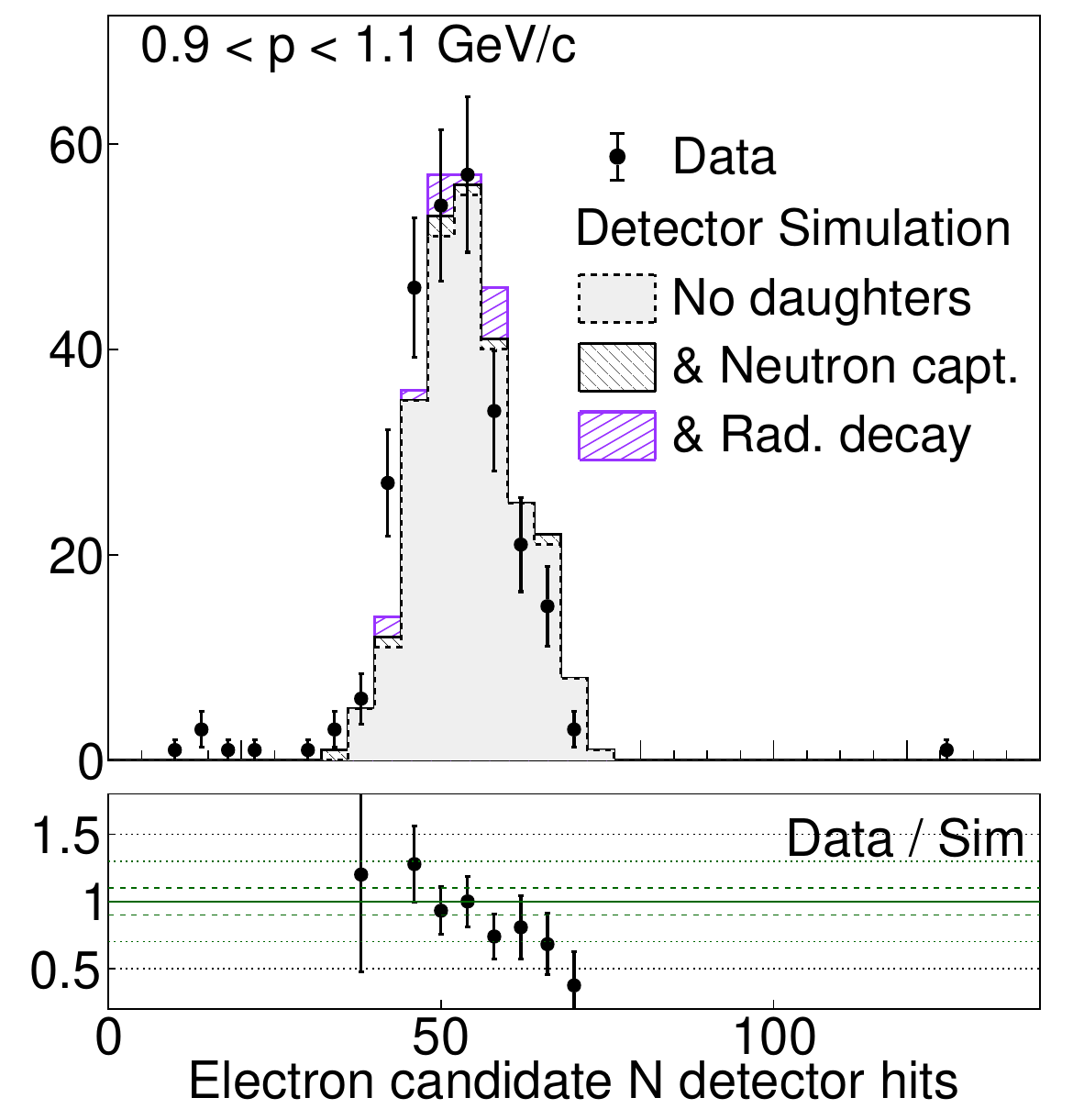}
    \caption{The numbers of detector hits in data compared with uncorrected Data-Seeded \geantfour simulation for (left) proton, (center) pion, and (right) electron candidates. The top row shows the momentum range $0.65-0.85$\,~GeV/c, while the bottom row shows the range $0.9-1.1$\,~GeV/c. The numbers of entries in each bin are shown on the $y$-axes. For protons, one bin corresponds to one detector hit, whereas for pions and electrons one bin corresponds to four detector hits. The simulation processes given in the legends are those provided by \geantfour, and are detailed in \cite{GEANT4}. Only processes that contribute $\geq1\%$ of the distribution have been given legend entries. Proton, pion, and electron candidates are identified using loose selection criteria similar to those given in \autoref{tab:detsel}.
    }
    \label{fig:detsim2}
\end{figure}

 The uncorrected Data-Seeded simulation resulted in identical momentum distributions for the reconstructed beamline tracks in data and the truth particles in simulation, as demonstrated in \autoref{fig:detsim}. Example distributions of the numbers of reconstructed detector hits in data and uncorrected Data-Seeded simulation are shown in \autoref{fig:detsim2}. The distributions generally agree better for pions and electrons than for protons, though some disagreements remain due to uncertainties in the underlying simulation of the detector and in particle interactions with detector materials; these will be probed in future analyses. 

The NOvA light model uses \geantfour's accounting of energy deposited in a detector cell as input, and has several tunable parameters. The number of photons, $N_{\gamma}$, produced in a detector cell is modeled as
\begin{equation}
    N_{\gamma} = F_{x,y}(Y_{s}E_{\rm Birks} + \epsilon_{c}C_{\gamma}),
\end{equation}
where $Y_{s}$ is the scintillation light yield (photons/MeV), $E_{\rm Birks}$  is the total energy deposited in the scintillator (corrected for quenching effects), $\epsilon_{c}$  is the efficiency of the scintillator absorbing and re-emitting Cherenkov light, and $C_{\gamma}$  is the number of photons from Cherenkov radiation.  Two scale factors, $F_{x,y}$, one for each of the horizontal and vertical views of the detector, account for the APD quantum efficiency and factors related to the brightness level and the fraction of photons collected. 
To account for light attenuation, $N_{\gamma}$ is multiplied by the fiber transmission function, $T_f$, which models attenuation as the sum of two exponentials,
\begin{equation}
    T_{f}=c_{0} e^{-x/L_{0}}+c_{1} e^{-x/L_{1}},
\end{equation}
where $L_{0}$  and $L_{1}$  are short and long fiber
attenuation lengths, $c_{0}$  and $c_{1}$  are the fractions that each component
contributes to the overall attenuation $(c_{0}+c_{1}=1)$, and $x$ is the distance from the
readout electronics.
The final simulated number of photons is taken from a Poisson distribution with a mean equal to the attenuated number of photons.

\section{Operations and Data Collection} 
\label{sec:operations}
\begin{figure}
\includegraphics[width=\textwidth]{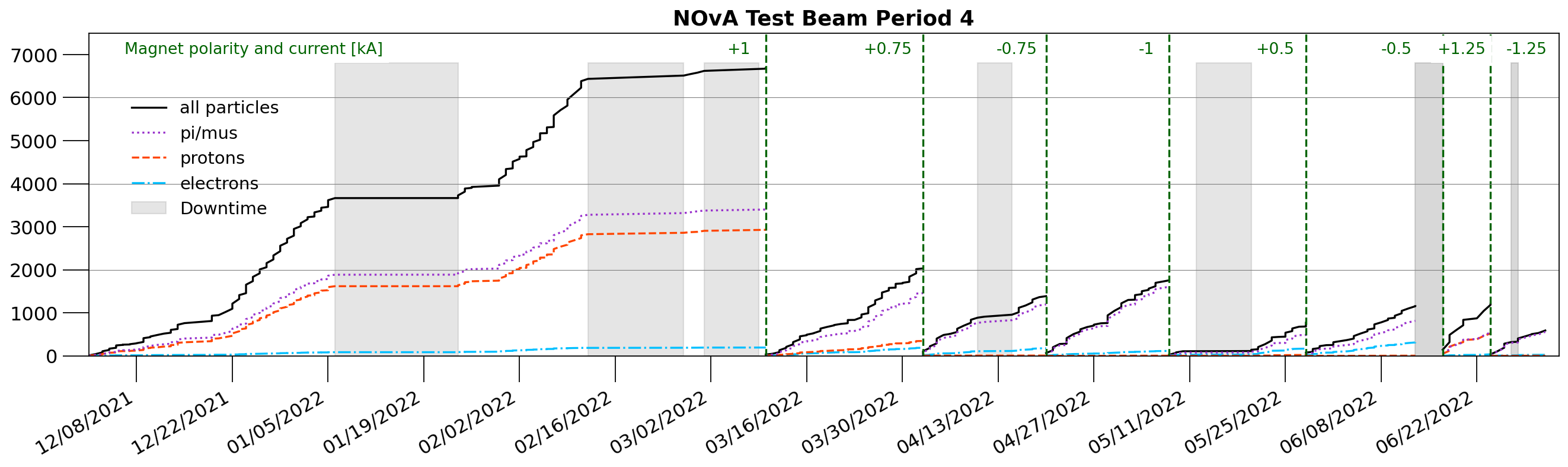}
\caption{Cumulative estimated particle counts during the 2022 Run (Period 4). The magnet polarity and current were regularly changed to facilitate collection of particles with various charges and momenta. The gray bands show the periods of beam downtime caused by various factors.}
\label{fig:cumulatives}
\end{figure}

\noindent
As indicated in \autoref{tab:DataTaking}, the data were collected in four periods. In total, close to 34,000 particle interactions were determined to pass preliminary ``analysis-quality'' cuts, including full reconstruction, complete beamline information, physical ToF and momentum measurements, and no indications of dead time in the NOvA data stream (see \autoref{sec:ops-dataquality}). These were collected at four analyzing-magnet settings and both polarities to accumulate positive and negative particles at central momenta of \SI{500}{MeV/c}, \SI{750}{MeV/c}, \SI{1}{GeV/c}, and \SI{1.25}{GeV/c}. The estimated cumulative counts are shown in \autoref{tab:particle-counts}, and the cumulative counts versus time for Period 4 are shown in \autoref{fig:cumulatives}. The loose selections used for these estimates are described in \autoref{sec:performance}.

\renewcommand{\arraystretch}{1.2}
\begin{table}
    \centering  
     \caption{Overview of the \datataking\ periods.}
    \begin{tabularx}{0.65\textwidth}{ l l c  }
      \toprule
\textbf{Period} & &\textbf{Date}\\
\midrule

\textbf{Commissioning:} &Period 1 &  29 Mar--7 Jul 2019\\

\textbf{2020 Run:} &Period 2 &  5 Dec 2019--20 Mar 2020\\

\textbf{2021 Run:} &Period 3 &  12 Jan--27 June 2021\\

\textbf{2022 Run:} &Period 4 &  30 Nov 2021--10 Jul 2022\\

 \midrule
\bottomrule
    \end{tabularx}
       
        \label{tab:DataTaking}
\end{table}

\begin{table}
\caption{Estimated particle counts during data-taking for each magnet polarity and current setting. These estimates were based on the momenta and times of flight of preliminary reconstructed tertiary particle candidates. Final reconstruction of the data and refinements to detector uptime, data quality, and particle identification are not included in these estimates.}
\centering
\begin{tabular}{p{0.15\textwidth}|p{0.16\textwidth}|p{0.1\textwidth}p{0.1\textwidth}p{0.1\textwidth}p{0.1\textwidth}p{0.1\textwidth}
}
\toprule
&\textbf{Magnet} & \multicolumn{5}{c}{\textbf{Particle Counts}}  \\
 &\textbf{Current (A)}& $\boldsymbol{e}$ & $\boldsymbol{\pi/\mu}$  &$\boldsymbol{K}$ & $\boldsymbol{p}$ &$\mathbf{Totals}$ \\
 \midrule
 \textbf{Period 2} & 
$+500$ & 29 & 100 & 1 & 1 & 131\\
&$+1000$& 105 & 1773 & 46 & 1399  & 3323 \\
\midrule
 \textbf{Period 3} & 
$+500$ & 360 & 1898 & 13 & 46  & 2317\\
&$+750$  & 132 & 1617 & 9& 222  & 1980\\
&$-750$ & 122 & 1546 & 2 & 1  & 1671\\
&$+1000$& 247 & 4880 & 121 & 3208  & 8456 \\
\midrule
 \textbf{Period 4} & 
$+500$ & 189 & 547 & 0 & 20  & 756 \\
&$-500$ & 313 & 814 & 4 & 4  & 1135\\
&$+750$ & 200 & 1567 & 18 & 367  & 2152 \\
&$-750$ & 175 & 1194 & 3 & 7 & 1379\\
&$+1000$& 193 & 3405 & 49 & 2937 & 6584 \\
&$-1000$& 115 & 1605 & 11  & 4 & 1735 \\
&$+1250$& 44 & 793 & 81 & 784 & 1702\\
&$-1250$& 28 & 618 & 5 & 3 & 654\\
 \midrule
 \textbf{All Periods} & (+) & 1499 & 16580 & 338 & 8984 & 27401 \\
  & (\,--\,) & 753 & 5777 & 25 & 19 & 6574 \\
  & Total & 2252 & 22357 & 363 & 9003 & 33975 \\
 \midrule
\bottomrule
\end{tabular}

\label{tab:particle-counts}
\end{table}

The data taken in Period 1 were used mostly for commissioning and to determine the optimal running conditions. Period 2 was a study period used to quantify and investigate the effects of the muon plume (\autoref{sec:ops-plume}), the various available front-end firmware versions, and the beam trigger conditions.  The Period 2 data were collected before the installation of the muon plume shielding so they are the most affected by data quality issues and were collected at a relatively low rate. Given the larger quantity of higher-quality data collected in Periods 3 and 4, Period 2 data will likely be excluded from most analyses, and possibly retained only in analyses that require an increase in statistics.  

All of Period 3 data were collected following the installation of the plume-mitigation shielding and in what was considered the best-practice configuration of front-end firmware and beamline trigger. The final optimizations were completed during this \datataking\ period, including improving the beam delivery and topping up the oil in the underfilled horizontal NOvA cells (\autoref{sec:PVC}). The data collected after 30 April 2021 (towards the end of Period 3 in \autoref{tab:DataTaking}) therefore had the optimal beam configuration, the plume shielding in place, a completely filled NOvA detector, and the preferred detector firmware and beam trigger. Future analyses will use these data and those from Period 4, which are the highest-quality datasets and comprise roughly half the total dataset.

\subsection{Beam Configuration and Delivery}
\label{sec:ops_beam}

\noindent
Period 3 operations were dedicated to optimization of the beam configuration and delivery.  Given the requirements of the integrated systems, in particular the readout of the NOvA TB detector, simply increasing the rate of particles incident on the MCenter target did not necessarily increase the volume of analysis-quality data.  Factors including the beam angle on the primary target and the configuration of the momentum-selecting collimator in the secondary beamline were crucial to data quality.  

Runs were taken with various beam intensities and collimator aperture settings.  Measurements from the beam control and monitoring system, the NOvA TB instrumentation and detectors, and additional auxiliary counters placed on the front face of the detector and strategically around the experimental enclosure were used to determine the optimal configuration.  The particle rates from the counters (correlated with beamline backgrounds) and the rate of beamline triggers that displayed no associated electronics dead time in the detector, referred to as ``good triggers,'' were particularly useful metrics. The studies were repeated for each tertiary-particle momentum setting. The optimal running modes were applied as the default settings for the remainder of the data-taking and corresponded to an intensity of $3\times 10^{9}$ to $6\times 10^{9}$ protons per pulse (depending on the selected tertiary-particle momentum), a 64\,\,GeV/c momentum for secondary particles, and a 10\,\,mm aperture for the secondary beamline collimator.

 \subsection{Data Quality }
\label{sec:ops-dataquality}

\noindent

\noindent
The primary data quality challenge for the NOvA TB experiment was dead time caused by electronics buffer saturation. To counter this, an offline algorithm for identifying affected events was developed.

\begin{figure}
\centering
    \includegraphics[width=0.9\linewidth]{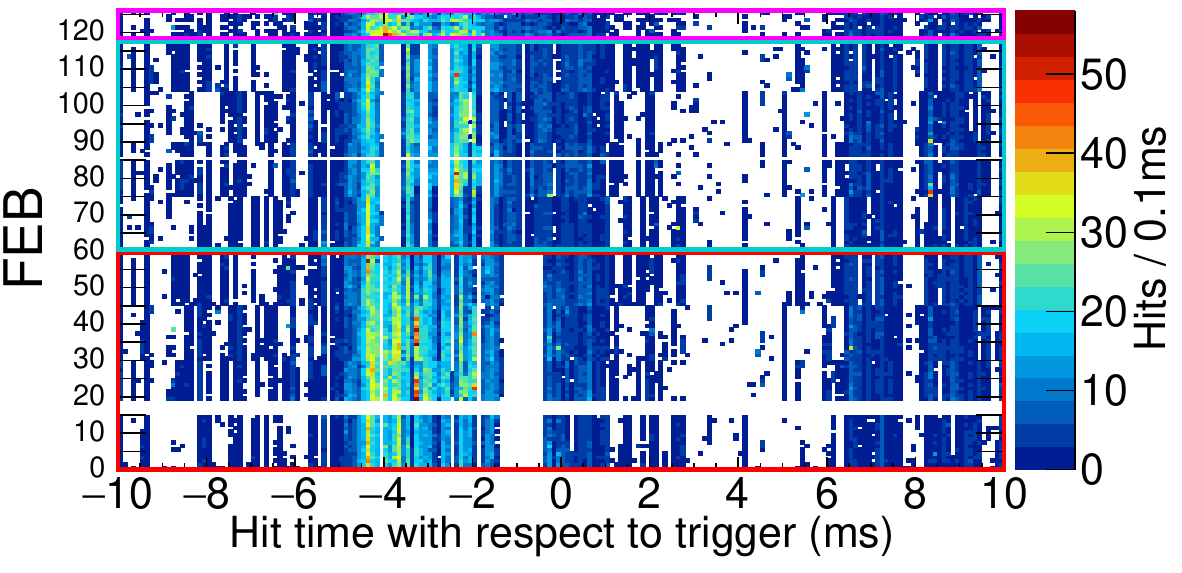}
    \caption{Hit map showing the effects of dead time on NOvA Test Beam data.  Individual hits representing light collected in each cell are counted for each FEB in a 20~ms time window around a single beam trigger.  The first 60 FEBs are aggregated by DCM1, the next 58 by DCM2, and the final 8 by DCM3; this is highlighted by the corresponding colored boxes.  FEB shutoffs, where the front-end buffer saturates, are seen on a few individual FEBs; these are characterized by a long period of dead time and affect the boards individually (in this case, FEBs 16--19 and 86 are down). DCM dead time extends across all FEBs on a given board (and is typically of shorter duration), seen for example on DCM1 from about \SI{1.5}{\milli\second} to \SI{0.5}{\milli\second} before the trigger and at multiple times slightly earlier on DCM2. The higher activity around \SI{-4}{\milli\second} and lower activity around \SI{4}{\milli\second} are due to the substructure of the secondary beam.
    }
    \label{fig:DeadTime}
\end{figure}

There were two mechanisms that led to detector dead time, one arising from the FEB, the other from the DCM.  Examples of both are shown in \autoref{fig:DeadTime}. The periods of FEB dead time were referred to as ``shutoffs," since the boards stopped collecting data upon becoming saturated.  The DAQ system re-enabled all the FEBs at a rate of \SI{10}{\hertz}, resulting in relatively long periods of downtime of up to \SI{100}{\milli\second}.  Since this saturation occurred at the very first stage of the readout, it impacted the FEBs individually and was most frequent in the high-activity part of the detector (the upper-west quadrant, where the muon plume was present). The FEB firmware set a bit in the header of the data format to signify when it was close to saturating. The FEB status bit saved by the firmware was found to be an unreliable means of flagging the shutoffs, particularly in the newer firmware versions; since no data could be sent when the FEB was down, capturing the change in the status bit required that there was activity in that board just before the shutoff occurred.
 
 The second stage of the readout where saturation occurred was at the DCMs, which aggregated data from multiple front-end boards. This affected groups of FEBs simultaneously, and was more sporadic and unpredictable than the FEB shutoffs.  DCM saturation typically lasted for a much shorter period of time while the buffer was cleared, but there were significant variations.  Additionally, DCM saturation did not necessarily affect data from all the FEBs identically, and was therefore more difficult to identify.  Unlike in the FEB case, there was no indication from the hardware that DCM saturation had occurred.  Since DCM3 interfaced with only 8 FEBs, each of which had faster timing, dead time was not observed in its data.

FEB shutoffs were relatively easy to identify by their hit rates given their very long periods of inactivity.  DCM dead time was identified in a similar manner, but due to its nature it was challenging to define its exact beginning and end.  Hand scans of several hundred events led to a ``dead-DCM" criterion of \SI{80}{\micro\second} of inactivity on all FEBs interfaced to a given DCM.
 This criterion led to occasional false positives and negatives, but its performance was sufficient for the vast majority of events.  Following the identification of affected FEBs and DCMs for a particular event, a decision was made as to how to proceed, with the typical course of action being to exclude the event from further processing and analysis. \subsection{Environmental Stability and Monitoring} 
\label{sec:ops-environment}
\noindent
The MC7 experimental hall in FTBF had limited insulation from the external environment. To counter this, a new large-capacity HVAC system was installed to ensure environmental stability, which was monitored by multiple temperature, humidity, and dew-point sensors installed in various locations of the experimental hall and detector. 

The variation in temperature was limited to about $\pm$\SI{4}{\celsius}. The typical temperature coefficient for signal delay through the cables used in the experiment (RG-58/U and Cat-6 twisted pair) is about \SI{60}{\pico\second\per\kilo\metre\per\celsius}.
The length of cables used to carry photodetector signals from the ToF and Cherenkov subsystems to the front-end electronics was limited to \SI{20}{m}. The corresponding shift in timing was less than \SI{1}{ps} and negligible compared to the width of the signals and electronic gates employed in the readout.

The front-end electronics boards that digitized the signals from the NOvA TB detector APDs and the wire chambers had very short signal paths (less than \SI{10}{cm}) and any shift in delay was negligible compared to the clock period of the corresponding digitizers. The clock system used to synchronize the NOvA TB detector FEBs has a period of \SI{15.6}{ns}.
The clock signals were delivered through cables shorter than \SI{3}{m} and changes in propagation delay for these signals due to temperature drift were also negligible.

Controlling the humidity was more difficult. Data-taking was automatically paused and the detector electronics shut down whenever the dew point in the experimental hall rose above \SI{10}{\celsius}, most often in late spring and summer. This measure was implemented to avoid noisy events from the NOvA APDs and protect the FEBs from potential damage due to condensation. %

\subsection{Data Processing}
\label{sec:ops_data}
\noindent
The first stage in data processing was converting the data from the raw format saved by the DAQ systems to the format used in analyses.  
While this approach utilized the existing data frameworks for the FD and ND, it required building additional machinery to accommodate the new NOvA TB instrumentation. Custom data formats were defined to hold the information required for offline processing and a new software framework was developed to interface with the DAQ-formatted files. During this first stage the separate data streams saved by the NOvA TB detector and the beamline were merged.  

Given that the systems operated completely independently, there was a possibility that data would exist for some events in one stream but not the other. The NOvA TB detector data are considered the primary data stream, which can be used without the instrumentation stream. Any instrumentation information that existed for each event was processed and added to that data stream. Some events had no associated data from the instrumentation; these were excluded from most analyses. Conversely, there were events with no NOvA TB detector data; these were omitted from the offline data stream irrespective of the presence of instrumentation information. Events in the NOvA TB detector and instrumentation were considered matched if they were within 8 ticks of the NOvA \SI{64}{\mega\hertz} clock; this allowed for a small offset in the times provided by the TDUs due to stochastic delays present in the system.  
\noindent

During operations, the NOvA TB detector DAQ saved data from other trigger sources.  The ``beam spill'' data stream, containing everything read out by the NOvA TB detector during the full \SI{4.2}{\second} spill, was converted to offline format without reconstruction being applied.  This was used in the dead-time analyses described in \autoref{sec:ops-dataquality}.  An ``activity'' data stream was defined by a data-driven trigger designed to determine if the level of activity in the detector met the threshold to be useful for detector calibration. This data stream was the primary source of cosmic ray particles saved during operations and was processed by calibration-specific software (see \autoref{sec:calibration}).

\section{Detector Calibration}
\label{sec:calibration}

\noindent
The calibration of the NOvA TB detector followed the same procedure as for the ND and FD. The procedure corrected for the attenuation of light along fibers, removed differences in response among detector elements, and provided a conversion factor from collected charge to physical energy units. Cosmic muons were used as the standard candle that allowed direct comparisons of deposited energy among the detectors. 

Each correction was calculated offline for a position $i$ within the detector, determined by the plane and cell number, and the distance $w$ along a cell measured from the readout end. The energy deposited in the scintillator, $E_{dep}$, is estimated as 

\begin{equation}\label{eq:edep}
    E_{dep} [\SI{}{\mega\electronvolt}]=
    \text{Signal}[\text{ADC}] \times g \times
R_i\left(t\right) \times A_{v}\left(t\right),
\end{equation}
\noindent
where Signal[ADC] is the ADC signal, $g$ is the APD gain factor, $R_i\left(t\right)$ is the relative calibration correction at position $i$, and $A_v\left(t\right)$ is the absolute calibration correction for view $v$.  

\subsection{Calibration Samples}
\label{sec:calib-intro}
\noindent
The calibration of the NOvA detectors was performed using cosmic ray muons.  For the ND and FD, NOvA uses the \textsc{CRY} Monte Carlo generator \cite{CRY} to create a sample of cosmic muons that are passed through the same reconstruction and selection as data. For the TB detector, a data-based simulation of cosmic muons was developed~\cite{TeresaThesis, RobertThesis} instead. The primary motivation for this development was to improve efficiency, in terms of processing time and disk space, by simulating only particles that would enter the detector volume. 

A sample of cosmic muons was selected from Period 4 data, and their reconstructed vertex positions and four-momenta were used to seed a \geantfour \cite{GEANT4, GEANT4dev, GEANT4devapp} simulation of the detector. The vertices were taken as the intersections of the cosmic tracks with the edges of the detector, and four-momenta were calculated using an adapted BreakPoint fitter (BPF) method \cite{BreakPointFitter,breakpoint}. As for the ND and FD simulations, the TB detector cells were divided into 12 equally populated brightness bins based on the uncorrected average response in their centers. The variations in brightness among cells, due to different scintillators and fiber quality, were included in the simulation using this binning.

 Prior to the determination of the relative calibration corrections, both the data and the simulation were corrected for known biases introduced by the APD thresholds (\autoref{sec:apd}) and by the self-shielding of the detector. The corrections for these effects were determined using the simulation and applied to both data and simulation as a function of position, $i$.
 
The threshold correction was necessary because photons produced far from the readout, which was positioned on the east side of the TB detector, were more likely to be attenuated below the APD threshold. The correction for this effect was calculated as the ratio between the PE recorded at the readout, $P\!E_{\rm rec}$, and the PE that would have been recorded in the absence of a threshold, $P\!E_{\lambda}$. The variation in this ratio as a function of the distance along each cell, $w$, was measured for all cells in the TB detector. The size of the correction factor was found to be less than 5$\%$ in all cases, with the largest deviations seen at the end of the cells farthest from the APDs.

 Cosmic muons used for energy calibration have different energy distributions at the top of the detector versus the bottom of the detector because muons lose energy while traversing the detector.  Since the distribution of energies of muons that make it to different locations in the detector changes, the energy deposited by those muons also changes. The size of this self-shielding effect was calculated as the ratio of the deposited energy, $E_{\rm true}$, which includes shielding effects, and a naive no-shielding approximation based on the path length through the cell, $E_{\rm MIP}$, where the subscript MIP stands for Minimum Ionizing Particle. The shielding correction was less than 2$\%$ for the TB detector, with negligible dependence on fiber brightness or cell position.

\subsection{Relative \& Absolute Calibration Corrections}
\label{sec:calib-rel}
\noindent 
The relative and absolute calibration corrections for the TB detector were calculated for each predefined time period, $t$, based on detector operations as described in \autoref{sec:operations}.  The reconstruction and selection of cosmic tracks for use in determining the calibration corrections are described in detail in~\cite{RobertThesis}.

\begin{figure}
    \centering
\includegraphics[width=.7\textwidth]{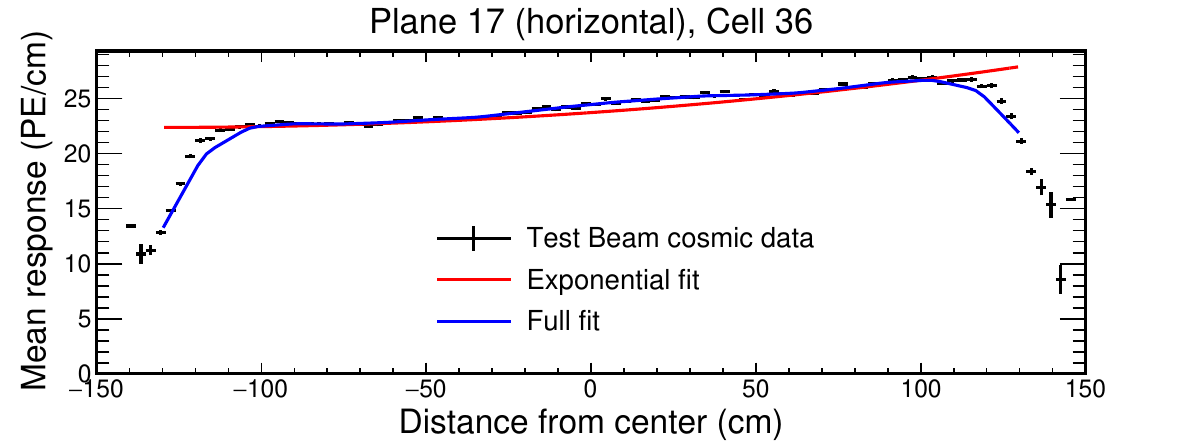}
\caption{An example attenuation fit for a single cell in the NOvA Test Beam detector. The readout is located at the right. The red curve shows the initial exponential fit and the blue curve the final full fit after the \textsc{lowess} correction, both described in the text.}
    \label{fig:CalibrationAttenuationFit}
\end{figure}

The relative calibration corrects for attenuation of scintillation photons as they travel through the fiber to the readout and for response differences among detector cells. The correction function for a given cell is equal to the ratio between the average energy response across the entire detector (a single, constant number) and a fit to the mean response of cosmic muon hits as a function of the distance along the cell ($w$). An example attenuation fit for one of the TB detector cells is shown in \autoref{fig:CalibrationAttenuationFit}. The fit was done in two steps; the first step was a three-parameter exponential fit,

\begin{equation}\label{eq:expfit}
  y=C+A\left(\exp\left(\frac{w}{X}\right)+\exp\left(-\frac{L+w}{X}\right)\right),  
\end{equation}

\noindent
where $L$ is the cell length and $C$, $A$, and $X$ are the fit parameters that represent the background, attenuation scale, and attenuation length, respectively. The second step was a full fit to a combination of the intermediate exponential fit and the data, with the LOcally WEighted Scatter plot Smoothing (\textsc{lowess}) method. This method is a nonparametric extension of a least-squares fit. The mean fractional deviation of the fit data is $\Delta^2$. Cells for which $\Delta^2\geq 0.2$ were marked as uncalibrated for subsequent reconstruction and analysis; fewer than $1\%$ of cells fell into this category. The relative calibration provided the correction factors denoted by $R_i\left(t\right)$ in \autoref{eq:edep}. The corrected response resulting from the relative calibration is called PECorr, indicating the number of photoelectrons.

\begin{figure}
    \centering
    \includegraphics[width=0.49\textwidth]{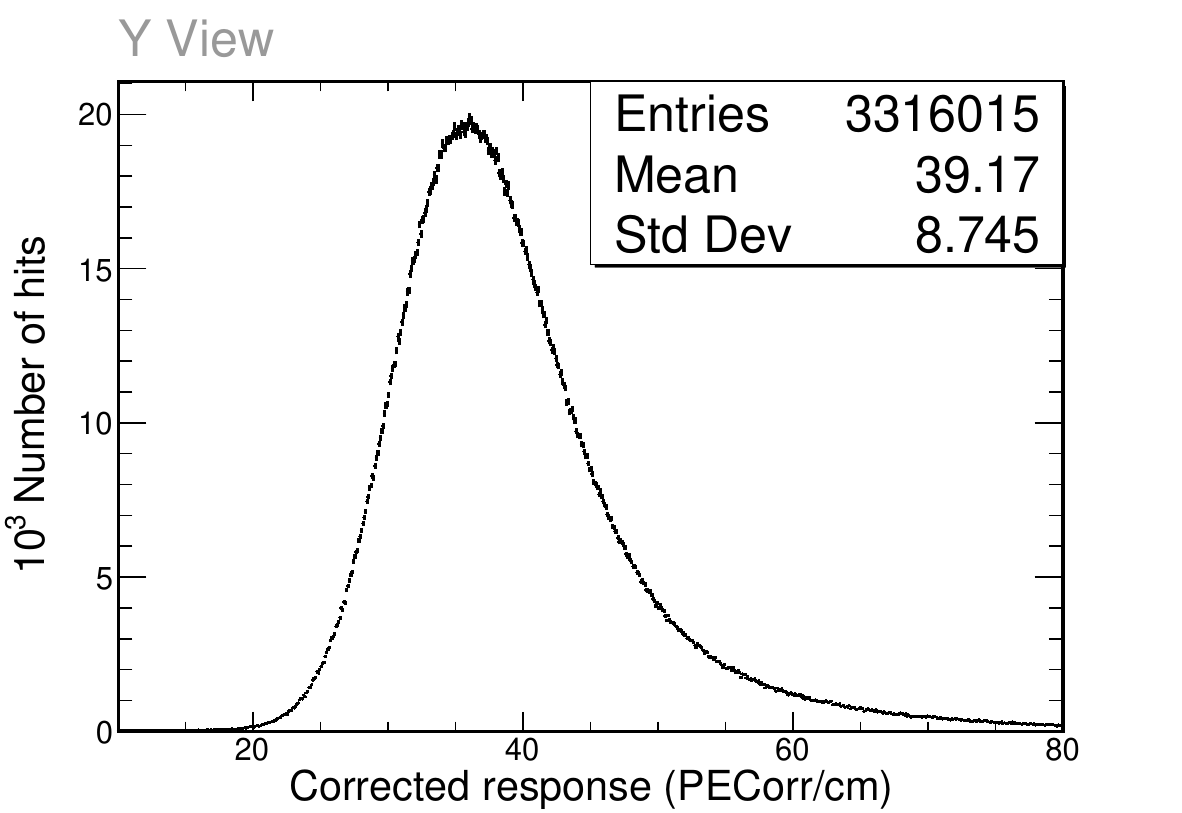}
\includegraphics[width=0.49\textwidth]{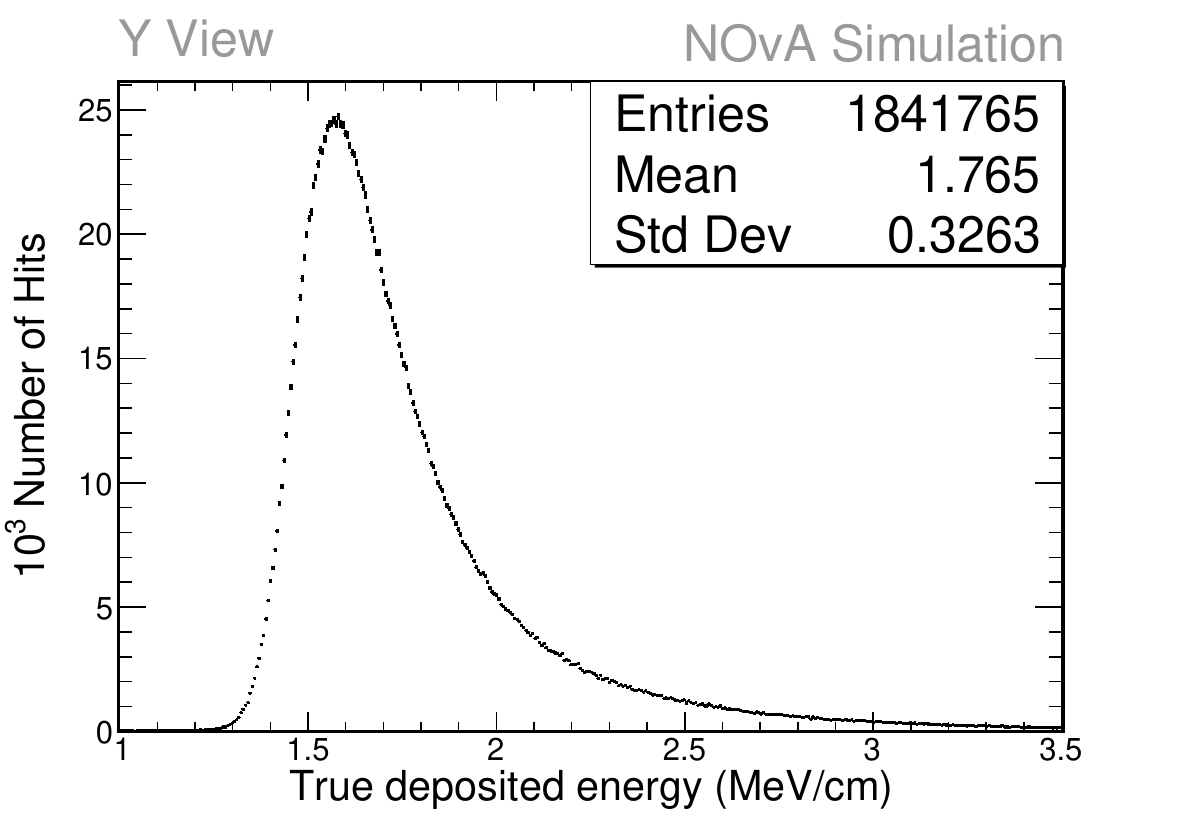}
\caption{The reconstructed response in the $y$-view for selected stopping muons in data, collected during Period 4 of \datataking, is shown in the left plot. The true deposited energy response in the $y$-view for selected stopping muons in simulation is shown in the right plot.}
    \label{fig:CalibrationAbsoluteEnergyScale}
\end{figure}

The absolute calibration used hits from stopping muons \SIrange{1}{2}{\meter} from the end of the tracks, as these stopping muons can be approximated as MIPs with well-known energy deposition behavior. Hits at the ends of each cell were removed to mitigate the impact of poorer track reconstruction and detector modeling at the detector edges. Stopping muons were identified by selecting muon tracks with Michel electrons. The distribution of the corrected response, PECorr, is shown on the left of \autoref{fig:CalibrationAbsoluteEnergyScale} for Period 4 of \datataking, in the $y$-view. The PECorr distribution is compared with the simulated true deposited energy distribution in the right of \autoref{fig:CalibrationAbsoluteEnergyScale}. The ratio of the means of these distributions was calculated for each view and data-taking period and used as the scale factor to convert from PECorr to \SI{}{\mega\electronvolt}, denoted by $A_v\left(t\right)$ in \autoref{eq:edep}.

\subsection{Test Beam Calibration Results}
\label{sec:calib-results}
\begin{figure}[!ht]
    \centering
\includegraphics[width=.48\textwidth]{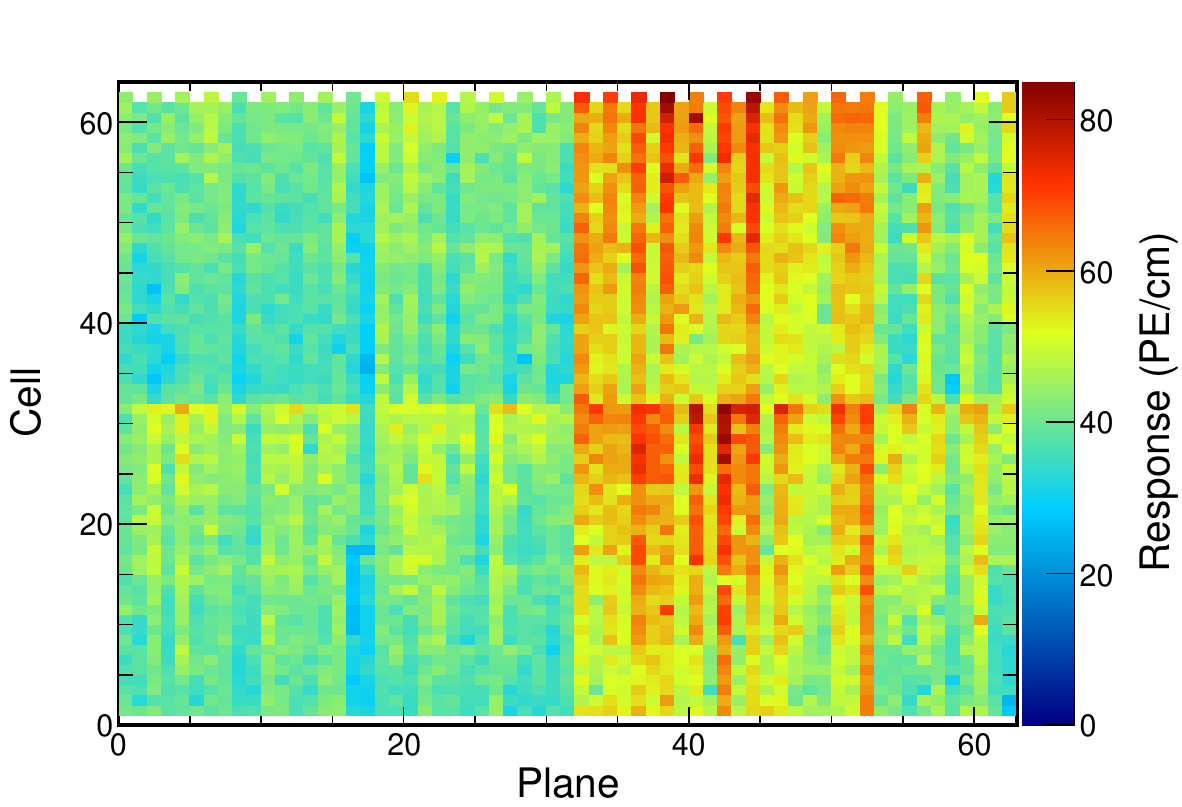}
    \includegraphics[width=.48\textwidth]{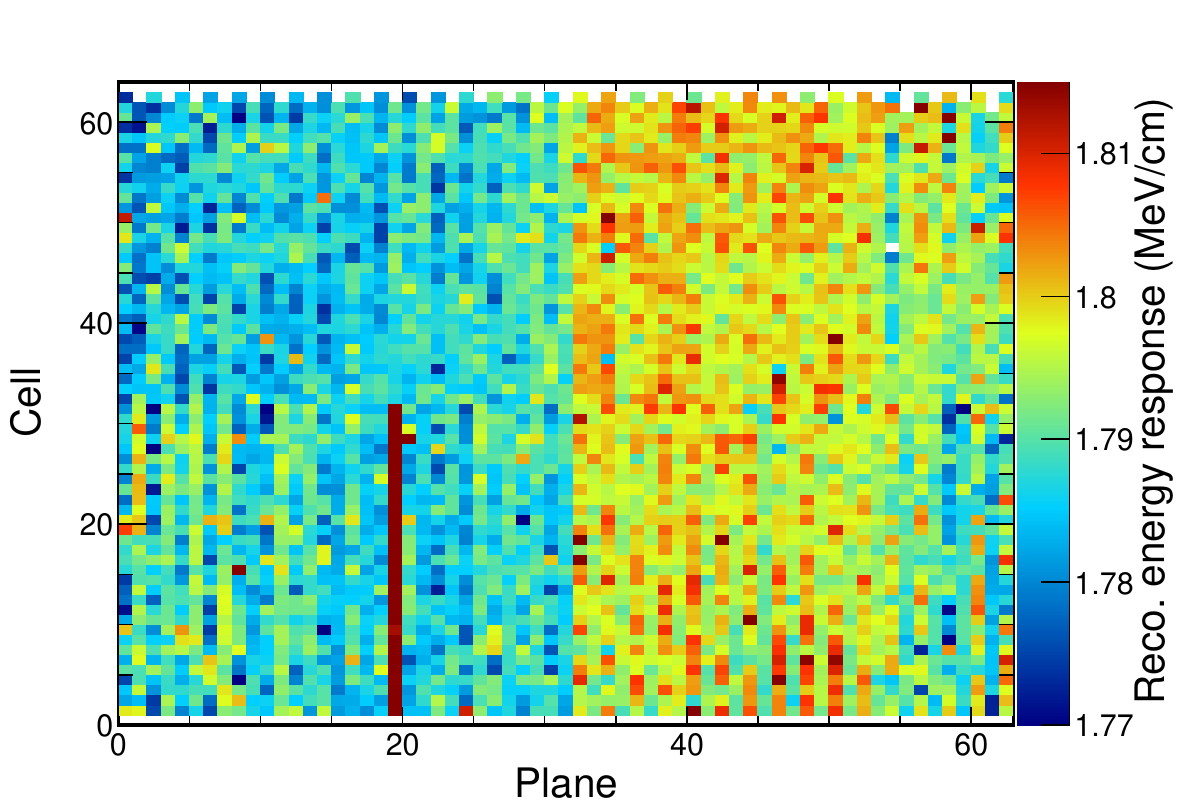}

\caption{The TB detector response to  cosmic muons before (left) and after (right) applying the results of the calibration. The right plot combines the effects of the relative calibration, which reduces the response variations to within $1\%$ across the detector (note the difference in color-scale), and the subsequent absolute calibration, which converts PECorr to MeV using a constant factor described in the text. The blank cells at the detector edges and in plane 59 are uncalibrated, due to either a low number of calibration hits or to statistical fluctuations. The higher light response in planes 32 to 52 is due to the different scintillators used, as described in \autoref{sec:scint}. The effect of a faulty FEB is visible on the right plot in plane 19; this is not  discernible in the left plot because the scale of the relative corrections is much larger than that of the faulty FEB. }
    \label{fig:CalibrationPECorrcmPlaneCellP4_XView}
\end{figure}

\noindent
The TB detector was significantly smaller in size than the other NOvA detectors, which lessened the effects of APD thresholds, self-shielding, and the attenuation of light along the length of a cell. 
Other notable features of the TB detector were the variety of scintillators (\autoref{sec:scint}) and readout electronics (\autoref{sec:detdaq}) used. These variations provided the potential to test the detector response and the calibration procedure in various scenarios. The effect of calibration is shown in \autoref{fig:CalibrationPECorrcmPlaneCellP4_XView}. The raw response to stopping muons, shown on the left of the figure, varies by a factor of two among detector cells, while after calibration, the mean energy responses for all cells fall within 1$\%$ of the mean across the detector. 

\section{Particle Reconstruction}
\label{sec:performance}
\noindent

\noindent
The primary goal of the NOvA TB program was to inform the detector-based systematic uncertainties in measurements of neutrino oscillation parameters. Achieving this requires measuring the energy response of the detector to identified particles over a range of momenta. This section briefly describes the steps taken to identify and measure electrons, protons, and pions/muons. Preliminary analyses of pions and electrons are detailed in~\cite{DavidThesis} and~\cite{DaltonThesis} respectively.
 \subsection{Momentum Reconstruction }
\label{sec:tracking-mom}
\begin{figure}   
 \includegraphics[width=0.48\textwidth]{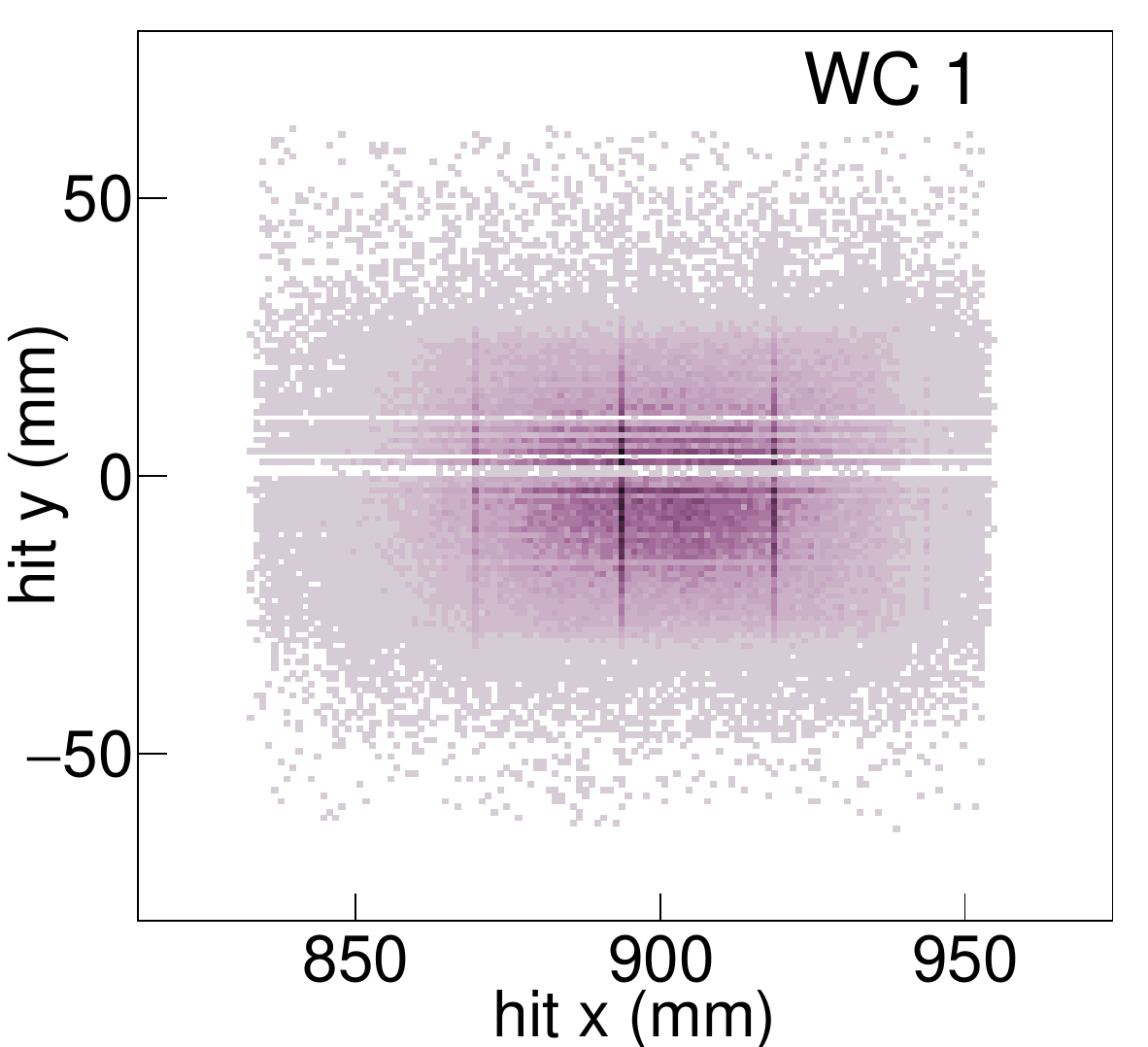}    \includegraphics[width=0.48\textwidth]{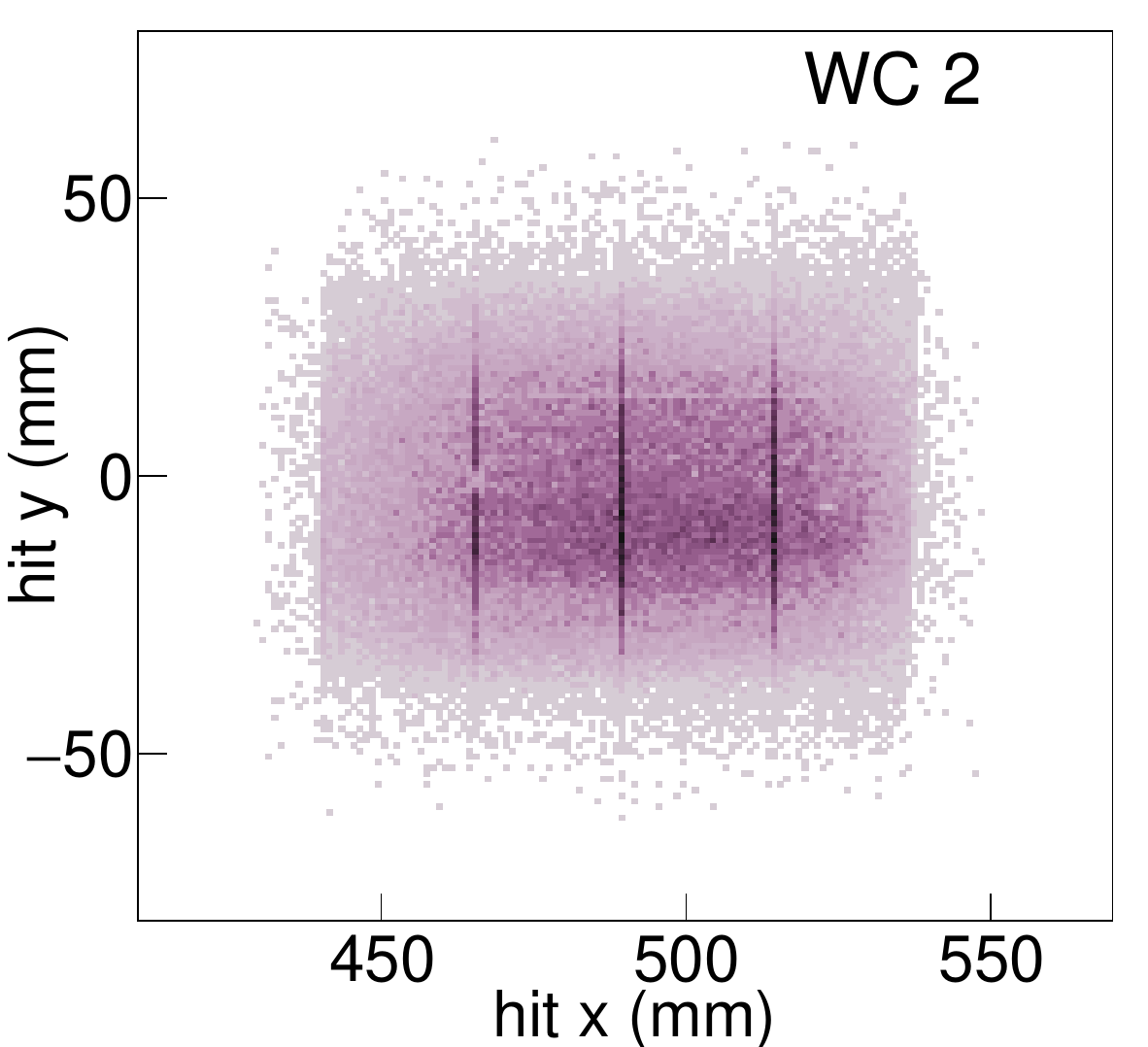}   \includegraphics[width=0.48\textwidth]{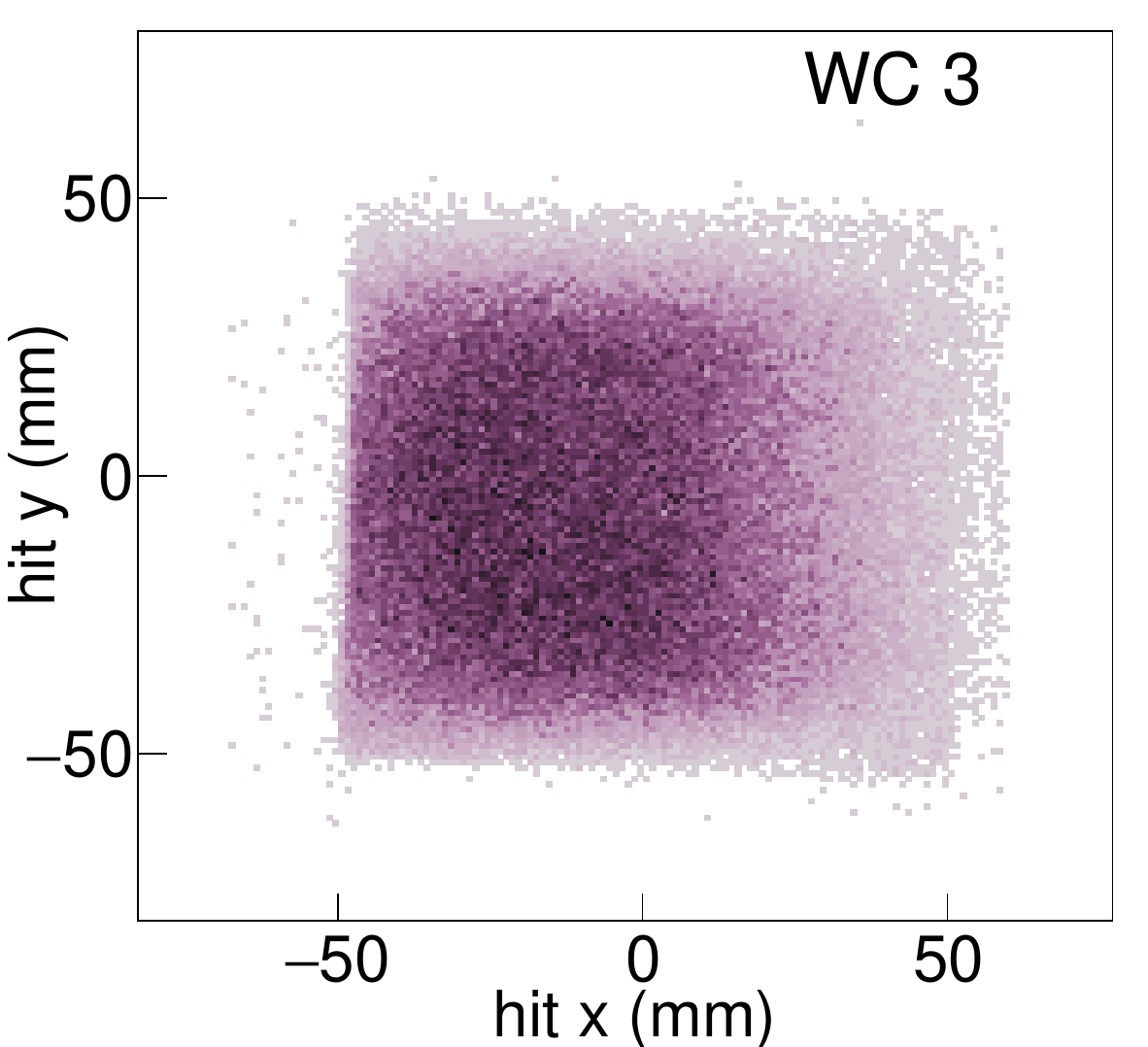}
  \includegraphics[width=0.48\textwidth]{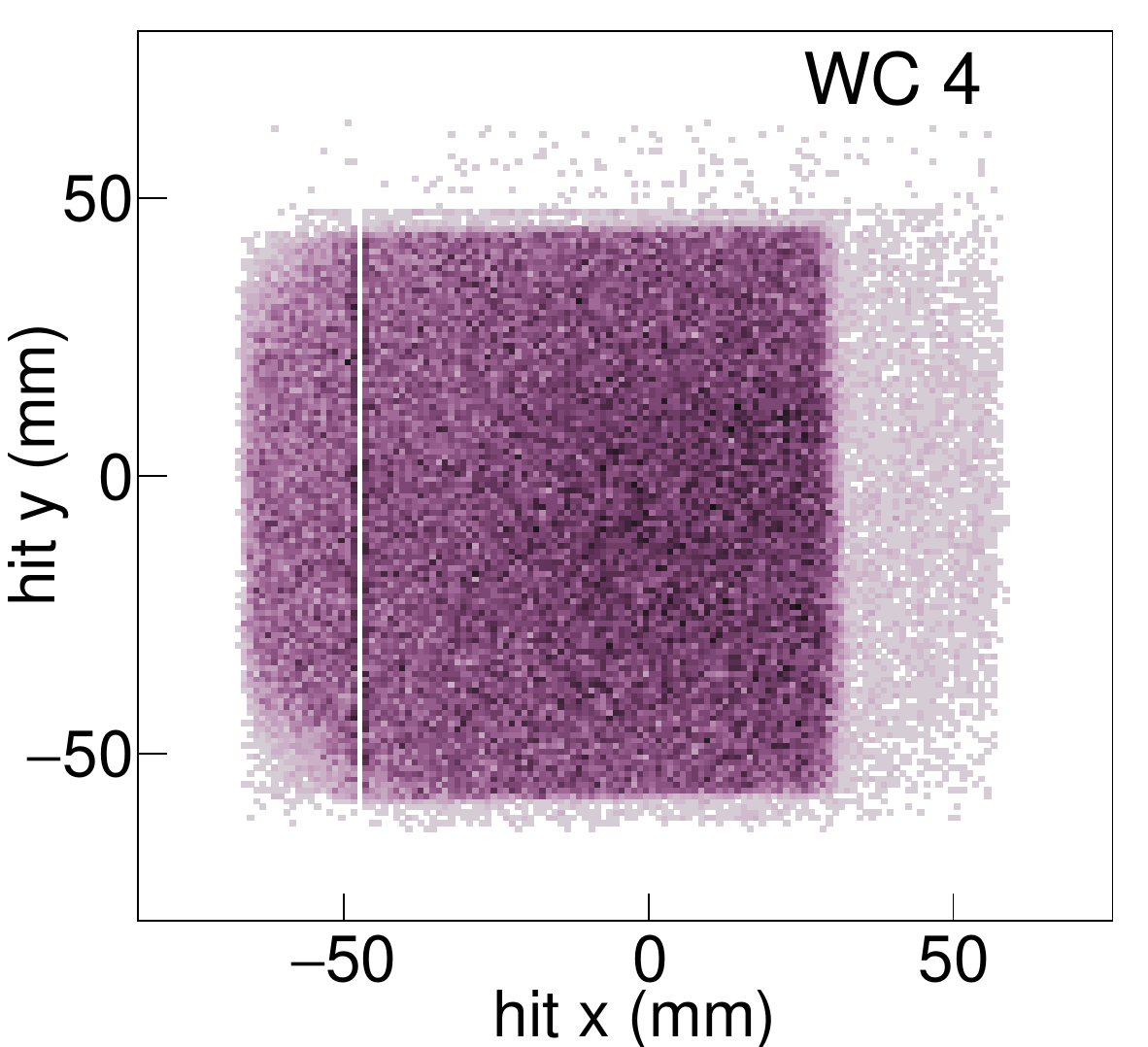}
   \caption[short]{The \wc\ track ``best hit'' positions in each of the four wire chambers, relative to the center of the front face of the NOvA TB detector.  These figures represent the state of the \wcs\ in the final period (Period 4) of \datataking, and show missing wires in WC1, noisy vertical wires in WC1 and WC2, a slight rotation in WC3, and one missing wire in WC4.  The coordinate system is defined in \autoref{fig:TertiaryModel}.}
   \label{fig_xyhits}
\end{figure}

\noindent
The \wc\ tracking algorithm required that each of the four \wcs\ (\autoref{sec:wirechambers}) had at least one hit in ($x,y$). For every possible combination of \wc\ hits, the straight line in the ($y,z$) plane through them was found by minimizing the sum of the squared residuals in a least-squares fit. The hits associated with the track with the smallest sum are called the ``best hits,'' and their spatial distributions are shown in \autoref{fig_xyhits} for all four \wcs\ in Period 4 of \datataking.

\begin{figure}[ht]
    \centering        \includegraphics[width=0.95\textwidth]{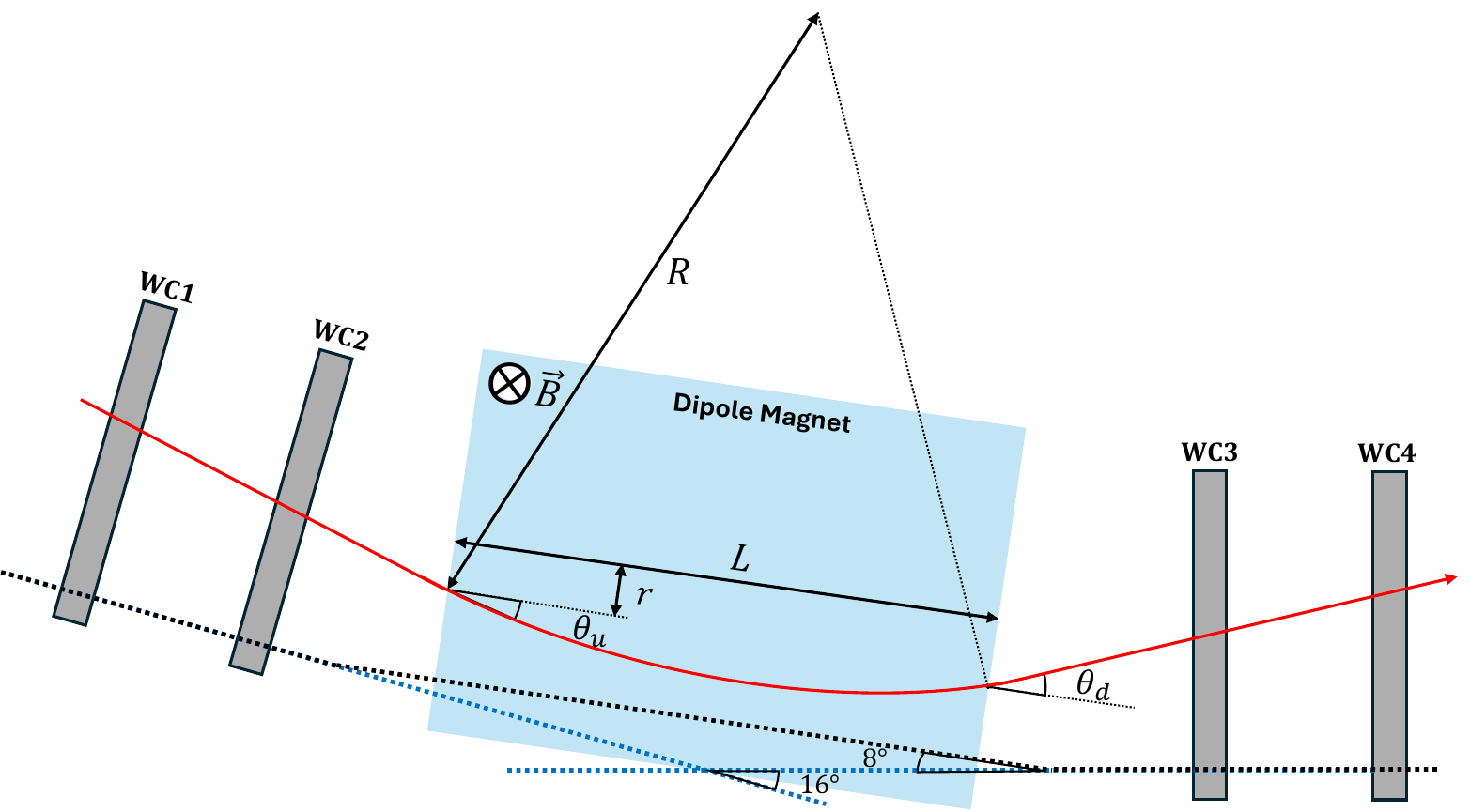}
    \caption{Top view of a positively charged particle entering the dipole magnet from the left at an angle $\theta_{u}$ and curving in the presence of the magnetic field, which is directed into the page. The angle of exit is $\theta_d$, the length of the field region is $L$, the radius of curvature is $R$, and the transverse distance between the track entry point and the center of the magnet face is $r$. The diagram also illustrates how the \wcs~determine the direction of the particle trajectory upstream (\wc~1 and \wc~2) and downstream (\wc~3 and \wc~4) of the dipole magnet. The dashed lines show the angles with respect to the direction of the secondary beamline. This is an illustrative diagram and is not to scale.}
    \label{fig:bfield-angles}
\end{figure}

The radius of curvature of the track within the $B$ field was calculated using the length of the field region, $L$ =\SI{106.7}{\cm} (see \autoref{sec:magnet}), and the upstream and downstream track angles, $\theta_{u}$ and $\theta_{d}$. The angles were measured in the ($x,z$) plane, in the magnet's frame of reference, as illustrated in \autoref{fig:bfield-angles}. The radius of curvature of the track is given by

\begin{equation}\label{eq:radius}
R = \frac{ L }{\sin{\theta_{u}} - \sin{\theta_{d}} }.
\end{equation}

The transverse momentum of the track, $p_T$, was computed using $p_T = qBR$, where $B$ is the effective value of the magnetic field (described in~\autoref{sec:magnet}). The angle $\phi$ of the track in the ($y,z$) plane was then used to find the total momentum, 

\begin{equation}\label{eq:mompt}
p = \frac{ p_T}{\cos{\phi}}.
\end{equation}

To account for the nonuniformity of the magnetic field, a correction scale factor $c_{B}(\textit{r})$ was calculated \cite{TeresaThesis} using field maps and applied to the track momenta based on the transverse distance $\textit{r}$ between the track's entry point into the magnetic field region and the center of the magnet face (see Fig.~\ref{fig:bfield-angles}). The correction was calculated as 

\begin{equation}\label{eqn:momentum_correction_params}
c_{B}(r) = ( \alpha_1 + \alpha_2 r ) \left(
 \frac{1}{ 1 + e^{  (\textit{r}-\alpha_3)  / \alpha_4} }  - 
 \frac{1}{ 1 + e^{  (\textit{r}-\alpha_5)  / \alpha_6  } } \right),
\end{equation}

\noindent where the parameters $\alpha_i$ were determined from a fit to simulation. The corrected momentum measurement for particles exiting the magnetic field is then

\begin{equation}\label{eqn:momentum_corrected}
p_{cor} = c_{B}(\textit{r}) \cdot p.
\end{equation}

Particles exiting the magnet continued through the downstream beamline instrumentation comprising a collimator, two \wcs, two trigger paddles, the Cherenkov detector, and two \tof\ scintillators. Simulation studies revealed that the energy loss in these downstream components occurred mainly in the ToF scintillator positioned directly before the NOvA detector, known as DS2 ToF. Additionally, simulations showed that particles may hit or miss this ToF. 

\begin{figure}
    \centering
    \includegraphics[width=0.6\linewidth]{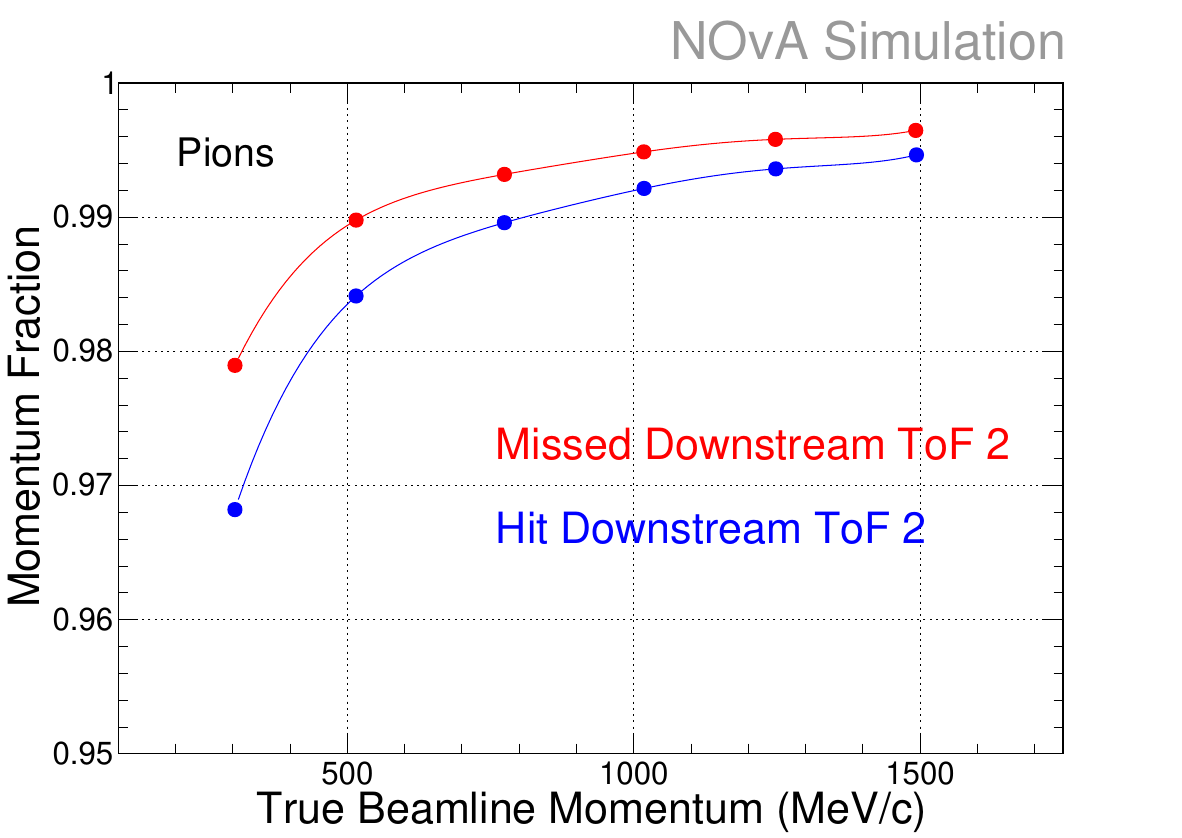}   \caption{
    Fraction of pion momentum remaining at the front face of the detector relative to the measured momentum, as predicted by the  \gfourbeamline code in the Tertiary Simulation mode described in \autoref{sec:tertiary-beamline-sim}. The simulated data were divided based on whether they passed through the downstream ToF or missed it, as this was identified in simulation as the point at which most of the pion momentum loss occurred.}
    \label{fig:pion-momcor}
\end{figure}

As discussed in \autoref{sec:nova-simulation}, determining the energy lost in the tertiary beamline components is crucial for an accurate estimate of the NOvA detector response. \autoref{fig:pion-momcor} shows a preliminary polynomial (order 5) interpolation function for modeling the fraction of pion momentum remaining at the front face of the detector relative to the momentum measured using the wire chamber tracks. The points show the most probable values of the remaining momentum fractions, obtained using \gfourbeamline simulation in Tertiary Simulation mode (\autoref{sec:tertiary-beamline-sim}) at six true momentum settings. These interpolations were used to correct pions for the momentum loss in analysis~\cite{DavidThesis}. Preliminary studies suggest the energy losses for protons may be estimated similarly.

 \subsection{Time-of-Flight Estimation and Resolution}
\label{sec:tracking-tof}
\noindent 
To reconstruct the \tof\ of each particle, the pulses measured in each ToF panel were first identified and characterized. After establishing a response baseline, and applying a Savitzky--Golay (SG)~\cite{SGsmoothing} filter of order 3 with a window of 15 points to the raw data to reduce overall noise, pulses were located and reconstructed in the waveforms as described below.

Pulses were located using a fixed-threshold leading-edge discriminator. Since the pulses are negative, the threshold was set at fifty times the baseline noise level, 50\,\,$\sigma_{\textrm{noise}}$, and a candidate was accepted only if at least four consecutive samples fell below it.  This consecutive sample requirement suppresses noise excursions and allows both multiple pulses and partially overlapping pulses within a single waveform to be resolved. 
The 50\,\,$\sigma_{\textrm{noise}}$ threshold was chosen to be large enough to significantly reduce the non-beamline background while being small enough to include the vast majority of beam particles. For each identified pulse, the approximate start, peak, and end of the pulse were located by scanning backward and forward without interpolation.
The arrival time of each pulse was defined by a constant-fraction ``pickoff'' timing point. The peak region was fitted with a third-order polynomial to determine the peak time and amplitude, and the pickoff was taken as the location where the leading edge crosses a fixed fraction of that peak amplitude, found by scanning backward from the peak with linear interpolation. The fraction was set to 0.7\% of the peak amplitude, a point near the foot of
the leading edge, which was chosen to minimize the width of the speed-of-light ToF peak. Setting the timing low on the fast, early part of the leading edge places the pickoff close to the true particle arrival and reduces biases from residual amplitude and shape-dependent jitter present later in the pulse.
 
\begin{figure}
\includegraphics[width=0.5\linewidth]{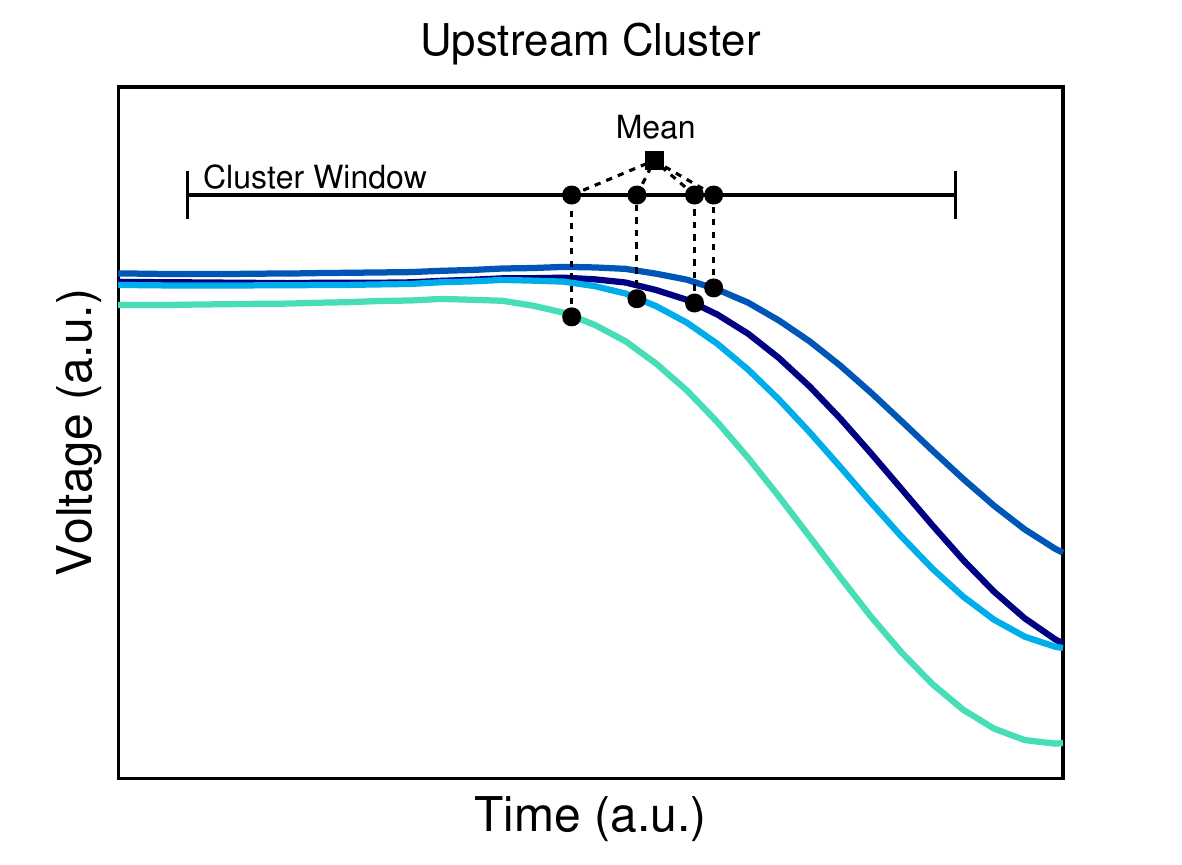}
\includegraphics[width=0.5\linewidth]{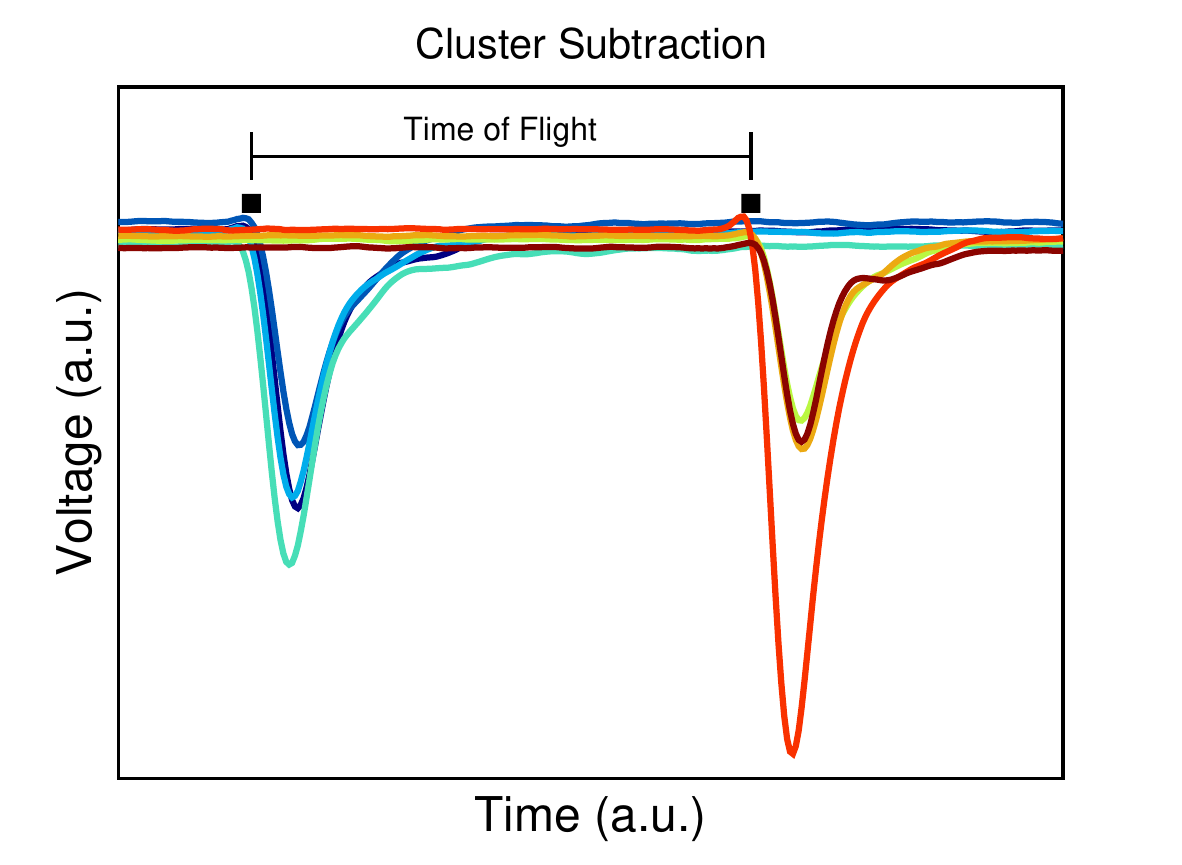}
\caption{Cluster and ToF reconstruction. Pulses were grouped into clusters by proximity in time. Cluster timestamps were calculated and then subtracted to determine the time-of-flight. Channel timing calibrations were applied before clustering. The upstream ToF pulses from the four PMTs are shown in shades of blue-turquoise, while the downstream ToF pulses are represented by shades of red-chartreuse.}
\label{fig:tofreco}
\end{figure}

Finally, in the reconstruction, the four photodetector pulses associated with each scintillator were grouped into clusters based on proximity in time. In a linear scan, the earliest time became the center of a new cluster, and all times within a \SI{5}{ns} window were added to that cluster. This algorithm was repeated on the remaining unclustered start times until all times were in a cluster. An illustration of this process is given in \autoref{fig:tofreco}. Pulses tended to occur within one nanosecond of each other, and in the case of multiparticle events the particles tended to be on the order of hundreds of nanoseconds apart. For each cluster, the characteristic time was defined as the mean of the start times of the pulses in that cluster. Since the average of $N$ normal variables with standard deviation $\sigma$ is normal with standard deviation $\sigma/\sqrt{N}$, taking the mean of the start times mitigated random error from individual start times by up to a factor of two. When exactly one cluster was found in each detector, with times $t_\textrm{US}$ in the upstream detector and $t_\textrm{DS}$ in the downstream detector, the ToF was found as their difference, $\tau = t_\textrm{DS} - t_\textrm{US}$. Events with multiple clusters in one detector were discarded.

\begin{figure}[!ht]
\centering
\includegraphics[width=0.32\textwidth]{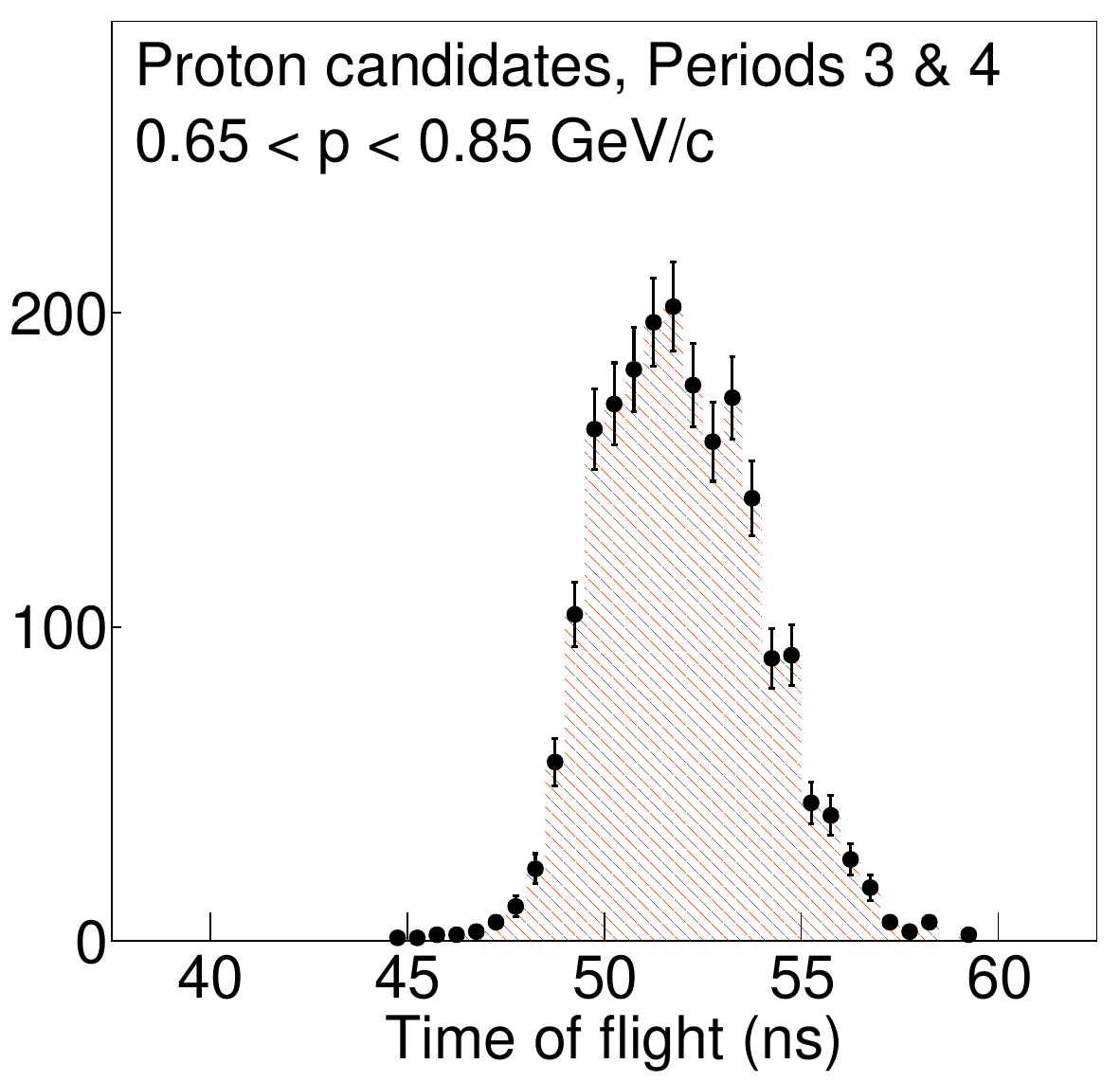}
\includegraphics[width=0.32\textwidth]{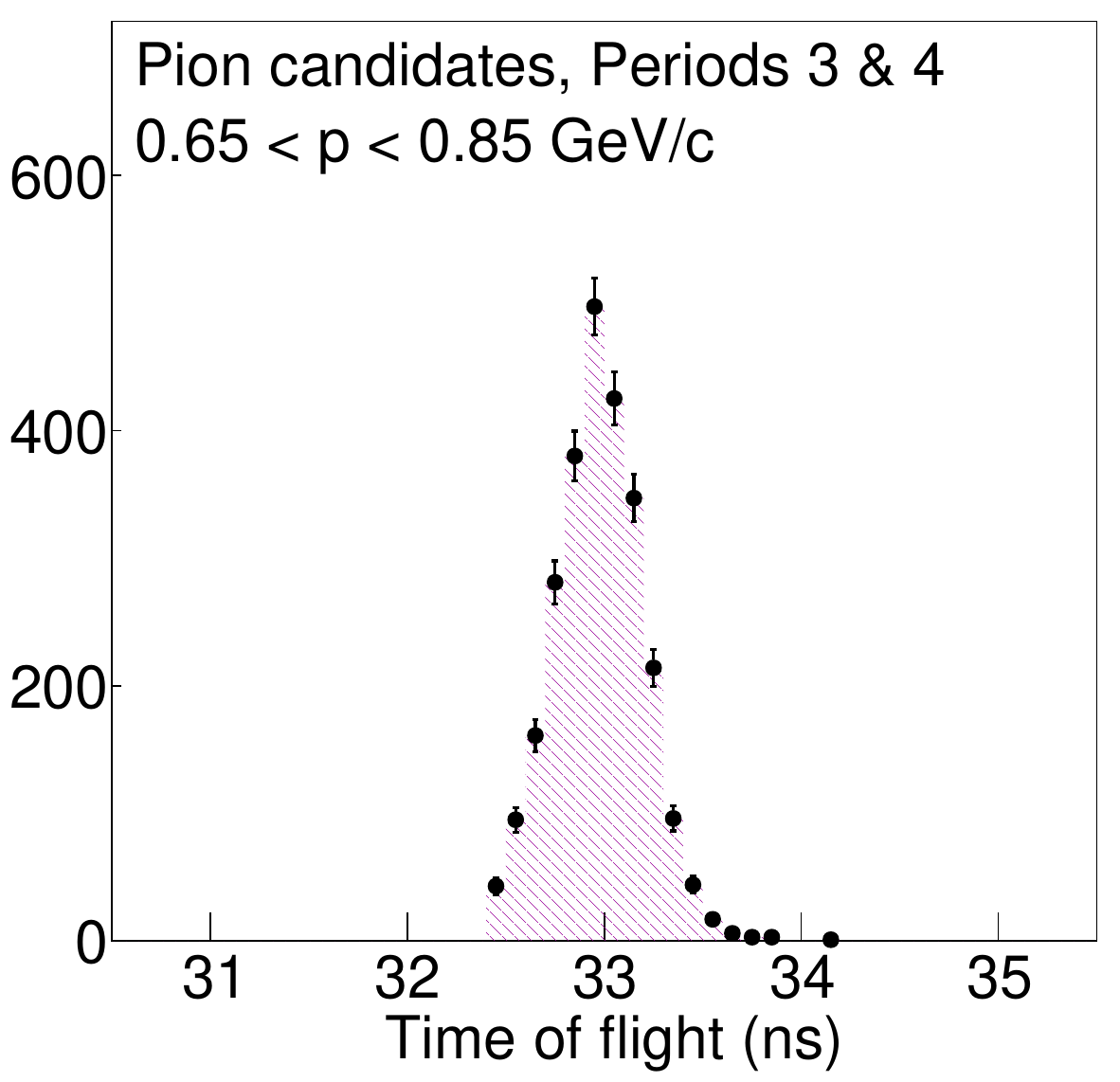}
\includegraphics[width=0.32\textwidth]{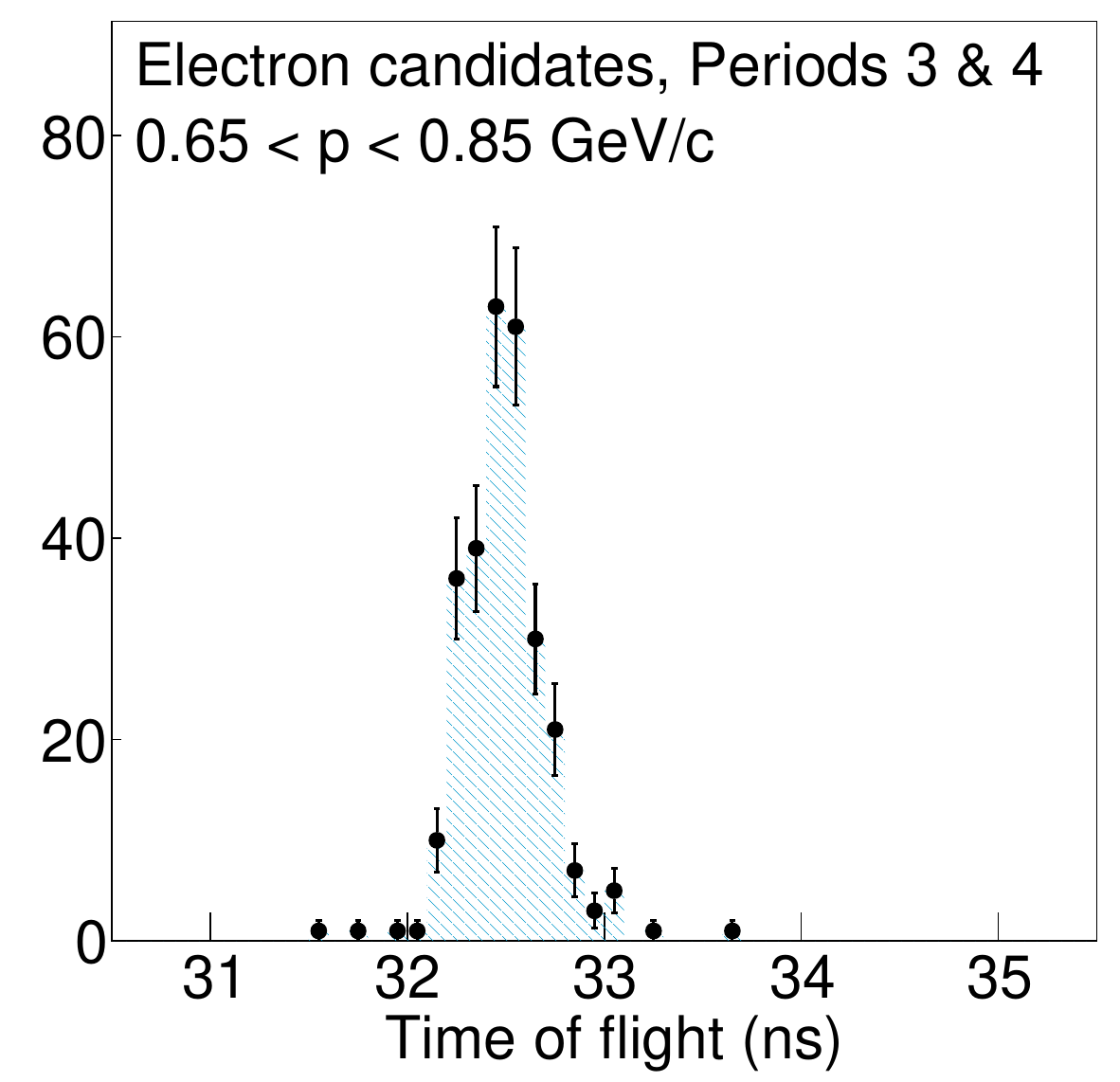}
\includegraphics[width=0.32\textwidth]{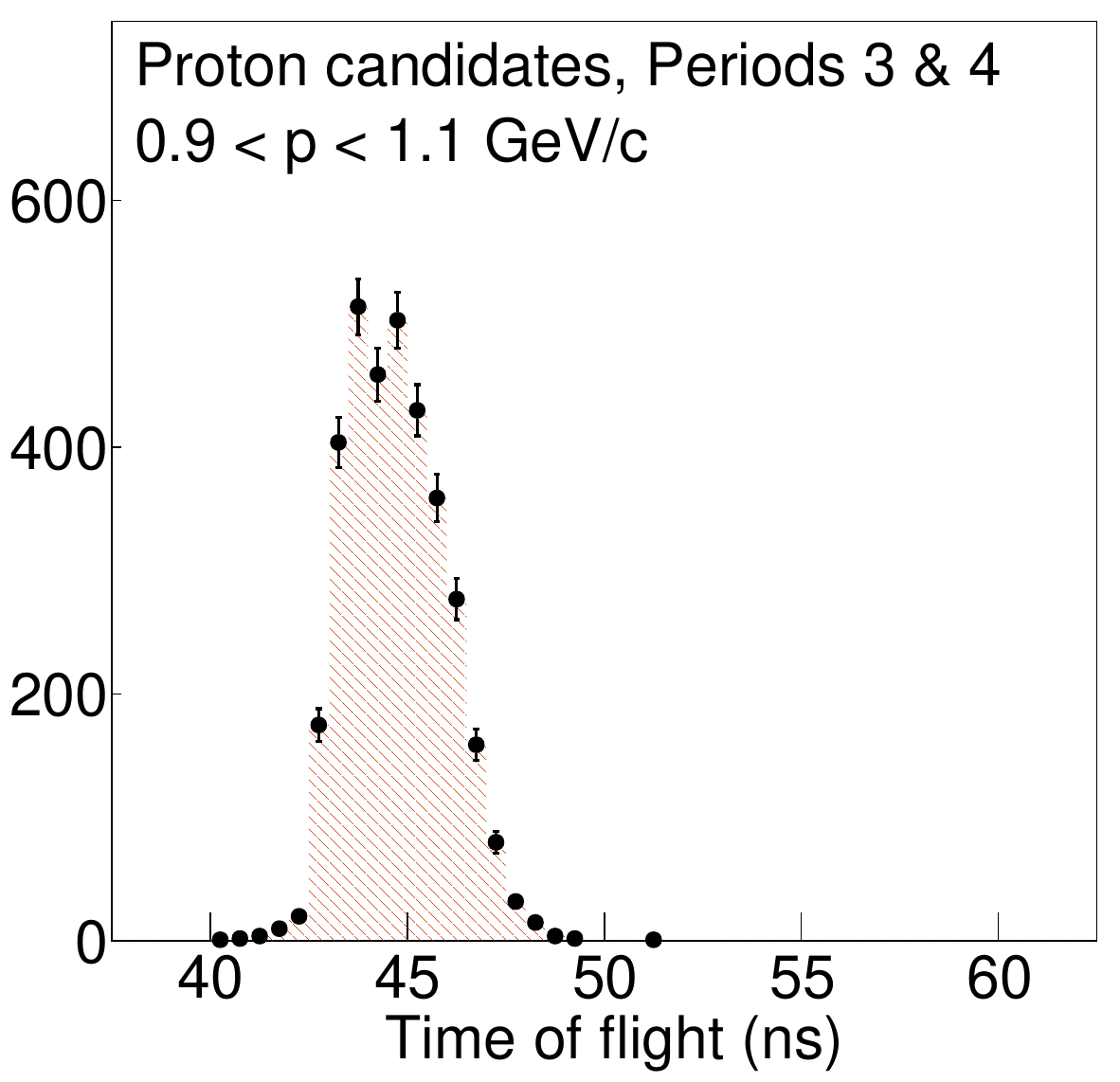}
\includegraphics[width=0.32\textwidth]{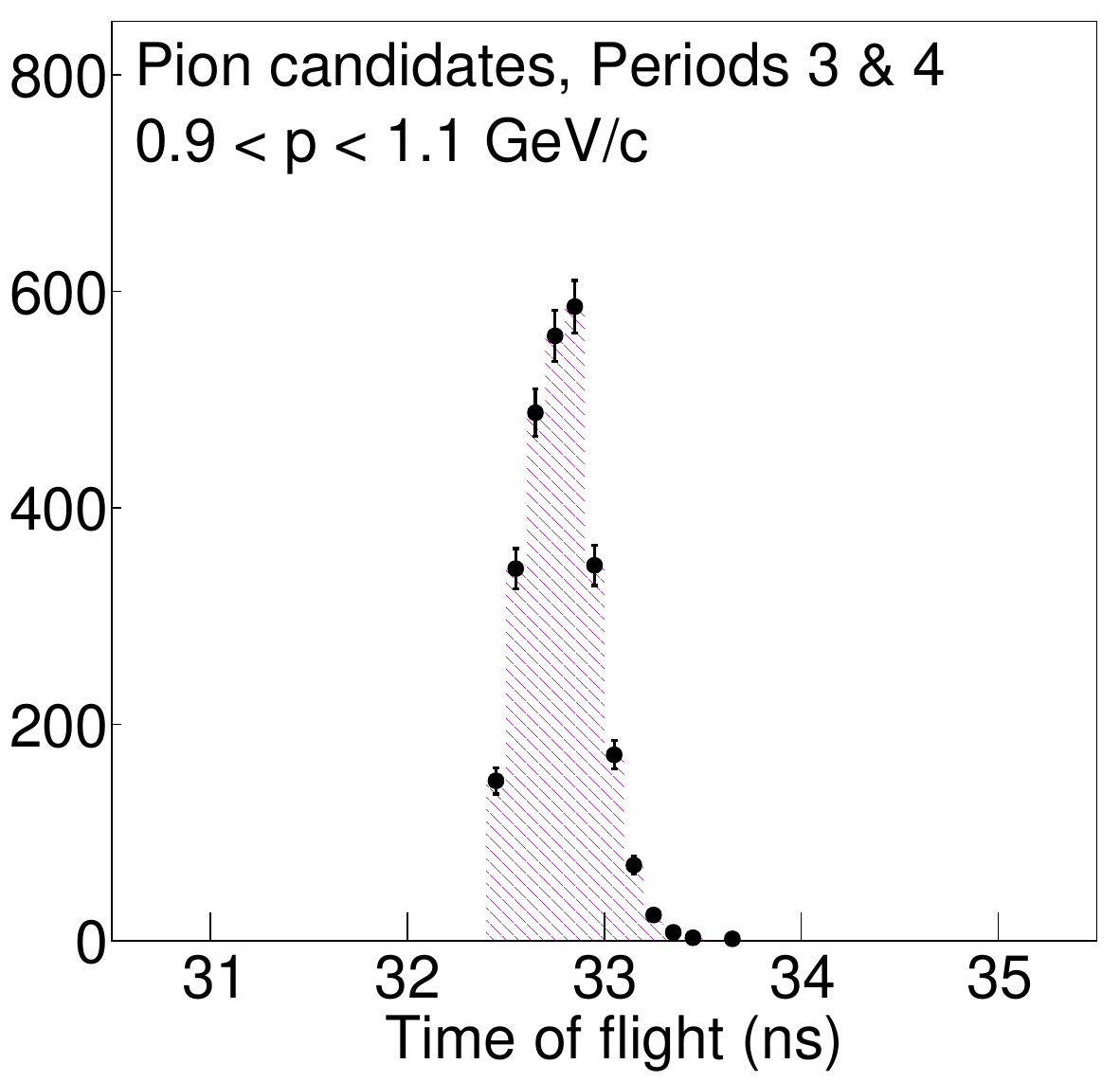}
\includegraphics[width=0.32\textwidth]{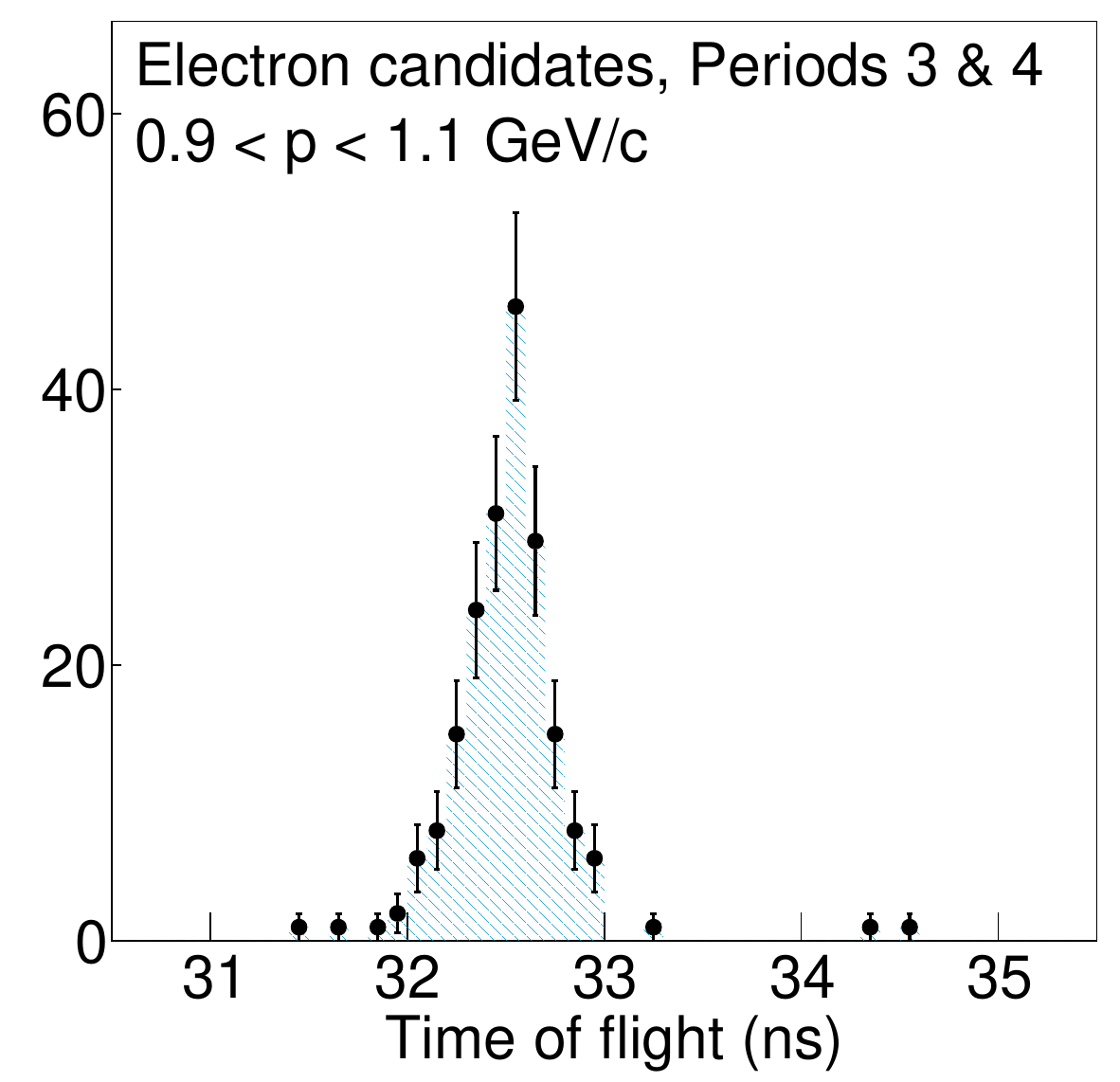}

\caption{The measured PMT ToF distributions for (left) proton, (center) pion/muon, and (right) electron candidates in Periods 3 and 4 of \datataking, during which the distance between PMT ToFs was $\approx$ \SI{9.7}{\metre}. The top row shows the momentum range $0.65<p<0.85$\,~GeV/c, while the bottom row shows the range $0.9<p<1.1$\,~GeV/c. The numbers of entries in each bin are shown on the $y$-axes. }
\label{fig:tof_perf}
\end{figure}

The reconstructed ToF peaks for protons, pions, and electrons are shown in \autoref{fig:tof_perf} for two momentum ranges. The ToF is independent of momentum to a good approximation for electrons, as their low mass means they travel close to the speed of light between the two ToF detectors. For higher-mass particles such as protons, the ToF becomes a strong function of momentum, making it a useful tool for particle identification.

 \subsection{Particle Identification}
\label{sec:tracking-pid}
\begin{figure}[!htbp]
    \centering

\includegraphics[width=.6\textwidth]{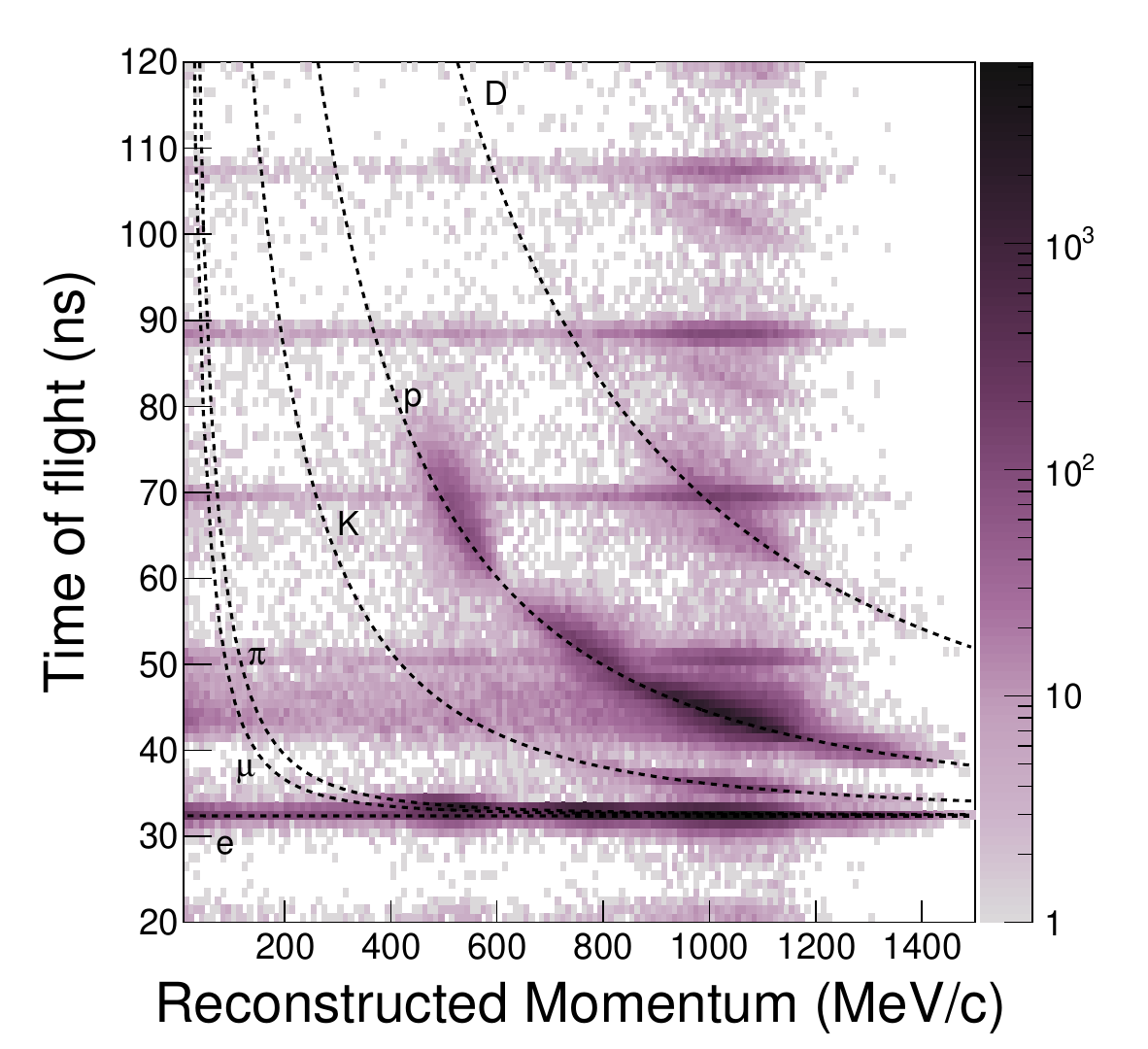}
    \caption{Reconstructed ToF versus momentum in data before data quality or selection requirements. The horizontal structures separated by \SI{18.83}{ns} correspond to ToF signal pairs from light particles from different main injector bunches (the main injector frequency is 53.1 MHz, as described in \autoref{sec:beamline}). The additional repeating sloped structures around \SI{1000}{MeV/c} are protons from different main injector bunches. Finally, the diffuse structures at around \SI{21}{ns} and \SI{42}{ns} are thought to be one light particle and one proton from different main injector bunches.}
    \label{fig:placeholder-momtof1}
\end{figure}

\noindent
The ToF versus reconstructed beamline track momentum for raw data contains particles of interest alongside ``accidentals'' as shown in \autoref{fig:placeholder-momtof1}. While data were being taken, two-dimensional box cuts on this distribution were used to get a rough idea of particle counts and an indication of when the counts were sufficient at a particular magnet current and polarity to allow for a switch in one or both. 

For offline analysis, particles can be more efficiently identified by reconstructing their mass using the momentum measurement from the \wc\ track and the speed $\beta$ from the ToF measurement, 
 
 \begin{equation}\label{eq:beta}
 \beta = \dfrac{L}{c \tau},
 \end{equation}
 \noindent
 where $L$ is the distance between the two ToF detectors providing a \tof\  measurement $\tau$.   The speed of light $c$ is used to normalize the velocity. 

\begin{figure}
    \centering
        \includegraphics[width=.49\textwidth]{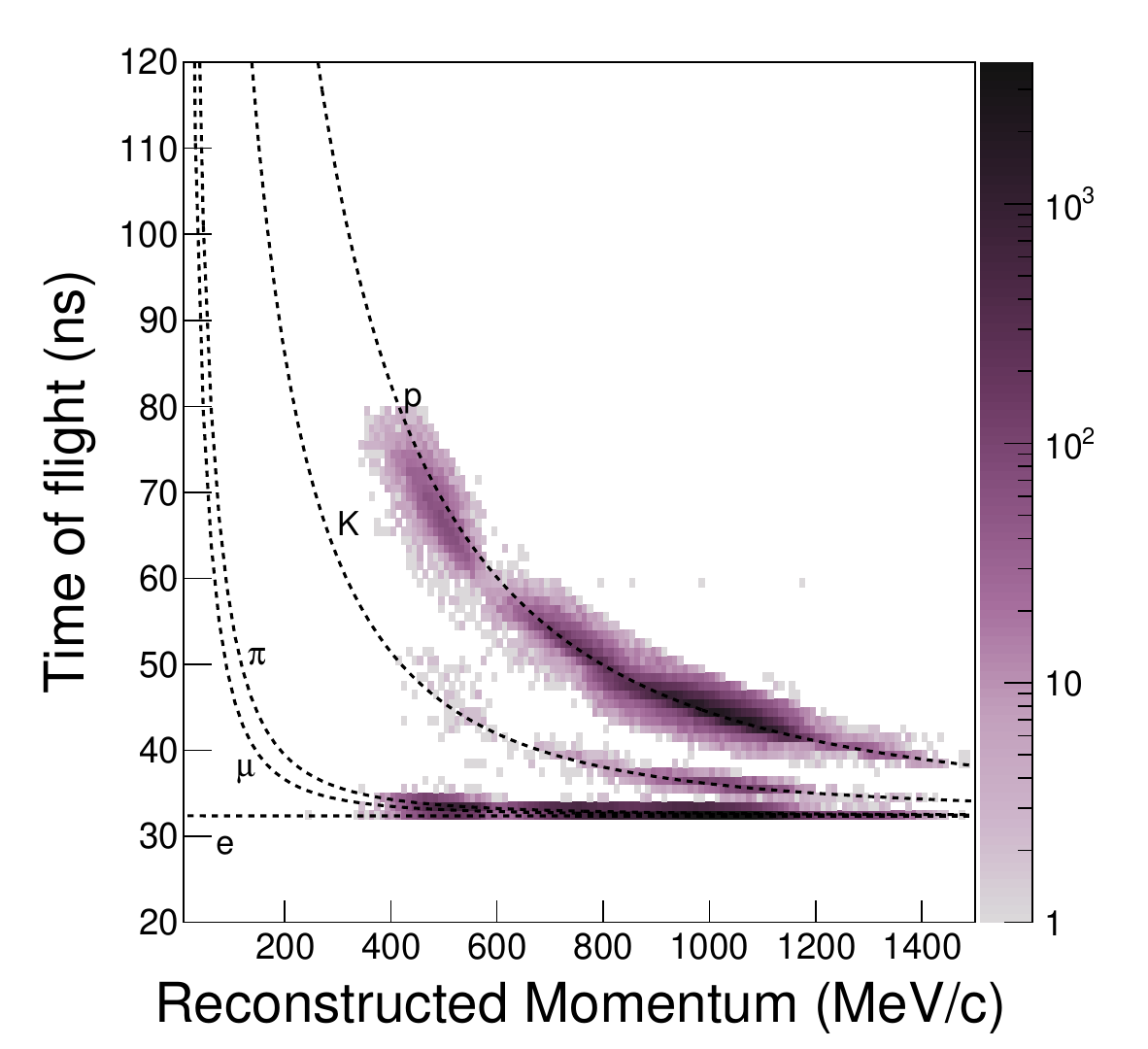}
    \includegraphics[width=.49\textwidth]{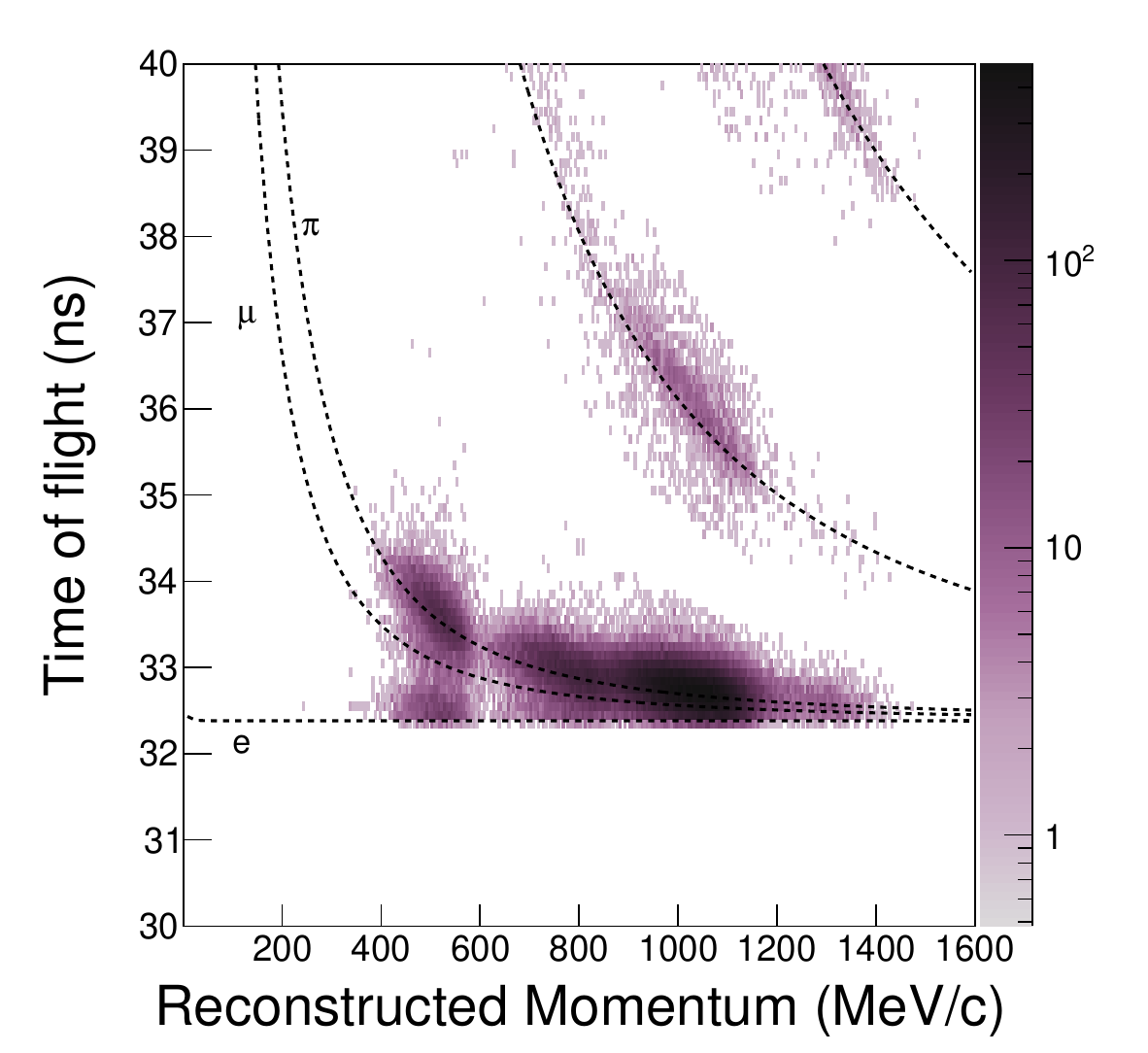}
    \caption{Reconstructed ToF versus momentum in data after applying beamline data quality and particle ID selections on mass and ToF as described in the text, for the full range of phase space (left), and zoomed in on the low ToF region (right), highlighting the need for a Cherenkov detector in the tertiary beamline to distinguish electrons from muons and pions.}
    \label{fig:placeholder-momtof}
\end{figure}

\begin{figure}
    \centering
    \includegraphics[width=.6\textwidth]{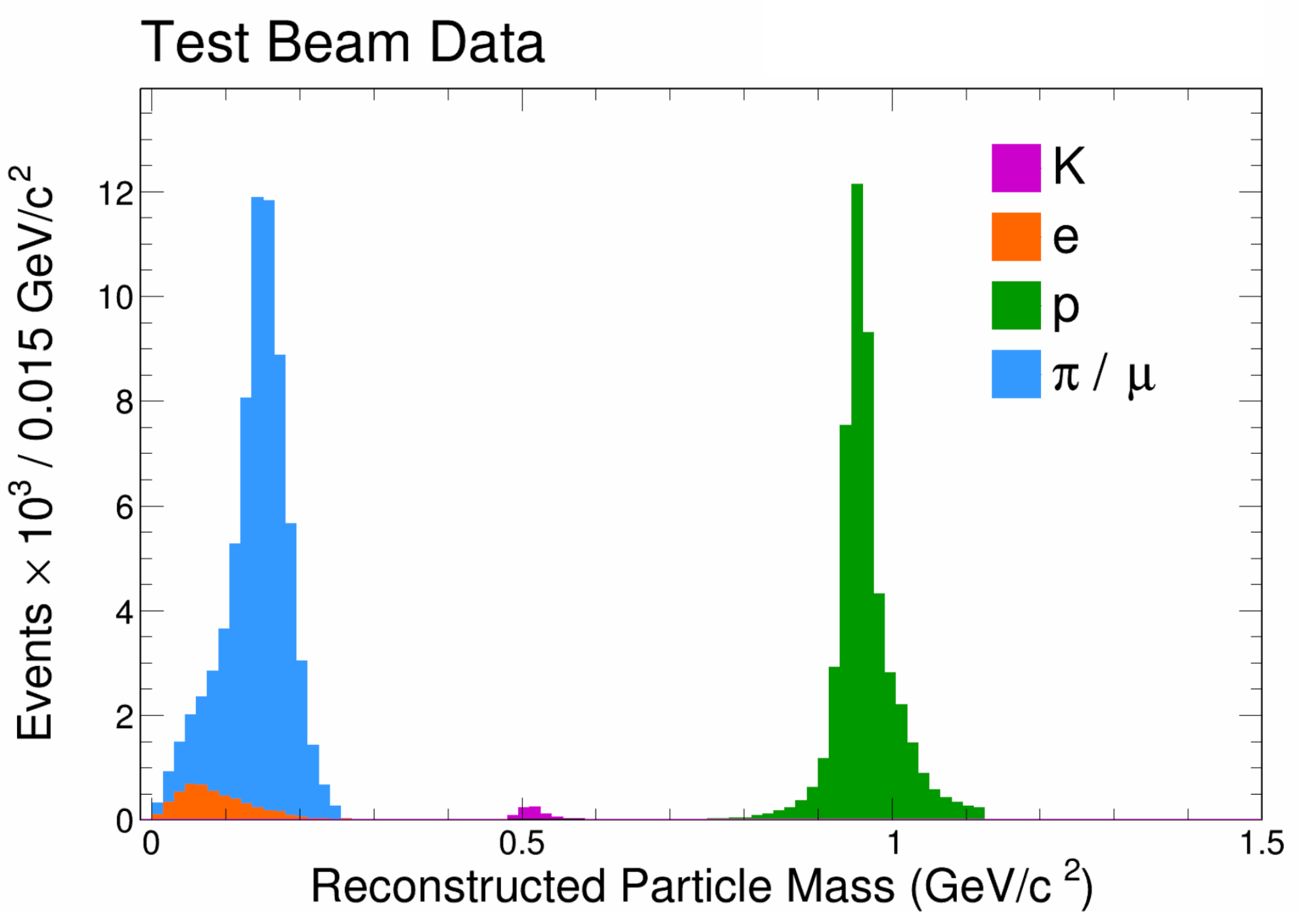}
    \caption{The reconstructed mass of particles passing the preliminary selections in \autoref{tab:detsel} using the reconstructed momentum and ToF. The mass of light particles using this technique does not reproduce the particle mass because the calculation relies on \tof\ measurements that cannot be resolved between pions and electrons --- they differ by only a fraction of a nanosecond over the distance of $\approx$\SI{9.7}{m} used in this experimental setup. Because of this, only an upper bound of \SI{300}{MeV} is used in the pion selection criteria.  The reconstructed masses of protons and kaons are a decent proxy for the mass of the particle and are used for particle ID. }
    \label{fig:placeholder-mass}
\end{figure}

The initial particle selection requirements are given in \autoref{tab:detsel}. These loose preselections were combined with some basic data quality requirements to define samples of particles for beamline tracking validation and momentum resolution studies. The ToF versus momentum distributions for particles satisfying loose mass cuts in addition to ToF cuts to reduce accidentals are shown in \autoref{fig:placeholder-momtof} and the mass distributions of preselected particles are shown in \autoref{fig:placeholder-mass}.

Particle candidates identified in the upstream beamline instrumentation were matched in time to NOvA detector information. Hits created in the NOvA TB detector were grouped in time and space into higher-level objects referred to as ``slices''  corresponding to individual particle interactions. The slicer algorithm~\cite{tdslicer,Dunn01011973} works by finding local maxima in the density of ToF-corrected hits. The algorithm was  originally designed to group hits from a single neutrino interaction, where the hits typically come from multiple daughter particles, but has several tuning parameters that can be adjusted to maximize the efficiency and purity of the slicer. These tuning parameters allowed the algorithm to be used effectively in the TB, ND, and FD.

Calibrated hit slices were reconstructed into tracks using algorithms that were similar to those used in the NOvA ND and FD; the main difference was that vertexing in the Test Beam NOvA detector was skipped, as the vertex was known to great accuracy from the projection of the \wc\ tracks (\autoref{sec:wirechambers}). Detector-based selection cuts are analysis-specific, but the baseline preselection cuts for electrons, pions, and protons of \autoref{tab:detsel} are a rough guide for the subsequent analysis of Test Beam detector particles.

To reconstruct particle trajectories, a 2-point Hough transform~\cite{osti_4746348} was applied to the slices.  The Test Beam vertex and the Hough Transform output were input to an adaptation of the FuzzyK algorithm~\cite{Bezdek1981} that created collections of hits with a vertex and direction, called a ``prong." Prongs were used as proxies for electron candidates in the detector. For pion and proton candidates, tracks were then formed using the BPF algorithm~\cite{BreakPointFitter,breakpoint} and these tracks were used to explore refinements to the particle candidate selection criteria. BPF tracks were not used for electron candidates, as the algorithm had not been optimized for electrons. 

\begin{table}[ht]
\caption{The preliminary particle ID selection cuts for Test Beam particles in the beamline instrumentation and in the NOvA detector. These selections form a loose baseline for later analyses. The beamline instrumentation cuts are on the reconstructed mass $m$, time of flight $t$, and Cherenkov detector activity ckov; the NOvA detector cut is on the number of detector hits, N hits. The \tof\ cuts differ in Period 2 because a different experimental setup was used, with a longer baseline between the \tof\ detectors (\autoref{sec:tof}). The preselected particle counts also include data quality selections on the particles' entry to the magnet (\autoref{sec:tracking-mom}) and a rectangular containment cut applied to the first six planes of the detector to remove muon backgrounds.}
\centering
    \begin{tabular}{l l l l  }
    \toprule
      \textbf{Selection} & $e$ & 
      $\pi/\mu$& 
      $p$ \\

    \midrule
    $m$ (GeV/c$^2$)& N/A  & $<0.3$ & $0.750 - 1.126$ \\
    $t$ (ns) Period 2 & $40 - 50$  & $40 - 50$ & $50 - 90$\\
   $t$ (ns) Periods 3--4 & $30 - 36$   & $30 - 40$  & $37 - 80$\\
    ckov & yes & no & no \\
N hits & $\geq$5 & $\geq$5 & $\geq$5 \\
      \midrule
     \textbf{Count} &  \textbf{1067} &  \textbf{9196} &  \textbf{8477}\\
    \bottomrule
    \end{tabular}
 
    \label{tab:detsel}
\end{table}

\begin{figure}
\centering
 \includegraphics[width=.64\textwidth]{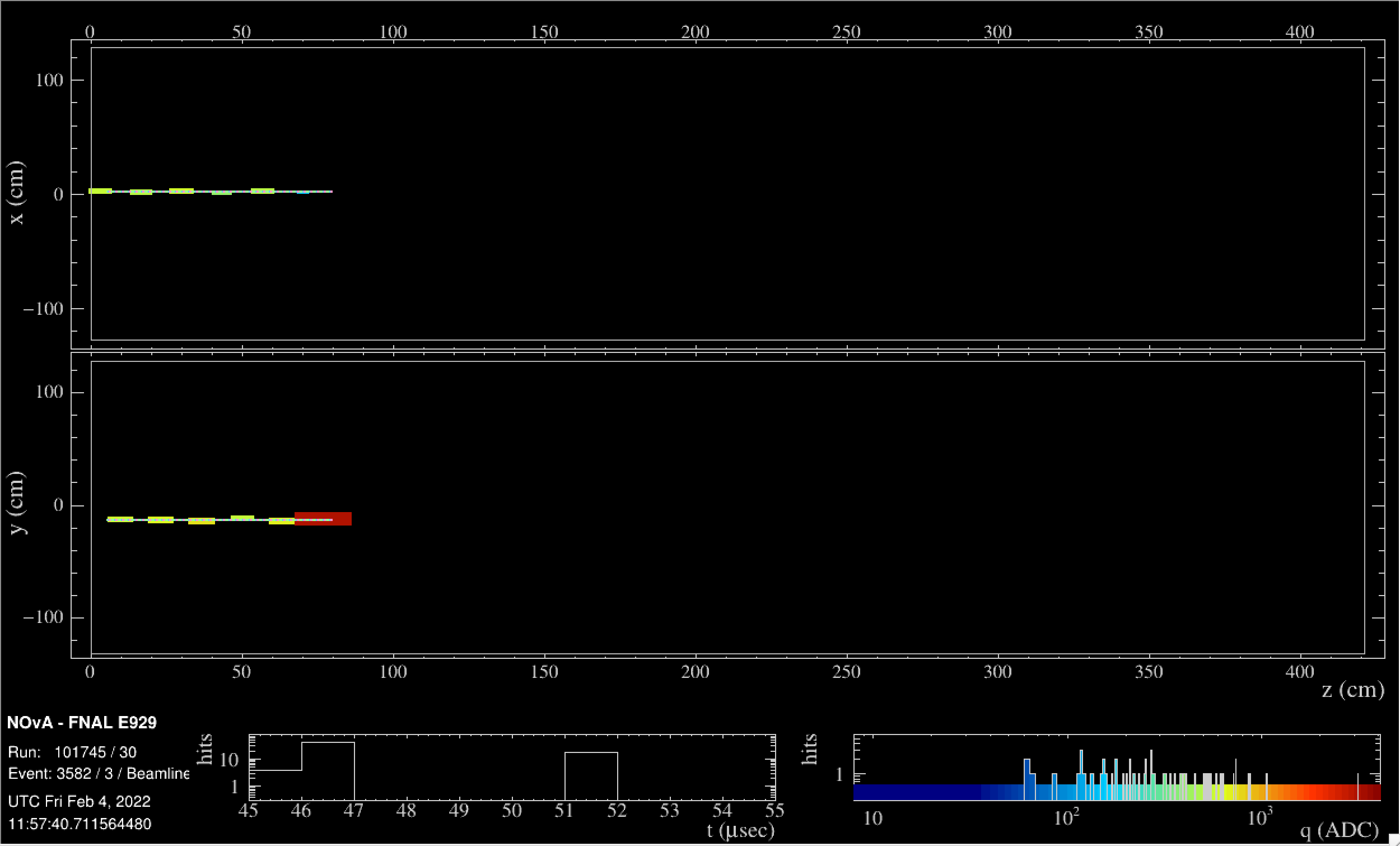} \\
  \includegraphics[width=0.64\textwidth]{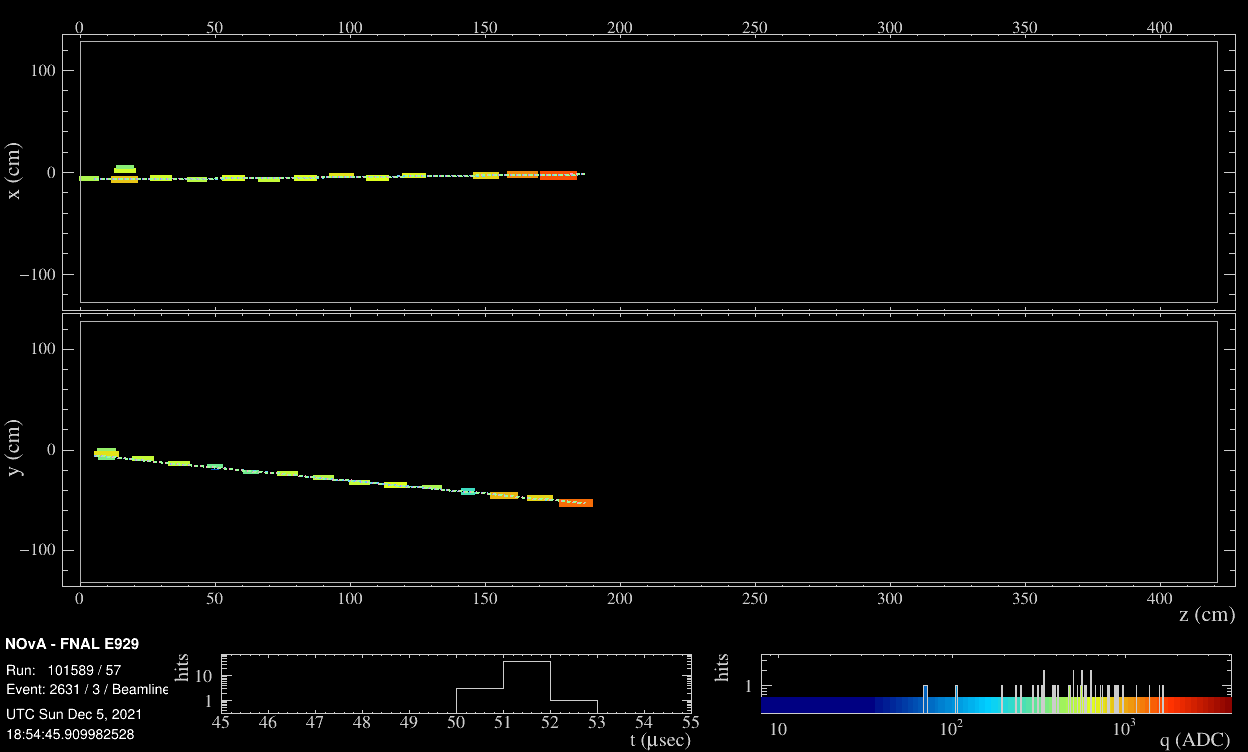}\\
  \includegraphics[width=0.64\textwidth]{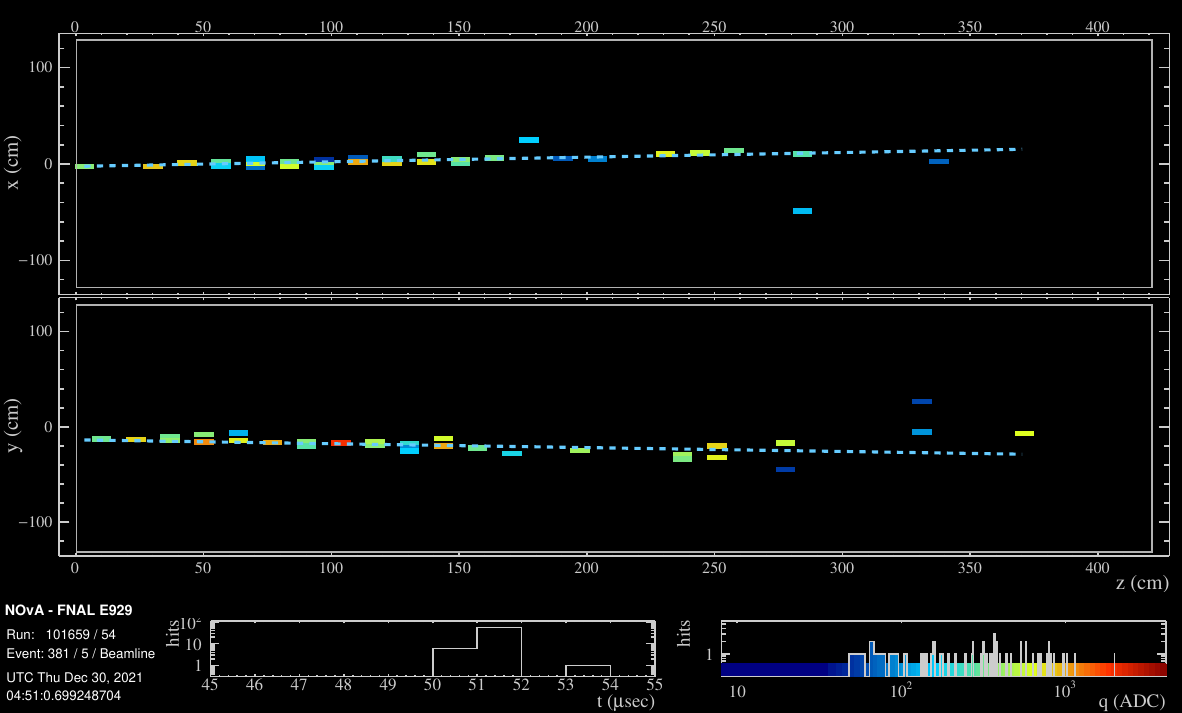}  
 \caption{Event displays of a proton (top), pion (middle), and electron (bottom) candidate entering the detector from the left. The top (bottom) panel in each display shows the view from above (from the side). Hits are shown as colored rectangles where color corresponds to the recorded charge of the hit (bottom-right histogram). The bottom-left histogram shows the hit times in nanoseconds.  The dashed line segments indicate a BPF track in the case of the proton and pion and a reconstructed prong in the case of the electron.}
  \label{fig:proton-ED}
\end{figure}

Event displays of some particles from Period 4 of \datataking\ are shown in~\autoref{fig:proton-ED} for events satisfying the loose selection criteria given in \autoref{tab:detsel}. The proton candidate has a reconstructed momentum of \SI{1.1}{GeV\per c} and \tof\ of \SI{42.4}{ns}. 
The $\pi/\mu$ candidate has a reconstructed momentum of \SI{1.1}{GeV\per c} and ToF of \SI{32.3}{ns}. The positron candidate has a Cherenkov hit, a reconstructed momentum of \SI{1.1}{GeV\per c}, and a reconstructed ToF of \SI{32.5}{ns}.

\section{Conclusions and Outlook}
\label{sec:outlook}
\noindent
The NOvA Test Beam program provides an invaluable opportunity for testing our calibration, reconstruction, and particle identification procedures.
This paper has summarized the design, operation, calibration, and measurement potential of the NOvA Test Beam detectors. The data processed as a result of this program are being analyzed to provide particle-dependent energy scales for electrons, pions, and protons; these will be used to extract a better estimate  of systematic uncertainties in future NOvA measurements, including those of the neutrino oscillation parameters.
 \section*{Acknowledgements}

This document was prepared by the NOvA collaboration using the resources of the Fermi National Accelerator Laboratory (Fermilab), a U.S. Department of Energy, Office of Science, HEP User Facility. Fermilab is managed by Fermi Forward Discovery Group, LLC, acting under Contract No. 89243024CSC000002.  This work was supported by the U.S. Department of Energy; the U.S. National Science Foundation; the Department of Science and Technology, India; the European Research Council; the MSMT CR, GA UK, Czech Republic; the RAS, the Ministry of Science and Higher Education, and RFBR, Russia; CNPq and FAPEG, Brazil; UKRI, STFC and the Royal Society, United Kingdom; and the state and University of Minnesota.  We are grateful for the contributions of the staffs of the University of Minnesota at the Ash River Laboratory, and of Fermilab. For the purpose of open access, the author has applied a Creative Commons Attribution (CC BY) license to any Author Accepted Manuscript version arising.

 
 \bibliographystyle{elsarticle-num} 
 \bibliography{references}

\end{document}